\documentclass[a4paper,12pt]{article}
\usepackage[utf8]{in
    putenc}
\usepackage{cancel}
\usepackage{ulem}
\usepackage{amsfonts}
\usepackage{amssymb}
\usepackage{graphicx}
\usepackage{amsmath,bm}

\usepackage{enumerate}
\usepackage{mathtools}
\usepackage{tikz} \usetikzlibrary{calc}
\usepackage{subfloat}
\usepackage{caption} 
\usepackage{subcaption}
\usepackage{xcolor}
\usepackage{float}
\usepackage{color}
\usepackage{here}
\usepackage{cite}
\usepackage{multirow}
\newcommand\tablex{2.5mm}
\newcommand\tabley{1mm}

\usepackage{mathrsfs}
\usepackage{float,epsfig}
\usepackage{dcolumn}
\usepackage{pgfplots}
\usepackage{graphicx}
\usepackage{bm}
\usepackage{amsmath,amssymb,amsthm}
\usepackage[colorlinks=true,linkcolor=violet,citecolor=red]{hyperref}
\usepackage{graphicx}
\usepackage{caption}
\usepackage{subcaption}
\usepackage{tikz}
\usepackage{pgfplots}
\usetikzlibrary{spy,calc}
\usepackage{hyperref}

\newif\ifblackandwhitecycle
\gdef\patternnumber{0}

\pgfkeys{/tikz/.cd,
    zoombox paths/.style={
        draw=orange,
        very thick
    },
    black and white/.is choice,
    black and white/.default=static,
    black and white/static/.style={ 
        draw=white,   
        zoombox paths/.append style={
            draw=white,
            postaction={
                draw=black,
                loosely dashed
            }
        }
    },
    black and white/static/.code={
        \gdef\patternnumber{1}
    },
    black and white/cycle/.code={
        \blackandwhitecycletrue
        \gdef\patternnumber{1}
    },
    black and white pattern/.is choice,
    black and white pattern/0/.style={},
    black and white pattern/1/.style={    
            draw=white,
            postaction={
                draw=black,
                dash pattern=on 2pt off 2pt
            }
    },
    black and white pattern/2/.style={    
            draw=white,
            postaction={
                draw=black,
                dash pattern=on 4pt off 4pt
            }
    },
    black and white pattern/3/.style={    
            draw=white,
            postaction={
                draw=black,
                dash pattern=on 4pt off 4pt on 1pt off 4pt
            }
    },
    black and white pattern/4/.style={    
            draw=white,
            postaction={
                draw=black,
                dash pattern=on 4pt off 2pt on 2 pt off 2pt on 2 pt off 2pt
            }
    },
    zoomboxarray inner gap/.initial=5pt,
    zoomboxarray columns/.initial=2,
    zoomboxarray rows/.initial=2,
    subfigurename/.initial={},
    figurename/.initial={zoombox},
    zoomboxarray/.style={
        execute at begin picture={
            \begin{scope}[
                spy using outlines={%
                    zoombox paths,
                    width=\imagewidth / \pgfkeysvalueof{/tikz/zoomboxarray columns} - (\pgfkeysvalueof{/tikz/zoomboxarray columns} - 1) / \pgfkeysvalueof{/tikz/zoomboxarray columns} * \pgfkeysvalueof{/tikz/zoomboxarray inner gap} -\pgflinewidth,
                    height=\imageheight / \pgfkeysvalueof{/tikz/zoomboxarray rows} - (\pgfkeysvalueof{/tikz/zoomboxarray rows} - 1) / \pgfkeysvalueof{/tikz/zoomboxarray rows} * \pgfkeysvalueof{/tikz/zoomboxarray inner gap}-\pgflinewidth,
                    magnification=3,
                    every spy on node/.style={
                        zoombox paths
                    },
                    every spy in node/.style={
                        zoombox paths
                    }
                }
            ]
        },
        execute at end picture={
            \end{scope}
            \node at (image.north) [anchor=north,inner sep=0pt] {\subcaptionbox{\label{\pgfkeysvalueof{/tikz/figurename}-image}}{\phantomimage}};
            \node at (zoomboxes container.north) [anchor=north,inner sep=0pt] {\subcaptionbox{\label{\pgfkeysvalueof{/tikz/figurename}-zoom}}{\phantomimage}};
     \gdef\patternnumber{0}
        },
        spymargin/.initial=0.5em,
        zoomboxes xshift/.initial=1,
        zoomboxes right/.code=\pgfkeys{/tikz/zoomboxes xshift=1},
        zoomboxes left/.code=\pgfkeys{/tikz/zoomboxes xshift=-1},
        zoomboxes yshift/.initial=0,
        zoomboxes above/.code={
            \pgfkeys{/tikz/zoomboxes yshift=1},
            \pgfkeys{/tikz/zoomboxes xshift=0}
        },
        zoomboxes below/.code={
            \pgfkeys{/tikz/zoomboxes yshift=-1},
            \pgfkeys{/tikz/zoomboxes xshift=0}
        },
        caption margin/.initial=4ex,
    },
    adjust caption spacing/.code={},
    image container/.style={
        inner sep=0pt,
        at=(image.north),
        anchor=north,
        adjust caption spacing
    },
    zoomboxes container/.style={
        inner sep=0pt,
        at=(image.north),
        anchor=north,
        name=zoomboxes container,
        xshift=\pgfkeysvalueof{/tikz/zoomboxes xshift}*(\imagewidth+\pgfkeysvalueof{/tikz/spymargin}),
        yshift=\pgfkeysvalueof{/tikz/zoomboxes yshift}*(\imageheight+\pgfkeysvalueof{/tikz/spymargin}+\pgfkeysvalueof{/tikz/caption margin}),
        adjust caption spacing
    },
    calculate dimensions/.code={
        \pgfpointdiff{\pgfpointanchor{image}{south west} }{\pgfpointanchor{image}{north east} }
        \pgfgetlastxy{\imagewidth}{\imageheight}
        \global\let\imagewidth=\imagewidth
        \global\let\imageheight=\imageheight
        \gdef\columncount{1}
        \gdef\rowcount{1}
        
    },
    image node/.style={
        inner sep=0pt,
        name=image,
        anchor=south west,
        append after command={
            [calculate dimensions]
            node [image container,subfigurename=\pgfkeysvalueof{/tikz/figurename}-image] {\phantomimage}
            node [zoomboxes container,subfigurename=\pgfkeysvalueof{/tikz/figurename}-zoom] {\phantomimage}
        }
    },
    color code/.style={
        zoombox paths/.append style={draw=#1}
    },
    connect zoomboxes/.style={
    spy connection path={\draw[draw=none,zoombox paths] (tikzspyonnode) -- (tikzspyinnode);}
    },
    help grid code/.code={
        \begin{scope}[
                x={(image.south east)},
                y={(image.north west)},
                font=\footnotesize,
                help lines,
                overlay
            ]
            \foreach \x in {0,1,...,9} { 
                \draw(\x/10,0) -- (\x/10,1);
                \node [anchor=north] at (\x/10,0) {0.\x};
            }
            \foreach \y in {0,1,...,9} {
                \draw(0,\y/10) -- (1,\y/10);                        \node [anchor=east] at (0,\y/10) {0.\y};
            }
        \end{scope}    
    },
    help grid/.style={
        append after command={
            [help grid code]
        }
    },
}

\newcommand\phantomimage{%
    \phantom{%
        \rule{\imagewidth}{\imageheight}%
    }%
}
\newcommand\zoombox[2][]{
    \begin{scope}[zoombox paths]
        \pgfmathsetmacro\xpos{
            (\columncount-1)*(\imagewidth / \pgfkeysvalueof{/tikz/zoomboxarray columns} + \pgfkeysvalueof{/tikz/zoomboxarray inner gap} / \pgfkeysvalueof{/tikz/zoomboxarray columns} ) + \pgflinewidth
        }
        \pgfmathsetmacro\ypos{
            (\rowcount-1)*( \imageheight / \pgfkeysvalueof{/tikz/zoomboxarray rows} + \pgfkeysvalueof{/tikz/zoomboxarray inner gap} / \pgfkeysvalueof{/tikz/zoomboxarray rows} ) + 0.5*\pgflinewidth
        }
        \edef\dospy{\noexpand\spy [
            #1,
            zoombox paths/.append style={
                black and white pattern=\patternnumber
            },
            every spy on node/.append style={#1},
            x=\imagewidth,
            y=\imageheight
        ] on (#2) in node [anchor=north west] at ($(zoomboxes container.north west)+(\xpos pt,-\ypos pt)$);}
        \dospy
        \pgfmathtruncatemacro\pgfmathresult{ifthenelse(\columncount==\pgfkeysvalueof{/tikz/zoomboxarray columns},\rowcount+1,\rowcount)}
        \global\let\rowcount=\pgfmathresult
        \pgfmathtruncatemacro\pgfmathresult{ifthenelse(\columncount==\pgfkeysvalueof{/tikz/zoomboxarray columns},1,\columncount+1)}
        \global\let\columncount=\pgfmathresult
        \ifblackandwhitecycle
            \pgfmathtruncatemacro{\newpatternnumber}{\patternnumber+1}
            \global\edef\patternnumber{\newpatternnumber}
        \fi
    \end{scope}
}

\definecolor{lime}{HTML}{A6CE39}
\newcommand{\orcidicon}{%
    \begin{tikzpicture}
    \draw[lime, fill=lime] (0,0)
        circle [radius=0.16]
        node[white] {{\fontfamily{qag}\selectfont \tiny ID}};
    \draw[white, fill=white] (-0.0625,0.095)
        circle [radius=0.007];
    \end{tikzpicture}   \hspace{-2mm}
}

\newcommand\orcidHasan{{\href{https://orcid.org/0000-0001-7408-0910}{\orcidicon}}}
\newcommand\orcidLahoucine{{\href{https://orcid.org/0000-0002-0143-5140}{\orcidicon}}}
\newcommand\orcidKarima{{\href{https://orcid.org/0000-0001-5419-8516}{\orcidicon}}}

\title{\bf  Signatures of the accelerating black holes with a cosmological constant from the  $\textrm{Sgr~A}^\star$ and $\textrm{M87}^\star$  shadow prospects  
}

\author{
L. CHAKHCHI\orcidLahoucine\!\!$^{1}$\thanks{lahoucine.chakhchi@edu.uiz.ac.ma} ,  
 H.  El Moumni\orcidHasan\!\!$^1$\thanks{h.elmoumni@uiz.ac.ma (Corresponding author)} , K. Masmar\orcidKarima\!\!$^2$\thanks{karima.masmar@gmail.com}
\\
{\small $^{1}$ LPTHE, Physics Department, Faculty of Sciences, Ibnou Zohr University, Agadir, Morocco. }\\
{\small $^{2}$Laboratory of  High Energy Physics and Condensed Matter
HASSAN II University,}\\{\small Faculty of Sciences Ain Chock, Casablanca, Morocco.}\\
}

\vspace{-3.6em}

\date{}
\begin{document} 
\maketitle
\begin{abstract}

Recently, the Event Horizon Telescope (EHT) achieved the realization of an image of the supermassive black hole $\textrm{Sgr~A}^\star$ showing an angular shadow diameter $\mathcal{D}= 48.7 \pm 7\mu as$ and the fractional deviation $\bm{\delta} = -0.08^{+0.09}_{-0.09}~\text{(VLTI)},-0.04^{+0.09}_{-0.10}~\text{(Keck)}$, alongside the earlier image of $\textrm{M87}^\star$ with angular diameter $ \mathcal{D}=42 \pm 3 \mu as$, deviation $\bm{\delta}=-0.01^{+0.17}_{-0.17}$ and deviations from circularity estimated to be $\Delta \mathcal{C}\lesssim 10\%$. In addition, the shadow radii are assessed within the ranges $3.38 \le \frac{r_{\text{s}}}{M} \le 6.91$ for $\textrm{M87}^\star$ and $3.85 \le \frac{r_{\text{s}}}{M} \le 5.72$ as well as $3.95 \le \frac{r_{\text{s}}}{M} \le 5.92$ for $\textrm{Sgr~A}^\star$ using the Very Large Telescope Interferometer (VLTI) and Keck observatories, respectively. These values are provided with $1$-$\sigma$ and $2$-$\sigma$ measurements.  Such realizations can unveil a better comprehension of gravitational physics at the horizon scale. In this paper, we use the EHT observational results for $\textrm{M87}^\star$ and $\textrm{Sgr~A}^\star$ to elaborate the constraints on parameters of accelerating black holes with a cosmological constant. Concretely,  we utilize the mass and distance of both black holes to derive the observables associated with the accelerating black hole shadow.

First, we compare our findings with observed quantities such as angular diameter, circularity, shadow radius, and the fractional deviation from the $\textrm{M87}^\star$ data. This comparison reveals constraints within the acceleration parameter and the cosmological constant. In fact, the acceleration parameter $\mathcal{A}$ is found to be in the interval $[0,0.475^{+0.0125}_{-0.0125}]$, the cosmological constant spans the range of $\Lambda\in[-14.5801^{+1.1051\times10^{-6}}_{-1.1051\times10^{-6}},\\ (1.0086\times10^{-6})^{+9.2121\times10^{-5}}_{-1.5396\times10^{-4}}]$.
Then by considering the $\textrm{Sgr~A}^\star$  data, the acceleration scheme persists with the same interval range, while, the cosmological constant shifts very slightly to the following domain $[-14.7266^{+1.1051\times10^{-6}}_{-1.1051\times10^{-6}},(4.6200\times10^{-5})^{+6.7890\times10^{-5}}_{-1.5888\times10^{-4}}]$ for the VLTI observatory. Moreover, within the Keck measurements, one can find $\Lambda\in[-14.7266^{+1.1051\times10^{-6}}_{-1.1051\times10^{-6}},(2.4149\times10^{-5})^{+7.2836\times10^{-5}}_{-1.5094\times10^{-4}}]$. Besides, for the others parameters, both black holes reveal the range of the spin parameter $a$ in $[0,1]$ and electrical/magnetic charges are $\{e,g\}\in[0,0.09]$.

Further, only the cosmological constant intervals are altered by $\frac{r_{\text{s}}}{M}$ measurements, precisely within the $\textrm{M87}^\star$ data, the upper bound shifts to $(\Lambda_{\text{1-}\sigma} = 1.1361\times10^{-4}, \Lambda_{ \text{2-}\sigma} = 1.7822\times10^{-4})$, the VLTI gives $(\Lambda_{\text{1-}\sigma} = 1.5457\times10^{-4}, \Lambda_{ \text{2-}\sigma} = 1.8208\times10^{-4})$, while the Keck results uncover $\Lambda_{\text{2-}\sigma} = 1.8015\times10^{-4}$.

Lastly, one cannot rule out the possibility of the negative values for the cosmological constant on the emergence of accelerated black hole solutions within the context of minimal gauged supergravity. This intriguing possibility raises the prospect of such solutions being plausible candidates for astrophysical black holes.



\end{abstract}

\tableofcontents

	\section{Introduction}
\label{sec:Intro}
\paragraph*{}Recently, there has been significant interest in the study of astrophysical black holes, driven by groundbreaking discoveries in the field. The Event Horizon Telescope collaboration, utilizing very long baseline interferometry, achieved a major breakthrough by reconstructing images of the event horizon scale of supermassive black hole candidates in the centers of giant galaxies such as $\textrm{M87}$ and the $\textrm{Milky Way}$ \cite{EventHorizonTelescope:2019dse,EventHorizonTelescope:2019ths,EventHorizonTelescope:2022wkp,EventHorizonTelescope:2022apq}. Additionally, the LIGO and VIRGO observatories made an early announcement of the detection of gravitational waves \cite{LIGOScientific:2016emj}, providing further evidence for the existence of black holes as predicted by Einstein's general relativity.
According to this theory, black holes possess a defining property at their event horizons and in the surrounding photon region, which creates a dark region in the observer's celestial sky. This dark region, known as a shadow \cite{Bardeen:1973tla,Falcke:1999pj}, is an optical phenomenon cast by a black hole when it appears in front of a distant luminous source.
These remarkable achievements have significantly advanced our understanding of black holes and provided direct confirmation of their existence, aligning with the predictions of Einstein's general relativity. The study of black holes continues to be a captivating and active area of research in astrophysics \cite{Fathi:2023ccx,Saadati:2023jym,Hoshimov:2023tlz,Gao:2023mjb,Gregoris:2023wrg,Zahid:2023csk,Ovgun:2023wmc,Ovgun:2023ego,Zubair:2023cor,Magos:2023nnb,Lambiase:2023zeo,Yang:2023agi}.

 The well-known vacuum C-metric, initially introduced in \cite{Weyl:1917gp}, has been extensively examined in subsequent works \cite{newman1961new, Robinson:1962zz}. Building upon this, a charged C-metric was developed in \cite{PhysRevD.2.1359}. The C-metric can be interpreted as two uniformly accelerating black holes influenced by forces originating from conical singularities \cite{Griffiths:2006tk}. Furthermore, an accelerating black hole with both charge and rotation was constructed in \cite{Plebanski:1976gy}.
The C-metric, with either a negative or positive cosmological constant, has been utilized as a background for investigating gravitationally and electromagnetically radiative fields near time-like or space-like infinity \cite{Podolsky:2003gm,Krtous:2003tc}. Numerous studies have further contributed to these investigation lines, often relying on the C-metric \cite{Kinnersley:1970zw,Dias:2002mi,Dias:2003xp,Griffiths:2005qp,EslamPanah:2022ihg,EslamPanah:2023ypz,EslamPanah:2024dfq}. This metric has been employed not only to study the pair creation of black holes \cite{Dowker:1993bt}, the splitting of cosmic strings \cite{Gregory:1995hd,Eardley:1995au}, but also to construct black rings in five-dimensional gravity, motivated by string theory and related models \cite{Emparan:2001wn}. Moreover, the thermodynamics of accelerating black holes \cite{Appels:2016uha}, regular solutions \cite{Astorino:2016ybm}, supersymmetric solutions \cite{Cassani:2021dwa}, and the generalization of C-metric to varying conical deficits, as well as their applications to holographic heat engines, have been extensively studied in literature  \cite{Anabalon:2018ydc,Anabalon:2018qfv, Zhang:2018vqs,Zhang:2018hms,Rostami:2019ivr,Belhaj:2020lai,Barrientos:2023tqb,Barrientos:2023dlf,Barrientos:2022bzm}.
Recent studies suggest that accelerating black holes could evolve into supermassive black holes \cite{Vilenkin:2018zol,Gussmann:2021mjj}. However, if these black holes played a role in structure formation, their velocity and consequently the acceleration should be small \cite{Vilenkin:2018zol}. 

In other words, the study of accelerating black holes and the C-metric is crucial for advancing our understanding of black hole physics and its applications in various branches of astrophysics. It helps us unravel the intricate dynamics and peculiarities of black holes under acceleration, leading to valuable insights into the fundamental nature of these fascinating cosmic objects.

Thanks to the AdS/CFT correspondence \cite{Maldacena:1997re}, in recent years there has been tremendous progress in theoretical physics. Especially, the class of supersymmetric black holes in AdS spacetime with dimension
$D \geq 4$, there was recognized pertinent progress in elucidating the microscopic degrees of freedom, Bekenstein-Hawking entropy, and statistical ensembles of the dual superconformal field theory \cite{Benini:2015eyy,Benini:2016rke,Benini:2018ywd}. Through such a framework, the issue of counting microstates within holography is essentially reformulated in terms of a supersymmetric field theory path integral in a background with sources \cite{Nian:2019pxj,Hosseini:2017fjo,Benini:2017oxt,Bobev:2019zmz}. In the addition to the crucial role of the cosmological constant in AdS/CFT correspondence, one can't forget its key imprint in the Universe expansion. Indeed, dark energy, one of the most essential puzzles in cosmology, is assumed to drive the current cosmic accelerated expansion. The cosmological constant, $\Lambda$, is among the favored candidates to explain such acceleration \cite{Peebles:2002gy}. Furthermore, in  black hole thermodynamics framework, a particular emphasis has been put on the interplay between such physics on Anti-de-Sitter (AdS) geometries and phase transitions. This has opened numerous ways for further research and provided many results \cite{Kubiznak:2012wp,Cai:2013qga,Belhaj:2015hha,Belhaj:2015uwa,Belhaj:2020nqy,ElMoumni:2018fml}.

The Event Horizon Telescope (EHT) is a Very Long Baseline Interferometry (VLBI) array with Earth-scale coverage \cite{EventHorizonTelescope:2019dse,EventHorizonTelescope:2019uob,EventHorizonTelescope:2019jan,EventHorizonTelescope:2019ths,EventHorizonTelescope:2019pgp,EventHorizonTelescope:2019ggy,EventHorizonTelescope:2021bee,EventHorizonTelescope:2021srq}. It has thus far been able to produce the horizon-scale images of the two supermassive black holes situated at the centers of both  $\textrm{M87}$  and our own Milky Way galaxy while observing the sky at $1.3mm$ wavelength. 

After the first EHT announcement, a variety amount of research has been devoted to investigating the gravitational lensing by a large class of black hole configurations and the confrontation with the extracted information from the EHT black hole shadow image of  $\textrm{M87}^\star$  \cite{Konoplya:2019xmn,Chael:2021rjo,Belhaj:2020okh,Belhaj:2020rdb,Ghasemi-Nodehi:2021ipd,Bacchini:2021fig,Atamurotov:2021hoq,Belhaj:2021rae,Afrin:2021imp,Konoplya:2021slg,Rayimbaev:2022mrk,Guo:2021bhr,Cimdiker:2021cpz,Okyay:2021nnh,Bronzwaer:2021lzo,Liu:2021yev,Shaikh:2021yux} especially, solutions with cosmological constant \cite{Zhao:2015fya,Ishihara:2016sfv,Faraoni:2016wae,He:2017alg,Belhaj:2020kwv,Firouzjaee_2019,Afrin:2021ggx,Hou:2021okc,Maluf:2020kgf}. Afterwards, 
by comparing the observed image of $\textrm{Sgr~A}^\star$ with theoretical predictions, researchers  probed the nature of  various spacetimes and further refine our understanding of  different gravity theories. \cite{Vagnozzi:2019apd,Allahyari:2019jqz,Chabab:2019kfs,Khodadi:2020jij,Chowdhuri:2020ipb,Vagnozzi:2022moj,Ghosh:2022kit,KumarWalia:2022aop,Banerjee:2022iok,Khodadi:2022pqh,Uniyal:2022vdu,Afrin:2022ztr,Shaikh:2022ivr,Banerjee:2022jog,Ghosh:2022mka}

This work aims to investigate the shadow aspects of accelerating black holes with a cosmological constant in four dimensions, utilizing the null geodesic equations of motion derived from the Hamilton-Jacobi formalism. By varying relevant parameters, the corresponding shadow behaviors  and observable quantities are analyzed. Furthermore, a crucial step involves comparing the theoretical findings of this model with the  $\textrm{Sgr~A}^\star$ and $\textrm{M87}^\star$ observational data provided by the Event Horizon Telescope (EHT). This confrontation between theory and observation allows for the validation and refinement of the model, construction of constraints on the black hole parameters, enabling insights into the fundamental properties of black holes and advancing our understanding of their dynamics in the context of acceleration and cosmological constant effects.

The structure of this work is as follows.
In Section \ref{sec2}, we provide a concise overview of accelerating black hole solutions within the framework of minimal gauged supergravity.
Moving on to Section \ref{sec3}, we delve into the geodesic photon equations associated with these solutions, employing Hamilton-Jacobi to draw their shadow silhouettes. Then,  
in Section \ref{sec4}, we explore the behavior of the shadow cast by this black hole and examine how each black hole parameter influences its size and shape.
Subsequently, we compare our findings with data from the Event Horizon Telescope, focusing on both the $\textrm{M87}^\star$ and $\textrm{Sgr\ A}^\star$ black holes. This comparison allows us to elucidate observational constraints related to the cosmological constant, acceleration, and rotation parameters.
The final section is dedicated to drawing conclusions and addressing open questions.

\section{Light behaviour around a charged rotating accelerating black hole with a cosmological constant. }\label{sec2}
The line element of the accelerating  black hole with a cosmological constant reads in Boyer–Lindquist coordinates as \cite{Cassani:2021dwa}
\begin{eqnarray}\nonumber
ds^{2}&=&\frac{1}{H^{2}}[-\frac{Q}{\Sigma}(\frac{1}{\kappa}dt-\chi d\phi)^{2}+\frac{\Sigma}{Q} dr^{2}+\frac{\Sigma}{P} d\theta^{2}\\&+&\frac{P}{\Sigma}(\frac{\chi}{\kappa} dt-(r^{2}+a^{2})\sin^{2}\theta d\phi)^{2}].
\end{eqnarray}
where

\begin{eqnarray}\label{mmetric}
\nonumber
\chi &= &a\sin^{2}\theta ,\\ \nonumber
P(\theta) &=&1-2\mathcal{A} m \cos\theta + (\mathcal{A}^{2}(a^{2}+e^{2}+g^{2})+\frac{\Lambda a^{2}}{3})\cos^{2}\theta ,\\ \nonumber
Q(r) &=&(r^{2}-2mr+a^{2}+e^{2}+g^{2})(1-\mathcal{A}^{2}r^{2})
-\frac{\Lambda r^{2}}{3}(a^{2}+r^{2}).
\end{eqnarray}
and 
\begin{eqnarray}
H(r,\theta) &=&1-\mathcal{A} r \cos\theta,\\ \nonumber
\Sigma(r,\theta) &=&r^{2}+a^{2}\cos^{2}\theta.
\end{eqnarray}

The gauge field is obtained to be 
\begin{eqnarray}
A&=&A_tdt+A_\phi d\phi\\ \nonumber
&=&-e\frac{r}{\Sigma}\left(\frac{1}{\kappa}dt-a\sin^2\theta d\phi\right)+q\frac{\cos\theta}{\Sigma}\left(\frac{a}{\kappa}dt-(r^2+a^2)d\phi\right).
\end{eqnarray}
 Such a solution is  parameterized by five quantities $m$, $e$, $g$, $a$ and $\mathcal{A}$, which stand for mass, electric charge, magnetic charge, angular momentum and acceleration parameter, respectively, additionally $\Lambda$ is the cosmological constant and the constant $\kappa > 0$  is a trivial constant that can be absorbed in a rescaling of the time coordinate  and help to normalized the Killing vector $\partial_t$ in order to get a first law of thermodynamics. The coordinate $\theta$ is allowed to vary in the $0\leq\theta\leq\pi$ interval, where the pole $\theta_-=0$ and $\theta_+=\pi$ indicate the location of the conical deficit. The horizon radius is controlled by  the constraint $0<r_h<1/\mathcal{A}$.

To investigate the black hole shadow in a specific spacetime, it is crucial to analyze the motion of test particles. In this context, the presence of a luminous background surrounding the black hole leads to gravitational lensing effects on photons emitted from a distant source, resulting in their deflection before reaching an observer \cite{Islam:2020xmy,Kumar:2020owy}. The behavior of these photons, depending on their energy, can involve scattering, capture, or unstable orbits within the gravitational field \cite{Chandrasekhar:1985kt,Kumar:2020yem}. By tracing the trajectories of these photons using Carter's separable method and solving the Hamilton-Jacobi equation, the null-like geodesic equation for photons in the accelerating black hole spacetime with a cosmological constant can be obtained \cite{Grenzebach:2015oea,Zhang:2020xub}

\begin{eqnarray}\nonumber
\frac{\Sigma}{H^{2}} \frac{dt}{d\tau} &=&\frac{\kappa\chi(L_{z}-\kappa\chi E)}{P(\theta) \sin^{2}\theta}+\frac{\kappa(\Sigma+a\chi)((\Sigma+a\chi)\kappa E-a L_{z})}{Q(r)},\\
\frac{\Sigma}{H^{2}} \frac{d\phi}{d\tau} &=&\frac{L_{z}-\kappa\chi E}{P(\theta) \sin^{2}\theta}+\frac{a((\Sigma+a\chi)\kappa E-a L_{z})}{Q(r)},
\end{eqnarray}
\begin{eqnarray}\nonumber
\left(\frac{\Sigma}{H^{2}}\right)^{2} \left(\frac{d\theta}{d\tau}\right)^{2} &=&P(\theta) K-\frac{(\kappa\chi E-L_{z})^{2}}{\sin^{2}\theta}=\Theta,\\ \nonumber
\left(\frac{\Sigma}{H^{2}}\right)^{2} \left(\frac{dr}{d\tau}\right)^{2} &=&[(\Sigma+a\chi)\kappa E-a L_{z}]^{2}-Q(r) \mathcal{K} = R.
\end{eqnarray}

In the equations provided, $\tau$ represents the affine parameter, while $E$ and $L_{z}$ represent the energy and angular momentum associated with the Killing vectors, respectively. The Carter constant $\mathcal{K}$ is also introduced \cite{Carter:1968rr}. The longitudinal effective potential $\Theta(\theta)$ and the radial effective potential $R(r)$ are defined. The motion of photons is physically constrained by $R(r)\geq 0$ and $\Theta(\theta)\geq 0$. Additionally, if the constraints $R'(r)=0$ and $\dot{\Theta}(\theta)=0$ are satisfied, the photon's motion is confined to a circular orbit with a constant latitudinal angle. Denoting the radius and latitude as $r_{p}$ and $\theta_{p}$, respectively, the circular orbit of the photon around the accelerating black hole with a cosmological constant in the domain of outer communication (where $Q(r)>0$) can be understood through the following conditions
 \begin{eqnarray}
R(r_{p})=0,\quad
R'(r_{p})=0\quad
\text{and}\quad
R''(r_{p})>0.
\\
 \Theta(\theta_{p})=0,\quad
\dot{\Theta}(\theta_{p})=0\quad
\text{and}\quad
\ddot{\Theta}(\theta_{p})<0.
\end{eqnarray}
The illustration of the black hole shadow geometries requires two
dimensionless impact parameters expressed as
\begin{equation}
 \eta=\frac{\mathcal{K}}{E^{2}}, \qquad\zeta=\frac{L_{z}}{E}.
\end{equation}

\section{ Shadow shapes of a charged, rotating, and accelerating black hole with a cosmological constant}\label{sec3}

In this section, we would like to construct the optical behavior of such black hole solutions and then analyze the effect of the relevant black hole and the observer parameters on the shadow silhouette.
To obtain the contour of the black hole shadow, which is also the critical curve that can be seen by an observer. We recall the following normalized and orthogonal tetrad \cite{Griffiths:2009dfa,Grenzebach:2015oea}
\begin{equation}
\begin{split}
 \hat{e}_{(t)} &= \sqrt{\frac{g_{\phi\phi}}{g_{t\phi}^{2}-g_{tt}g_{\phi\phi}}}\left(\partial_{t} -\frac{g_{t\phi}}{g_{\phi\phi}}\partial_{\phi}\right),\quad
 \hat{e}_{(r)} = \frac{1}{\sqrt{g_{rr}}}\partial_{r}\\
 \hat{e}_{(\theta)} &= \frac{1}{\sqrt{g_{\theta\theta}}}\partial_{ \theta},\quad
 \hat{e}_{(\phi)} = \frac{1}{\sqrt{g_{\phi\phi}}}\partial_{ \phi},
 \end{split}
 \end{equation}
 For the static observer situated at a finite distance $(r_O,\theta)$, where $r_O$ represents the radial position and $\theta$ denotes the inclination angle between the direction of the rotation axis of the accelerating black hole and the static observer, we can employ the ZAMO (Zero Angular Momentum Observer) reference frame. In this tetrad frame, the vector $\hat{e}_t$ is timelike, while the remaining vectors are spacelike within the domain of outer communication. Using this ZAMO tetrad, we can project the four-momentum as measured by the locally static observer in the following manner

\begin{equation}
 p^{(t)}=-p_{\mu}\hat{e}_{(t)}^{\mu},\quad 
 p^{(i)}=p_{\mu}\hat{e}_{(i)}^{\mu},\hspace{7 mm}i=r,\theta,\phi.
\end{equation}
Since the photon is a massless, the linear momentum $\vec{P}$ as a three vector 
relating with $p^{(i)}$ obeys 
$\lvert \vec{P}\rvert=p^{(t)}$ in the observer's frame, and the observation angles $(\alpha,\beta)$can be recalled \cite{Cunha:2016bpi}
\begin{equation}
\begin{split}
p^{(r)}&=\lvert \overrightarrow{P} \rvert\cos\alpha \cos\beta,
\quad
p^{(\theta)}=\lvert \overrightarrow{P} \rvert\sin\alpha,
\quad
p^{(\phi)}=\lvert \overrightarrow{P} \rvert\cos\alpha \sin\beta.
\end{split}
\end{equation}
Which lead to

\begin{equation}
 \sin\alpha=\frac{p^{(\theta)}}{p^{(t)}}\quad \text{ and }
 \tan\beta=\frac{p^{(\phi)}}{p^{(r)}}.
\end{equation}
Using the above relations, we define the Cartesian coordinates $(x,y)$ associated with the apparent position on the plane of the sky for the observer as
\begin{equation}
 x\equiv -r_{O}\beta, \hspace{7mm}y\equiv r_{O}\alpha.
\end{equation}
 
\begin{figure}[!ht]
 \center
  \includegraphics[scale=1]{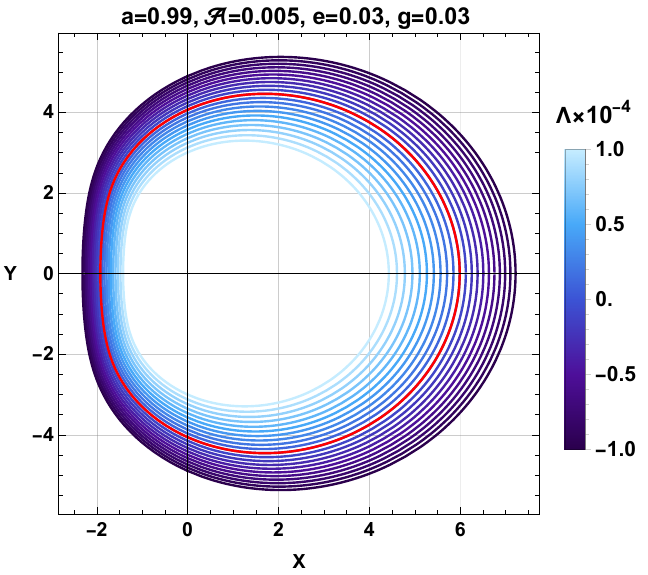}\\
  \caption {\it \footnotesize Shadows contours of the accelerating black hole, as seen by an observer at $(r_O = 100)$ for different values of the cosmological constant $\Lambda$. The red curve is associated with the non-acceleration solution ($\mathcal{A}=0$).}
\label{fig1}
\end{figure}

A detailed examination reveals that the shadow geometrical behaviors can depend on the involved parameters describing the accelerating black holes. In these solutions, the cosmological constant ($\Lambda$) will be considered as a relevant parameter. The corresponding shadow properties are represented in the $(x, y)$ plane by using expressions for $x$ and $y$. At this level, certain parameter values and positions have been fixed. Concretely, the observer is located at $r_O = 100$ and $\theta = \frac{\pi}{2}$ is chosen. It is noted that $m = 1$ and $a=0.99$ have been taken. The associated behaviors are plotted in Figure \ref{fig1}. Concretely, we illustrate the shadow aspects by varying the cosmological constant $\Lambda$.

\begin{figure}[!ht]\centering
\includegraphics[scale=.62]{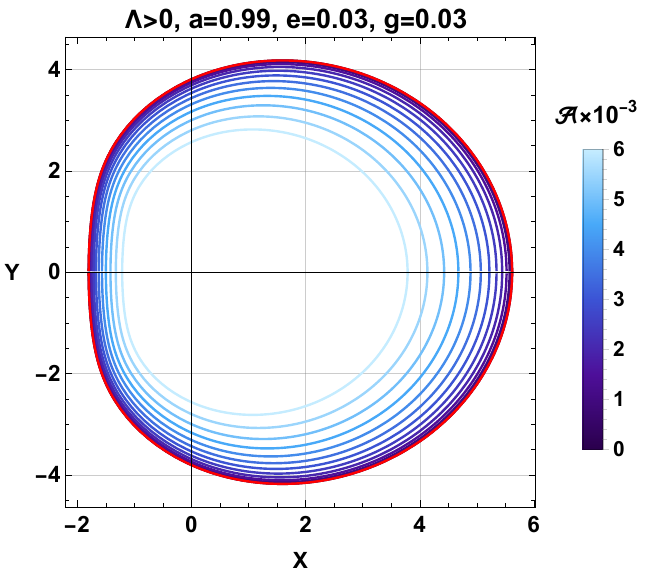}
\includegraphics[scale=.62]{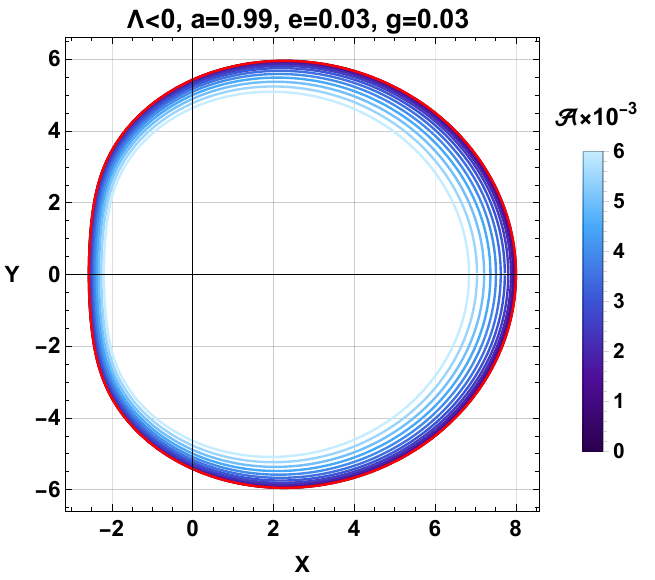}
\includegraphics[scale=.62]{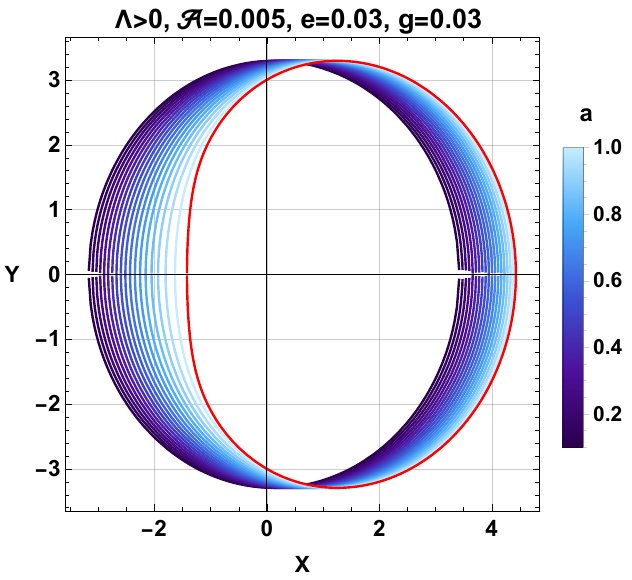}
\includegraphics[scale=.62]{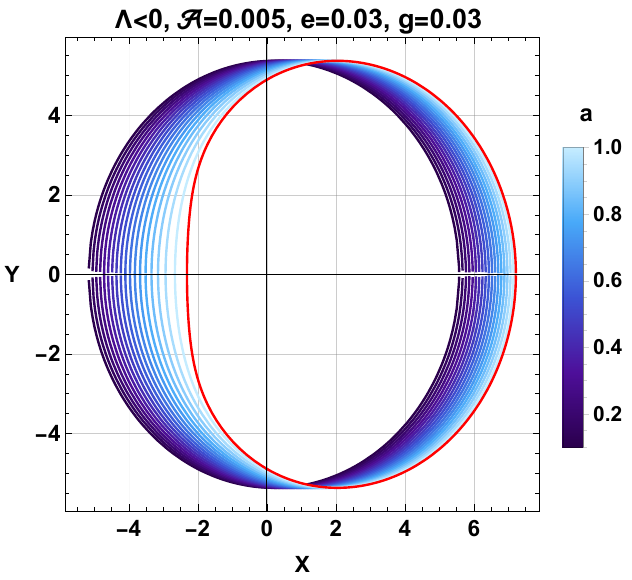}
\includegraphics[scale=.62]{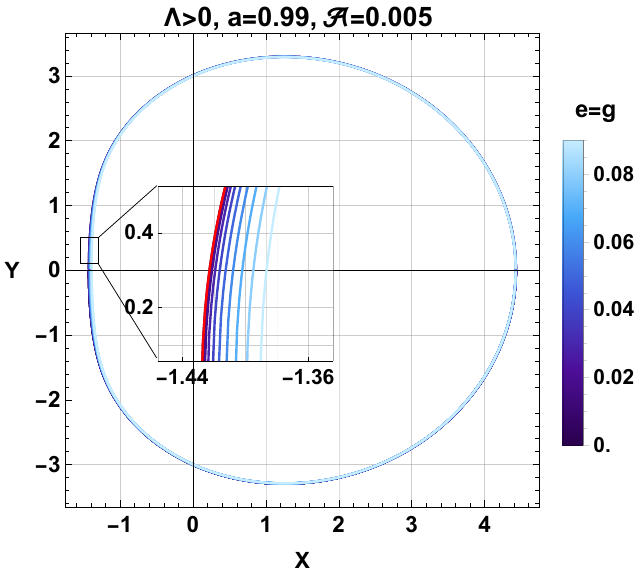}
\includegraphics[scale=.62]{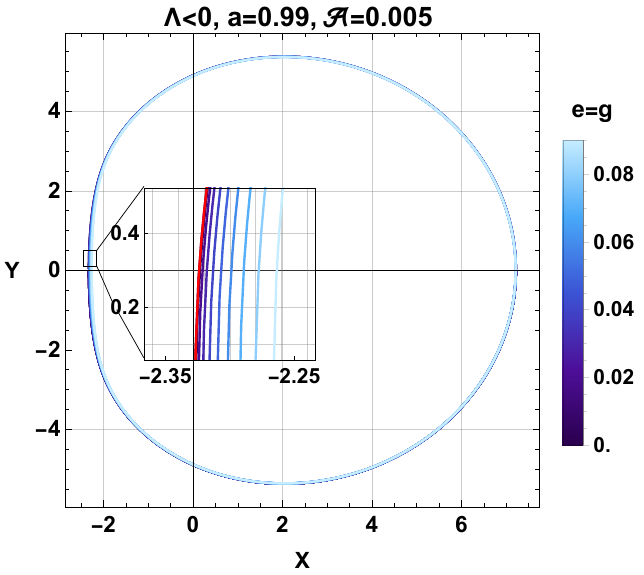}
\caption{\it \footnotesize The observable appearance of the accelerating black hole for different values of the acceleration parameter ({\bf top}), rotation ({\bf middle}), and charge ({\bf bottom}), as perceived by an observer located at $r_O=100$. The panels in the {\bf left} column correspond to a positive cosmological constant ($\Lambda$), while those on the {\bf right} pertain to a negative cosmological constant. In all panels, the red curves represent the non-acceleration solutions ($\mathcal{A}=0$).}
\label{fig2}
\end{figure}

Choosing the rotation parameter $a=0.99$, the black hole shadow exhibits distortion and takes on a characteristic D-shape form. Remarkably, the size of the shadow is found to depend on the sign and magnitude of the cosmological constant ($\Lambda$). Specifically, as the cosmological constant increases, the shadow size decreases. Notably, the sign of the cosmological constant not only affects the spacetime configuration of the black hole but also impacts the observed shape of the shadow. Specifically, shadows associated with negative values of $\Lambda$ (anti-de Sitter) are more prominent compared to those associated with positive values (de Sitter). The red line in the figure corresponds to the case of vanishing cosmological constant, $\Lambda=0$.
As mentioned above, the considered black hole is controlled by many parameters, in Fig.\ref{fig2}, we illustrate the effect of each parameter on the shape and size of the shadow.

By analyzing the first line of Figure \ref{fig2}, it is evident that varying the acceleration parameter $\mathcal{A}$ has a notable impact on the shadow size. Increasing this parameter leads to a decrease in shadow size for both AdS/dS black hole configurations. Interestingly, the largest shadow size is observed in the case of vanishing acceleration parameter $\mathcal{A}$, as indicated by the red curves.

Moving to the second line of the figure, it is observed that the shadow size decreases as the rotating parameter $a$ increases. Additionally, for small values of $a$, the shadows exhibit circular behaviors. However, as the rotation parameter approaches $a\simeq0.7$, the shadows begin to take on a distinct D-shape form, as indicated by the red curves associated with $a=0.99$.

Upon examining the functions $P(\theta)$ and $Q(r)$ in Eq. \eqref{mmetric}, it becomes evident that the electric charge $e$ and magnetic charge $g$ contribute similarly. To investigate their effect on the black hole shadow, we consider the case where $e=g$ in the bottom line of Figure \ref{fig2}. Initially, it may appear that the charges have a negligible impact on the black hole shadow. However, upon closer inspection, zooming on the curves reveals a slight decrease in the shadow size with increasing charges. While the effect is not significant, it indicates a weak subtle influence of the charges on the size of the black hole shadow. Therefore, in what follows, we not will be interested in the effect of the charges.

 Having illustrated the effect of each black hole parameter on shadow size and shape, naturally, we move to investigate the impact of the observer location.
\begin{figure}[!ht] 
 \center
  \includegraphics[scale=.8]{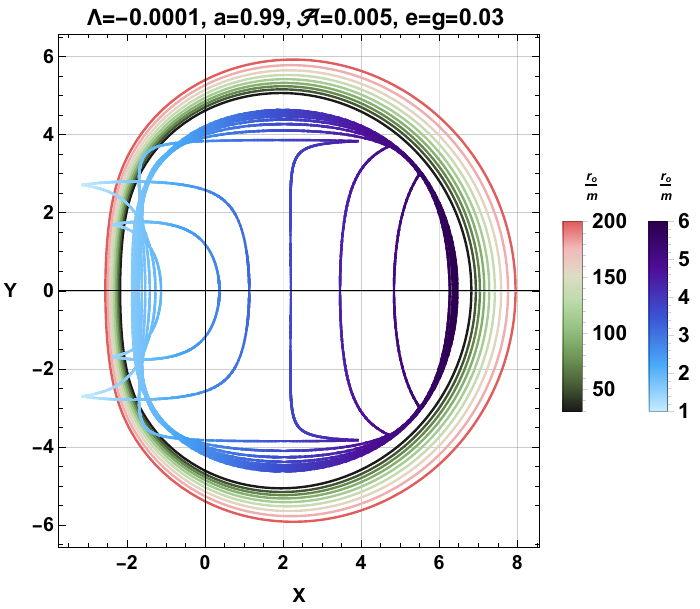}\\
  \caption {  \it \footnotesize 
  Apparent shape of charged-rotating-accelerating-AdS black hole with varying observer distance.}
\label{Shobsrotlmdnaccch}
\end{figure}
For fixed values of black hole parameters, from Fig.\ref{Shobsrotlmdnaccch}, the shadow is found to decrease in size as the static observer advances and becomes increasingly deformed and finally disappears, in this last situation the observer sky becomes totally dark. 

  \section{Analysis of the geometrical observables}\label{sec4}

\begin{figure} [!ht]
		\begin{center}
		\centering
			\begin{tabbing}
			\centering
			\hspace{9.5cm}\=\kill
			\includegraphics[scale=.8]{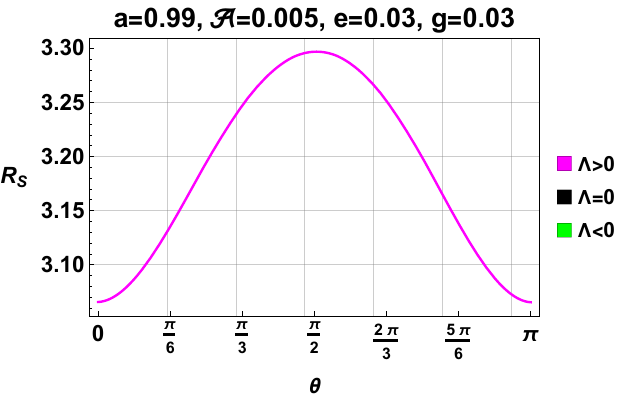}\>
			\includegraphics[scale=.8]{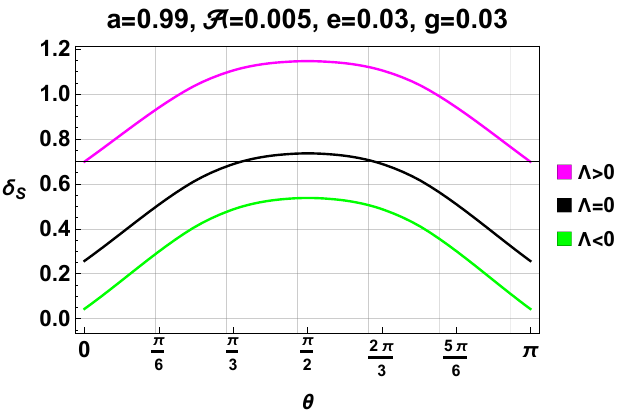} 	
		   \end{tabbing}
\caption{ \it\footnotesize  The behaviour of both size ({\bf left}) and distortion ({\bf right}) of the charged rotating accelerating black hole shadow concerning the observer's inclination angles across various cosmological constant values. The observable $R_s$ is insensible to $\Lambda$ variation, and thus the green, black and magenta are superposed.}\label{fig3}
\end{center}
\end{figure}

To probe the black hole parameters based on observational data, it is essential to investigate the geometrical observables associated with the black hole shadow. The two primary quantities of interest are the size and distortion of the black hole shadow. The size, denoted by $R_s$, represents the radius of a reference circle that corresponds to the black hole shadow, while the quantity $\delta_s$ quantifies the distortion of the black hole shadow in comparison to the reference circle. In the study by Hioki et al., \cite{Hioki:2009na}, the shadow silhouette was approximated using a reference circle defined by the top ($X_t$, $Y_t$), bottom ($X_b$, $Y_b$), and right ($X_r, 0$) extremes of the shadow. The observables were then determined based on this approximation as
\begin{equation}
R_{s} = \frac{(X_{t}-X_{r})^{2}+Y_{t}^{2}}{2(X_{r}-X_{t})},
\text{ and } \delta_{s}
 =\frac{X_{l}-\tilde{X_{l}}}{R_{s}}.
\end{equation}
In which the shadow symmetry concerning the $X$ axis implies $(X_b,Y_b)=(X_t,-Y_t)$. Furthermore, the reference circle and leftmost edge of the shadow silhouette intersect with the $x$ axis at ($\tilde{X}_l$, 0) and ($X_l$, 0), respectively. These geometrical quantities will be discussed graphically in Fig.\ref{fig3}, where their associated calculations are illustrated.

From this figure, it is obvious that both observables, $R_s$ and $\delta_s$, increase as the angle $\theta$ increases, reaching a maximum around $\pi/2$, and then decrease as $\theta$ approaches $\pi$. Notably, the change in the cosmological constant $\Lambda$ does not significantly affect the observable $R_s$. However, the distortion measurement $\delta_s$ is sensitive to variations in $\Lambda$. Specifically, the maximum distortion $\delta_{\text{max,dS}}$ for the de-Sitter case is greater than $\delta_{\text{max,}\Lambda=0}$ (without $\Lambda$) and $\delta_{\text{max,AdS}}$ (anti-de Sitter). This observation aligns with the findings presented by Zhang et al. \cite{Zhang:2020xub}, which indicate that the acceleration of the black hole causes deviations from the equatorial plane and alters the properties of the black hole shadow.

In the Kerr black hole scenario,
 the maximum values of $R_s$ and $\delta_s$ are observed at the equatorial plane ($\theta=\pi/2$). However, the introduction of the black hole's acceleration leads to a deviation of the maximum values of these observables from the $\theta=\pi/2$.

\begin{figure}[!ht]\centering
\includegraphics[scale=.73]{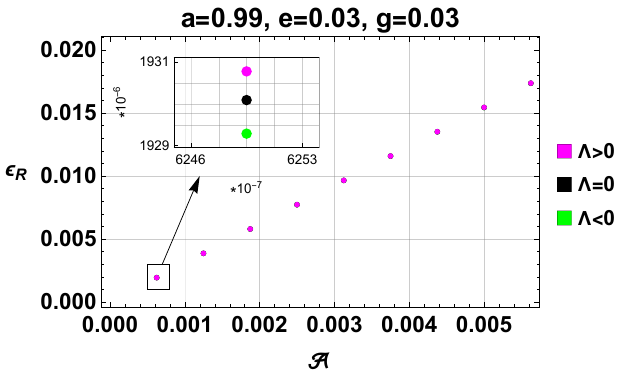}
\includegraphics[scale=.73]{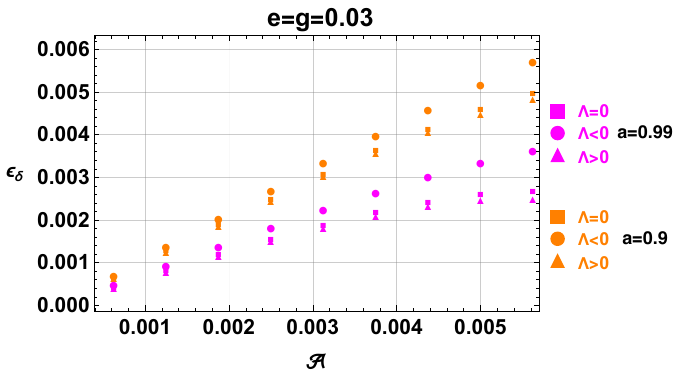}
\caption{\it\footnotesize  {\bf Top:} The variation of the degree of deviation of the inclination angle, which maximizes the shadow radius $R_{s}$ in terms of the acceleration of the black hole. {\bf Bottom: } Graph of the deviation degree of the inclination angle that maximizes the shadow distortion $\delta_s$ as a function of the black hole acceleration.}
\label{fig:BH_pop}
\end{figure}

To quantify this deviation, the introduction of additional parameters becomes necessary\begin{eqnarray}
 \epsilon_{R}&=&\theta_{m}(\mathcal{A}\ne0,R_{s}=R_{max})-\frac{\pi}{2},
\\
 \epsilon_{\delta}&=&\theta_{m}(\mathcal{A}\ne0,\delta_{s}=\delta_{max})-\frac{\pi}{2},
\end{eqnarray}
where $\theta_m$ is the inclination angle of the observer where the observable is a maximum. The variation of such quantities in terms of the accelerating parameters $\mathcal{A}$ is depicted in Fig.\ref{fig:BH_pop}.
The deviation of the shadow radius and the deviation of the shadow distortion increase with increasing black hole acceleration, as distinguished in Fig.\ref{fig:BH_pop}. 
Furthermore, the quantities $\epsilon_R$ and $\epsilon_\delta$ are positive, implying that the black hole acceleration causes the critical inclination angle to deviate from the equatorial plane to one close to the black hole south pole located at ($\theta=\pi$). In addition,  in the bottom diagram of Fig.\ref{fig:BH_pop}, we illustrate just the curves associated with $a=0.99$, due to the very tiny difference between the $a =0.99$ case and $a=0.9$, leading to an arduous distinction of them. The same difficulty is observed when the sign of the cosmological constant is considered, despite the zoom.
 
However for the deviation of the distortion in the bottom diagram, the curve with greater $a$  is greater than the one with a smaller one; Besides such a deviation quantity decreases, passing from the negative to the positive value of the cosmological constant, another key remark is that 
the separation between the plotting points becomes significant
as the acceleration parameter $\mathcal{A}$ grows.
In other words, a higher angular momentum corresponds to a reduced degree of deviation, whereas an increased acceleration of the black hole, along with a negative cosmological constant, results in a greater degree of deviation.

Having studied the shadow of the accelerating black holes with positive, vanishing, and negative cosmological constant configurations, the subsequent section will delve into a more detailed analysis by incorporating relevant data from the Event Horizon Telescope (EHT). This approach will allow for a more comprehensive investigation of the agreement between theoretical predictions and observational data.

\section{Signatures of accelerating black hole within  the 
 $\textrm{M87}^\star$  and $\textrm{S\lowercase{gr}~}A^\star$
 data}

Based on the analysis conducted previously, it is evident that the size and shape of the black hole shadow are influenced by various factors such as the cosmological constant, rotation/acceleration parameters, and charges, albeit to a minor extent. Given this observation, it becomes natural to employ the black hole shadow as a means to constrain the black hole parameters, as the shadow characteristics are intrinsically linked to these parameters. Therefore, we turn our attention to the observables associated with the black hole shadow, namely the shadow angular diameter $\mathcal{D}$, the deviation from circularity $\Delta\mathcal{C}$ and the deviation from Schwarzschild ${\bm\delta}$. By comparing our results with the observational measurements obtained by the Event Horizon Telescope for $\textrm{M87}^\star$ and $\textrm{Sgr\ A}^\star$, we aim to derive constraints on the black hole parameters.

We begin by considering  the angular diameter $\mathcal{D}$ of the shadow, which is the key observable and it is defined as \cite{Banerjee:2019nnj}
 \begin{align}
\mathcal{D}=\frac{G m}{c^2}\left(\frac{\Delta Y }{d}\right)~,
\label{Eq1}
\end{align}
where, $d$ is the distance of $\textrm{M87}^\star$, or $\textrm{Sgr~A}^\star$ from Earth and $\Delta Y$ corresponds to the maximum length of the shadow along the $Y$ direction in the $(X,Y)$ plane, in a direction orthogonal to $X$. In this perspective, one can also introduce another additional  observable, namely, the axis ratio, which obtained as
\begin{align}
\Delta A =\frac{\Delta Y }{\Delta X }.
\label{Eq2}
\end{align}

Similarly to $\Delta Y $, $\Delta X $ represents the maximum separation between two points on the boundary of the black hole shadow along the $X$ direction in the $(X,Y)$ plane, which is orthogonal to $Y$.
Furthermore, we define an additional observable, the dimensionless circularity deviation, which measures the distortion of the shadow from a perfect circle characterized by polar coordinates $\big(\ell(\varphi),\varphi\big)$ \cite{Johannsen:2010ru,Bambi:2019tjh}
\begin{align}
\Delta \mathcal{C}= \dfrac{1}{R_{\rm avg}}\sqrt{\dfrac{1}{2\pi}\int_{0}^{2\pi}d\varphi\left\{ \ell (\varphi)-R_{\rm avg}\right\}^{2}}
\label{Eq4-3}
\end{align}
where, the average radius $R_{\rm avg}$ is defined as,
\begin{align}
R_{\rm avg}=\sqrt{\dfrac{1}{2\pi}\int_{0}^{2\pi}d\phi\, \ell^{2} (\varphi )}
\label{Eq4-4}.
\end{align}
In addition, $\ell(\phi)^{2}=\left\{X(\phi)-X_{i} \right\}^{2}+Y^{2}(\phi)$ being the distance between the geometrical centre ($X_i,Y_i$) of the shadow and any point ($X,Y$) on the shadow surface. 
Besides, The EHT collaboration used $\bm\delta$ as a quantification of deviation between the infrared shadow radius and the Schwarzschild one \cite{EventHorizonTelescope:2022apq}
\begin{equation}\label{ddelta}
{\bm \delta}=\frac{r_\text{s}}{r_\text{Sch}}-1.
\end{equation}

\subsection{Contact with $\textrm{M87}^\star$ data}

According to the reports from the Event Horizon Telescope (EHT) collaboration \cite{EventHorizonTelescope:2019dse,EventHorizonTelescope:2019pgp,EventHorizonTelescope:2019ggy}, the observed parameters for the supermassive black hole $\textrm{M87}^\star$ at the center of the galaxy $\textrm{M}87$ are as follows: an angular diameter $\mathcal{D}=42 \pm 3 \mu$as, a deviation from circularity $\Delta \mathcal{C}\lesssim 10\%$, an axis ratio $\Delta A\lesssim 4/3$, and a fractional deviation from the Schwarzschild black hole shadow diameter ${\bm \delta}$. The values obtained for ${\bm \delta}$ by the EHT \cite{EventHorizonTelescope:2022xqj} are consistent with each other, irrespective of the specific telescope, image processing algorithm, or simulation used. For further details, Tab.\ref{m87_bounds} presents the fractional deviation parameter ${\bm \delta}$ derived from the black hole image of $\textrm{M87}^\star$ \cite{EventHorizonTelescope:2021dqv} in the first column.
 \begin{table}
	\centering
	\begin{tabular}{lccr} 
	& \multicolumn{3}{c}{$\textrm{M87}^\star$ estimates}\\[1mm]\hline\hline\\[-3.5mm]
		\hline
		 & Deviation ${\bm \delta}$ & $\text{1-}\sigma$ bounds & $\text{2-}\sigma$ bounds\\
		\hline
		EHT & $-0.01^{+0.17}_{-0.17}$ & $4.26\le \frac{r_{\text{s}}}{M}\le 6.03$ &  $3.38\le \frac{r_{\text{s}}}{M}\le 6.91$\\
		\hline
	\end{tabular}
	\caption{\it \footnotesize The $\textrm{M87}^\star$  bounds on the deviation parameter $\bm\delta$ and dimensionless shadow radius $\frac{r_\text{s}}{M}$.}
	\label{m87_bounds}
\end{table}
 
The central value of ${\bm \delta}$ is found to be closer to zero, but the larger errors are attributed to the uncertainties in the mass and distance measurements of $\textrm{M87}^\star$. Nevertheless, by utilizing the expression and value of ${\bm \delta}$ from Eq. \eqref{ddelta} and Tab.\ref{m87_bounds}, it is possible to impose constraints on black hole parameters via the dimensionless quantity $r_\text{s}/M$. Furthermore, an inclination angle of $17^\circ$ is predicted within the range of angles that the jet axis makes with the line of sight, assuming the axis of rotation corresponds to the jet axis. According to the EHT collaboration, the measured distance to $\textrm{M87}^\star$ is $d=16.8\pm 0.8~\textrm{Mpc}$, and the mass of the source is $m=(6.5\pm 0.7)\times 10^{9}~M_{\odot}$.

\begin{figure*}\centering
\begin{tikzpicture}
\node[align=left] {
\includegraphics[width=0.33\textwidth]{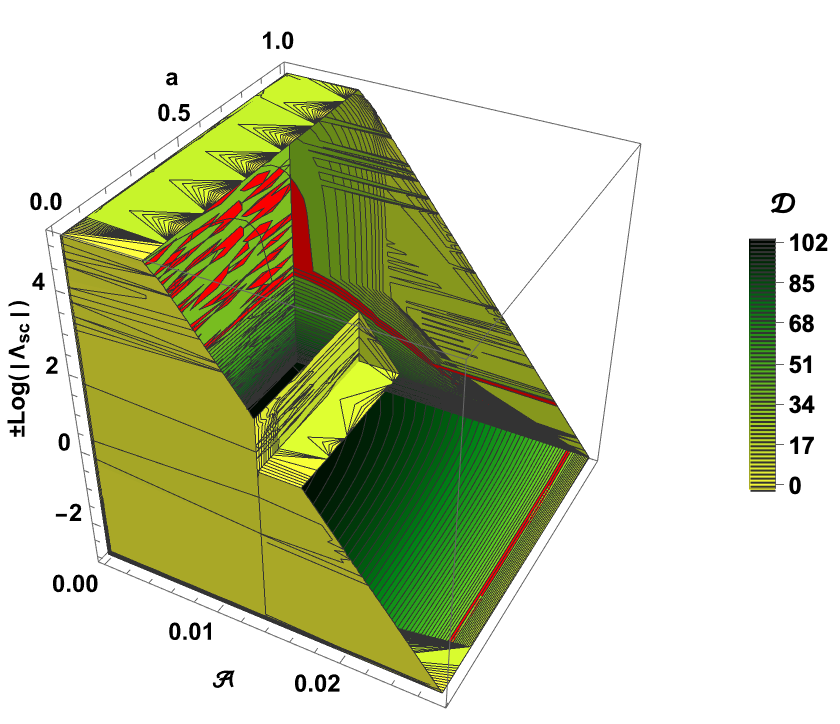}
\includegraphics[width=0.33\textwidth]{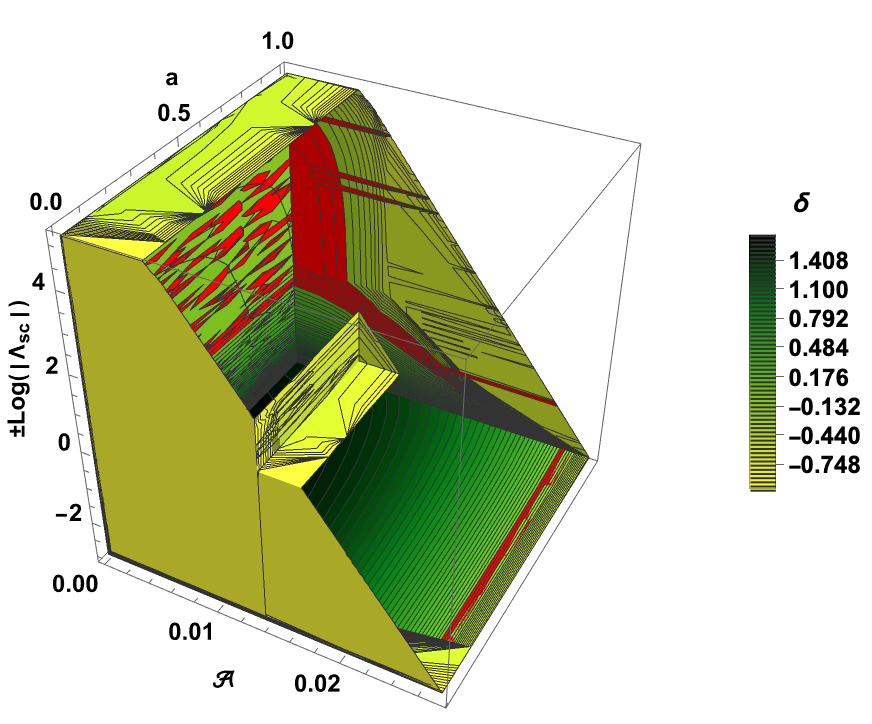}
\includegraphics[width=0.33\textwidth]{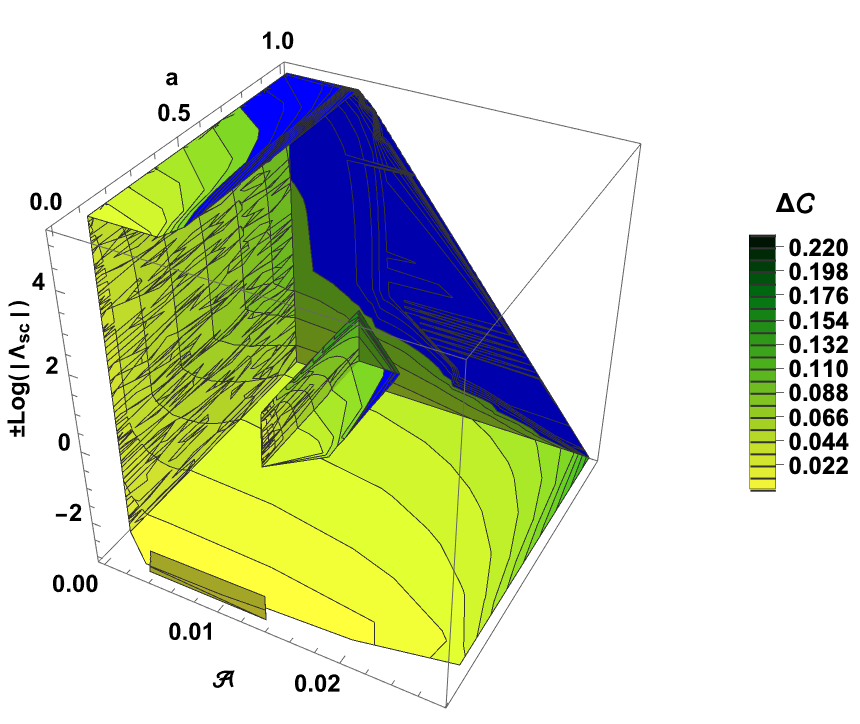}\\
\includegraphics[width=0.33\textwidth]{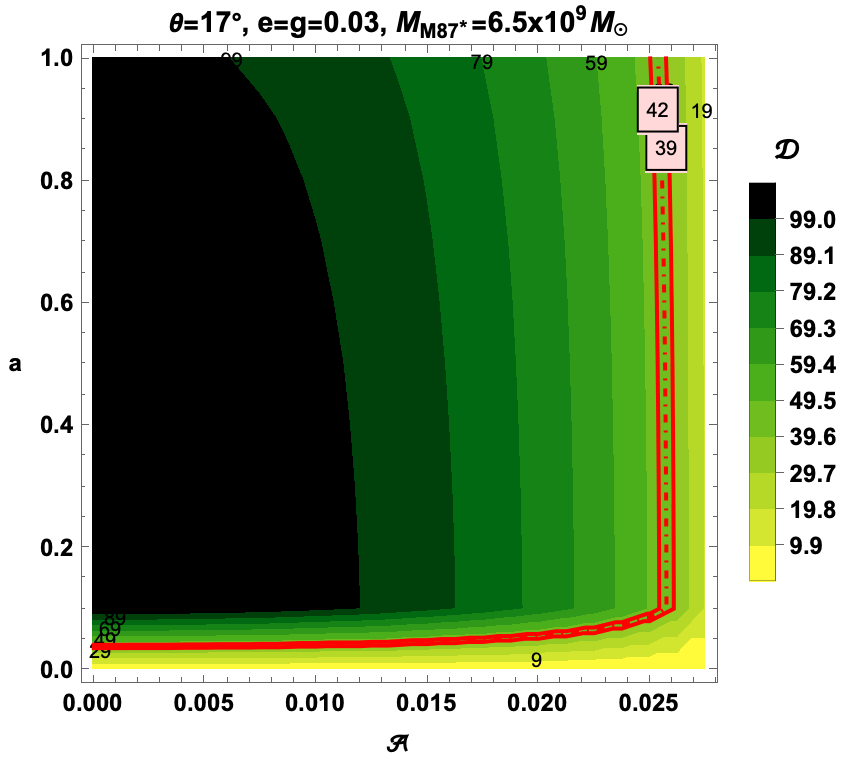}
\includegraphics[width=0.33\textwidth]{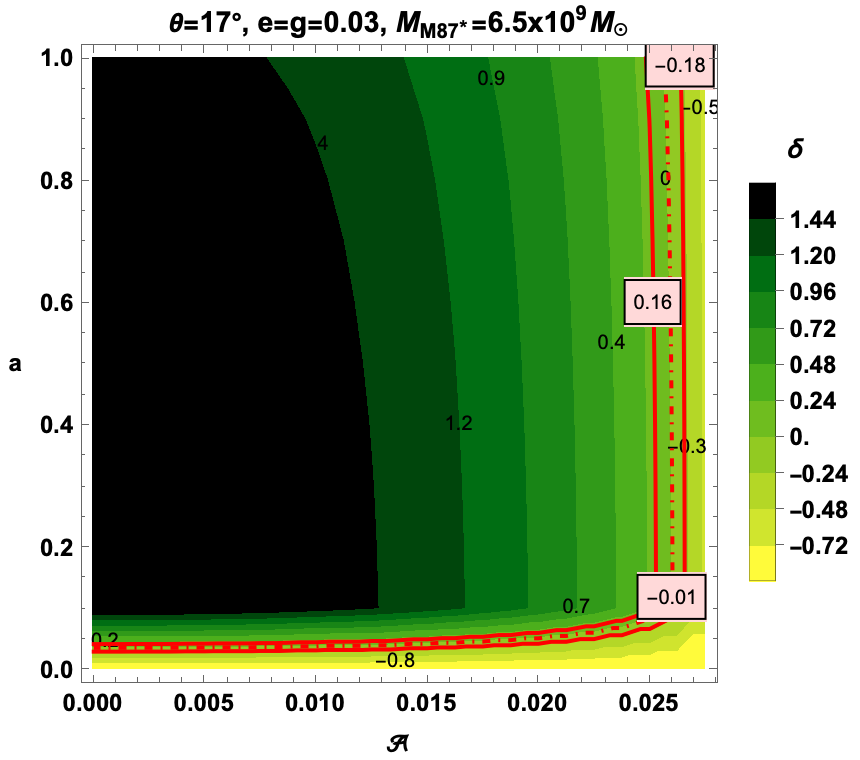}
\includegraphics[width=0.33\textwidth]{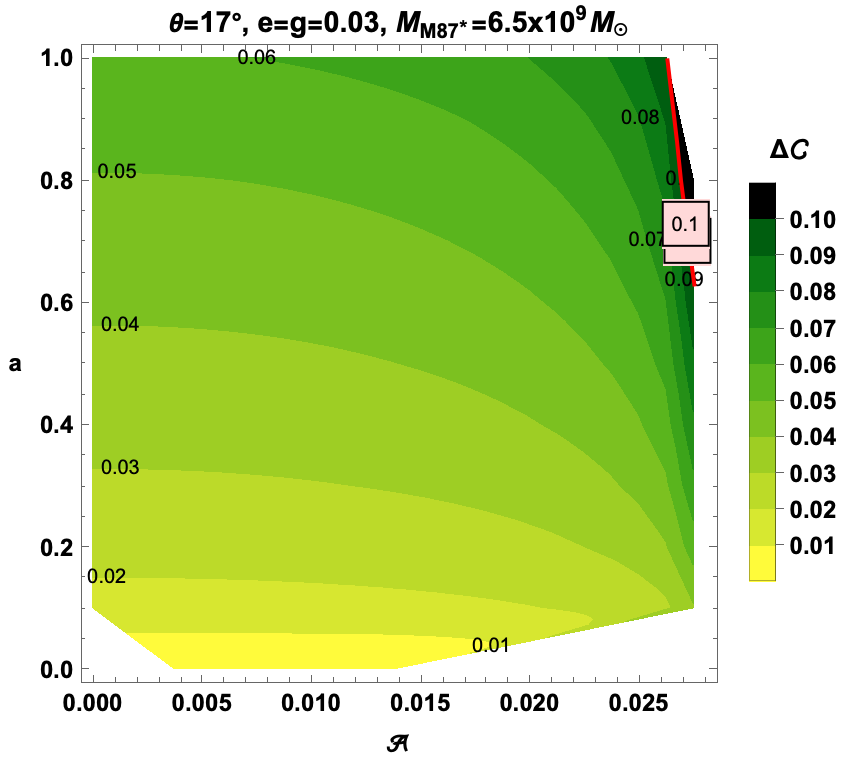}\\
\vspace{.2cm}
\includegraphics[width=0.33\textwidth]{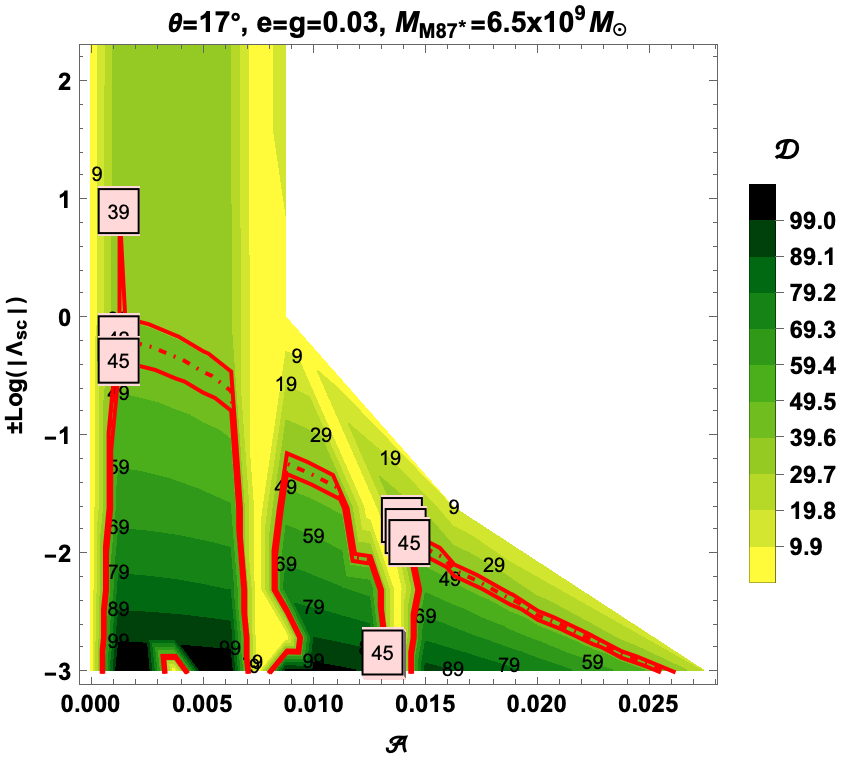}
\includegraphics[width=0.33\textwidth]{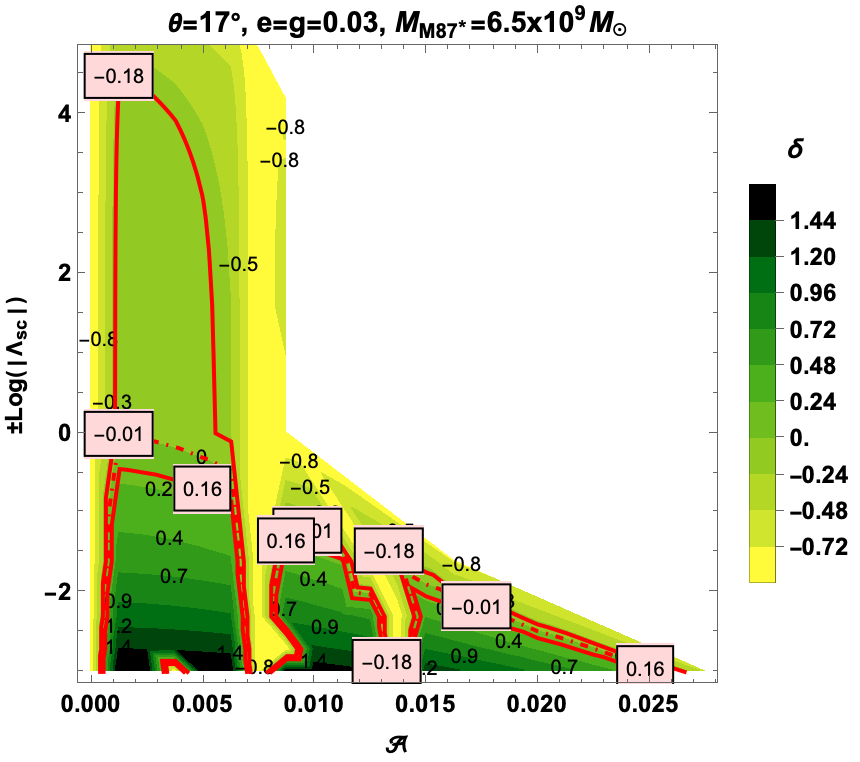}
\includegraphics[width=0.33\textwidth]{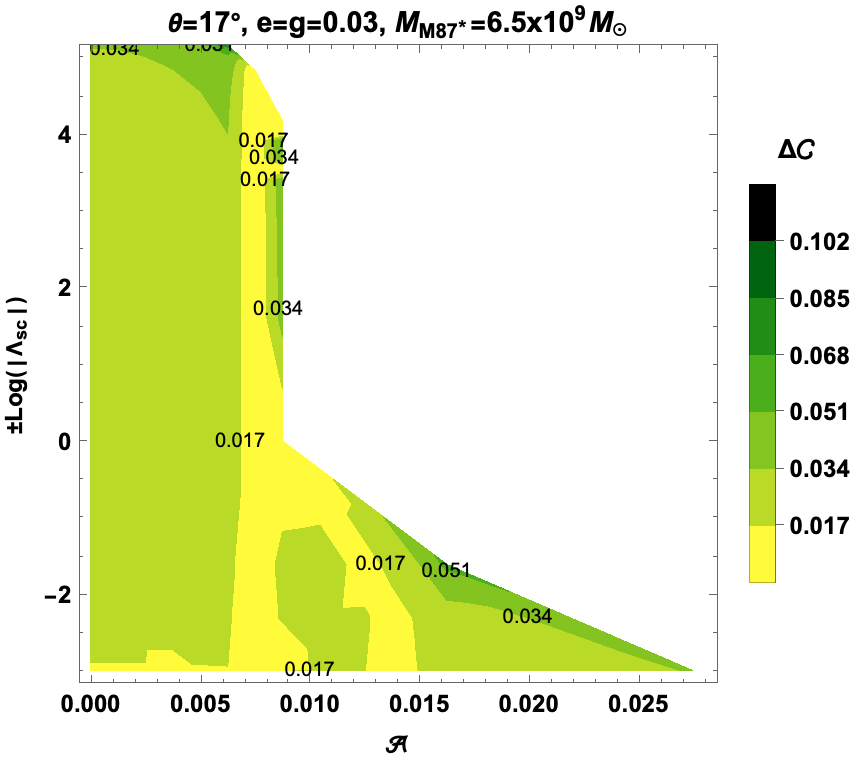}\\
\vspace{.2cm}
\includegraphics[width=0.245\textwidth]{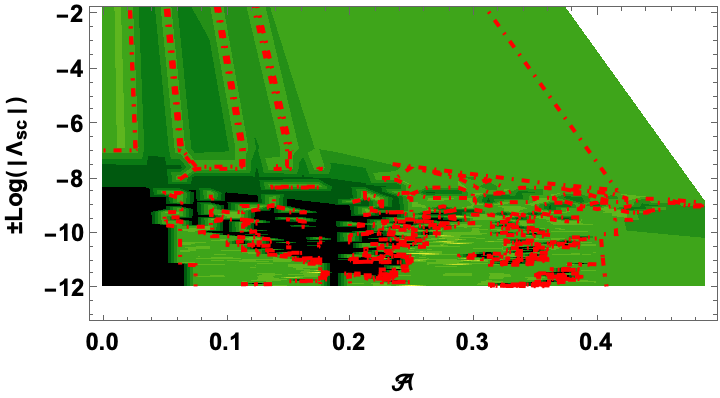}
\includegraphics[width=0.245\textwidth]{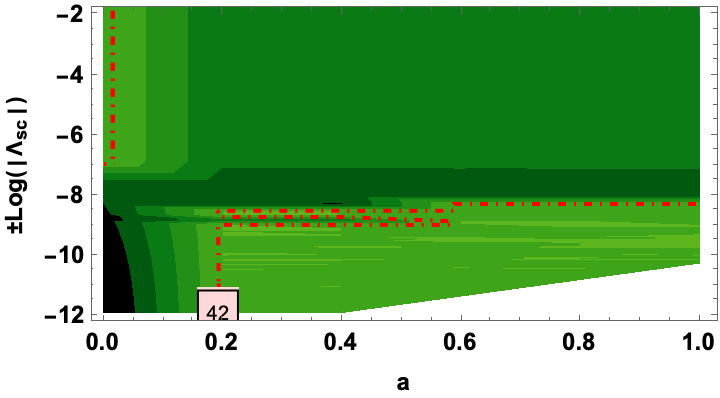}
\includegraphics[width=0.245\textwidth]{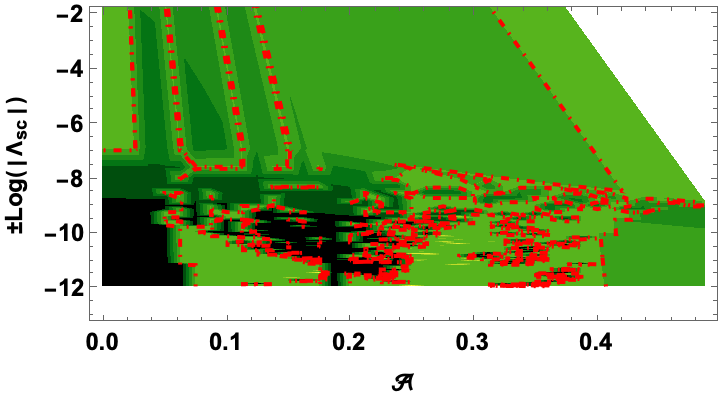}
\includegraphics[width=0.245\textwidth]{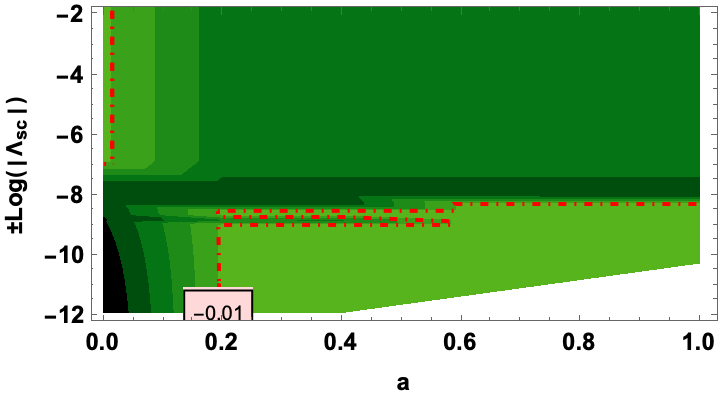}\\

\includegraphics[width=0.33\textwidth]{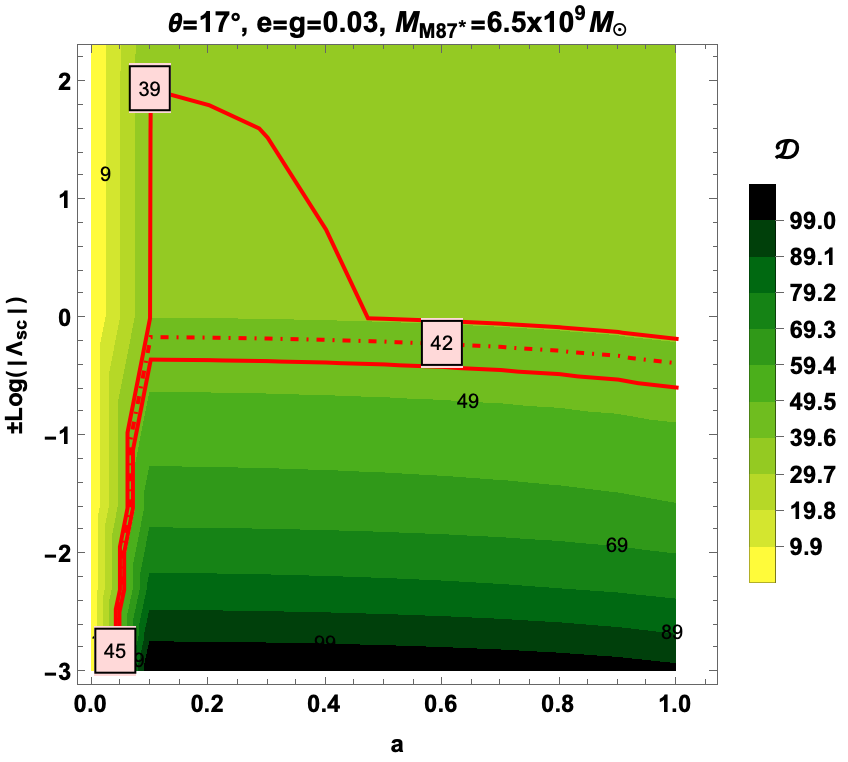}
\includegraphics[width=0.33\textwidth]{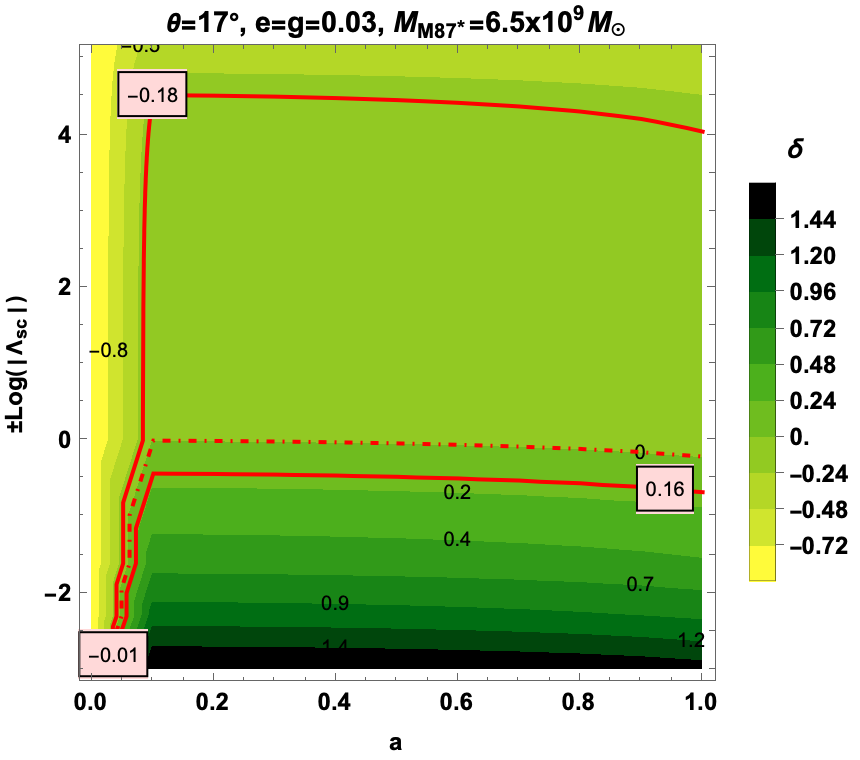}
\includegraphics[width=0.33\textwidth]{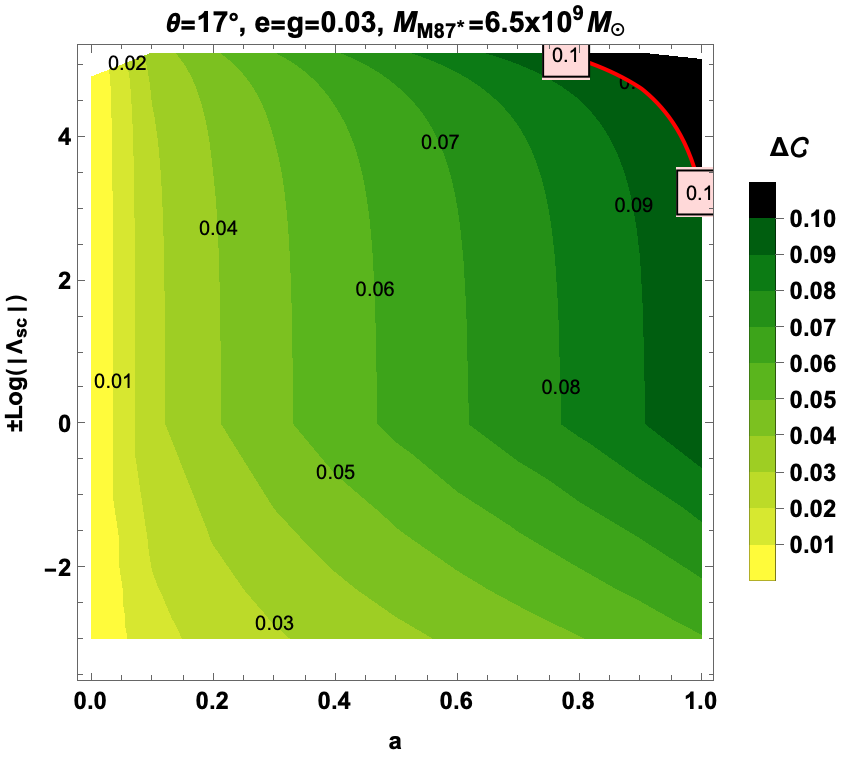}
};

\draw[violet, dashed,very thick] (-8.2,-2.8) rectangle (-3.4,-3.2);
\draw[cyan, dashed,very thick] (-8.2,-10.4) rectangle (-3.4,-10.8);
\draw[orange, dashed,very thick] (-2.6,-2.8) rectangle (2.,-3.2);
\draw[magenta, dashed,very thick] (-2.6,-10.4) rectangle (2.,-10.8);

\draw[violet, thick] (-8.34,-3.6) rectangle (-4.2,-6.);
\draw[cyan, thick] (-4.08,-3.6) rectangle (-0.02,-6.);
\draw[orange, thick] (0.1,-3.6) rectangle (4.2,-6.);
\draw[magenta, thick] (4.3,-3.6) rectangle (8.38,-6.);
\end{tikzpicture}
\caption{\it\footnotesize  Contours illustrating the dependence of the angular diameter $\mathcal{D}$, the deviation from circularity $\Delta\mathcal{C}$ and the fractional deviation $\bm \delta$ of the  $\textrm{M87}^\star$  on different planes of the accelerating-rotating black hole parameter. The dashed red line is associated with the EHT measurement and the solid lines correspond to estimated bounds. 
}
\label{fig87contours}
\end{figure*}

 In Fig.\ref{fig87contours}, we display the contours representing the angular diameter $\mathcal{D}$, the deviation from circularity $\Delta\mathcal{C}$, and the fractional deviation $\bm{\delta}$ of $\textrm{M87}^\star$ as functions of the accelerating parameter $\mathcal{A}$, rotation parameter $a$ and the cosmological constant $\Lambda$, first in the $(\mathcal{A},a, \log(\Lambda_{sc}))$ and then the projection on the each plan of this $3D$ space.
 The contours provide a visual representation of how these observables vary across different values of these parameters with a special focus on the negative region of the cosmological constant in the dashed panels. 

Before proceeding, it is important to recall that the estimated angular diameter of the shadow is $42\pm 3 \mu\text{as}$. However, due to a $10\%$ offset between the image and shadow diameters, the actual size of the shadow cannot be smaller than $37.8\pm 2.7 \mu\text{as}$.
In Fig.\ref{fig87contours}, each column represents a distinct observable. The contour plot illustrates the variations of the angular diameter, fractional deviation and deviation from circularity in the $(\log(\Lambda_{sc}),\mathcal{A},a)$ space. For a more detailed analysis, the subsequent rows depict the projections of these observables onto the planes $(\mathcal{A},a)$, $(\mathcal{A},\log(\Lambda_{sc}))$ and $(a,\log(\Lambda_{sc}))$, respectively\footnote{For the sake of convenience and to better capture the subtle variations of $\Lambda$, we employ the logarithmic representation of the scaled cosmological constant:
\begin{equation}
\log\left(\Lambda_{sc}\right)=\begin{cases}
 -\log \Big(10^4 .| \Lambda| \Big) & \Lambda<0 \\
 \Lambda & \Lambda=0 \\
 \log \Big(10^6 .\Lambda\Big) & \Lambda>0
\end{cases}.
\end{equation}}. 
For a more detailed illustration and given the sensitive nature of cosmological variations, a zoom is applied specifically for negative values of $\Lambda$. The olive shading represents the range of variation for the rotation, acceleration and cosmological constant parameters. The red color indicates the region that satisfies the observational constraints from $\textrm{M87}^\star$ data, while the violet color represents the excluded region.

Analyzing Fig.\ref{fig87contours}, several observations can be made :
\begin{itemize}

\item In the $(a,\mathcal{A})$ plane, It has been observed that for black holes (BHs) with low rotation values (i.e., $a\lesssim 0.1$), distinct behaviors manifest concerning the acceleration parameter. Specifically, for BHs with very-slow acceleration ($\mathcal{A}\lesssim 0.01$), the angular diameter $\mathcal{D}$ and the fractional deviation $\bm{\delta}$ tend to increase with increasing spin parameter $a$. In contrast, for BHs with fast acceleration values ($\mathcal{A}> 0.01$), the increase in the two observables is primarily following the decrease in $\mathcal{A}$ and $a$. In the case of high rotating BHs with $a> 0.1$, the increase in the two observables is mainly associated with the decrease in $\mathcal{A}$.
 Within the angular diameter, and when the slow rotating scheme is considered $a\in[0.04^{+0.003}_{-0.003},0.1]$, the experimental band unveils the range of acceleration values $\mathcal{A} \in[0,0.02573^{+0.00033}_{-0.00033}]$. While for the range $a\in[0.1,1]$, the experimental allowed band shows $\mathcal{A} \in[0.02573^{+0.00033}_{-0.00033},0.02535^{+0.00037}_{-0.00035}]$. For the fractional deviation the specific ranges are  
$a\in[0.038^{+0.0065}_{-0.0059},0.1]$ and $a\in[0.1,1]$ which correspond to $\mathcal{A} \in[0,0.026^{+0.00055}_{-0.00073}]$ and $\mathcal{A} \in[0.026^{+0.00055}_{-0.00073},0.02564^{+0.00071}_{-0.00083}]$ respectively.

Concerning the deviation from circularity $\Delta\mathcal{C}$, its upward trend is noticeable with increasing values of both $a$ and $\mathcal{A}$. However, it consistently stays below the $10\%$ threshold throughout the entire spectrum of acceleration and rotation parameters when considering the constraints imposed by $\mathcal{D}$ and $\bm{\delta}$.

Overall, such results provide valuable insights into the behavior of the angular diameter, fractional deviation, and circularity deviation in the $(a,\mathcal{A})$ plane, offering valuable information about the effects of rotation and acceleration on these observables.

\item In the $(\Lambda, \mathcal{A})$ plane, both observables $\mathcal{D}$ and $\bm{\delta}$ exhibit consistent increasing piecewise behavior as the acceleration parameter and the cosmological constant decrease.
 Moreover, we notice that the apparent diameter of the black hole falls within the experimentally observed range ($42 \pm 3 \mu as$) for some domains of acceleration values  $\displaystyle\mathcal{A}\in[(4.75\times10^{-4})^{+3.50\times10^{-5}}_{-4.56\times10^{-5}}, (6.99\times10^{-3})^{+4.10\times10^{-5}}_{-3.00\times10^{-5}}] \cup [(7.99\times10^{-3})^{+4.00\times10^{-5}}_{-4.00\times10^{-5}},0.01321^{+4.00\times10^{-5}}_{-4.0\times10^{-5}}] \cup [0.01430^{+4.00\times10^{-5}}_{-4.00\times10^{-5}},0.02572^{+3.40\times10^{-4}}_{-3.10\times10^{-4}}]$. Interestingly, the value of the corresponding cosmological constant decreases as the acceleration value increases and the corresponding bands are $\Lambda\in[-1.98\times10^{-3}, 2.46\times10^{-6}]$, $\Lambda\in[-1.98\times10^{-3}, -3.14\times10^{-4}]$ and $\Lambda\in[-1.98\times10^{-3}, -5.58\times10^{-4}]$ respectively.
 Notably, both Anti-de Sitter (AdS) and de Sitter (dS) space configurations remain consistent with EHT observational data for slowly accelerating black holes ($\mathcal{A}\in[0, 0.0015]$). However, beyond the special value of $\mathcal{A}\simeq 0.0015$, only the AdS black hole type continues to align with the observational measurements.
Similarly, observable $\bm{\delta}$ follows a behavior similar to $\mathcal{D}$ for the acceleration parameter. Throughout the corresponding ranges of cosmological constant are $\Lambda\in[-1.98\times10^{-3}, 8.80\times10^{-5}]$, $\Lambda\in[-1.98\times10^{-3}, -2.60\times10^{-4}]$ and $\Lambda\in[-1.98\times10^{-3}, -4.37\times10^{-4}]$. 
 Furthermore, the special point shifts, expanding the allowed acceleration interval to $\mathcal{A}\in[0, 0.0055]$, where both AdS and dS spaces are permitted. Outside this range, only the AdS space remains consistent. In addition, both observables $\mathcal{D}$ and ${\bm \delta}$ unveil more negative ranges of the cosmological constant $\Lambda\in[-0.1002, -0.0006]$, $\Lambda\in[-14.5801, -0.3866] \cup [-0.2980, -0.0006]$ and  $\Lambda\in[-14.5801, -0.0006]$ within special acceleration values, evaluated to  $\mathcal{A}\in[0,0.0252]$, $\mathcal{A}\in[0.0488,0.1812]$ and $\mathcal{A}\in[0.1924, 0.4867]$ respectively.

Therewith the contours of deviation from circularity  $\Delta\mathcal{C}$  cover up all the ranger of the cosmological constant and the acceleration parameters.

\item  Herein, in the $(\Lambda,a)$ plane, the observations unveil two discernible schemes within black holes (BHs) characterized by distinct spin parameters and cosmological constants. Both angular diameter $\mathcal{D}$ and fractional deviation $\bm{\delta}$ demonstrate an increasing trend with rising spin parameter $a$ and cosmological constant $\Lambda$ in low rotations ($a\lesssim 0.1$) scheme. While,  for significant rotations ($a> 0.1$), the trend of these two observables primarily results from the decrease in the cosmological constant $\Lambda$.

Comparing our computational results with experimental data reveals that, in terms of angular diameter, for slowly rotating black holes $a\in[0.039^{+0.002}_{-0.003},0.092^{+0.008}_{-0.008}]$, the cosmological constant aligns within the range $\Lambda\in[-1.98\times10^{-3},-1.42\times10^{-4}]$. While, for all other conceivable values of the rotation parameter $a\in[0.1,1]$, the corresponding range of the cosmological constant is found to be $\Lambda\in[(-1.16\times10^{-4})^{+1.23\times10^{-4}}{-2.48\times10^{-5}},\\(-1.45\times10^{-4})^{+2.71\times10^{-5}}{-3.34\times10^{-5}}]$.
 
 In examining the fractional deviation, it has been observed that for slowly rotating black holes, characterized by a range of $a$ values spanning from $0.036^{+0.006}_{-0.006}$ to $0.083^{+0.017}_{-0.017}$, the cosmological constant spans a range of $\Lambda$ values from $-1.98\times10^{-3}$ to $-1.52\times10^{-4}$, respectively. For $a\in[0.1,1]$, the corresponding range of the cosmological constant is observed to be between $\Lambda = (1.00\times10^{-6})^{+9.21\times10^{-6}}_{-1.53\times10^{-4}}$ and $\Lambda = (-1.19\times10^{-4})^{+1.81\times10^{-4}}_{-7.15\times10^{-5}}$.

Additional probes in negative cosmological domains uncover  within $\mathcal{D}$ and ${\bm \delta}$, $\Lambda\in[-0.1012 , -0.0006]$ and $\Lambda\in[-14.2914, -0.3827]$ for $a\in[0,0.015]$ and $a\in[0.19, 1]$ respectively.

Whereas some excluded regions associated with the $\Delta\mathcal{C}$ observable appear in this plan. The imposition of this constraint has slightly reduced the minimum uncertainty associated with the cosmological constant$\Lambda$ (relatively $20\%$ near $a\approx 1 $), particularly concerning the observable ${\bm \delta}$ in the context of extremely rotating de Sitter (dS) black holes with angular momentum parameter $a$ falling within the range of $[0.96,1]$.
\end{itemize}

The last observable i.e. the axis ratio $\Delta A$ cannot provide any significant funding because the maximum value is less than the EHT observation of $\Delta A \lesssim 4/3$ \cite{EventHorizonTelescope:2019dse}.

\begin{figure*}\centering
\begin{tikzpicture}[
    zoomboxarray
]
    \node [image node] { \includegraphics[width=0.52
\textwidth]{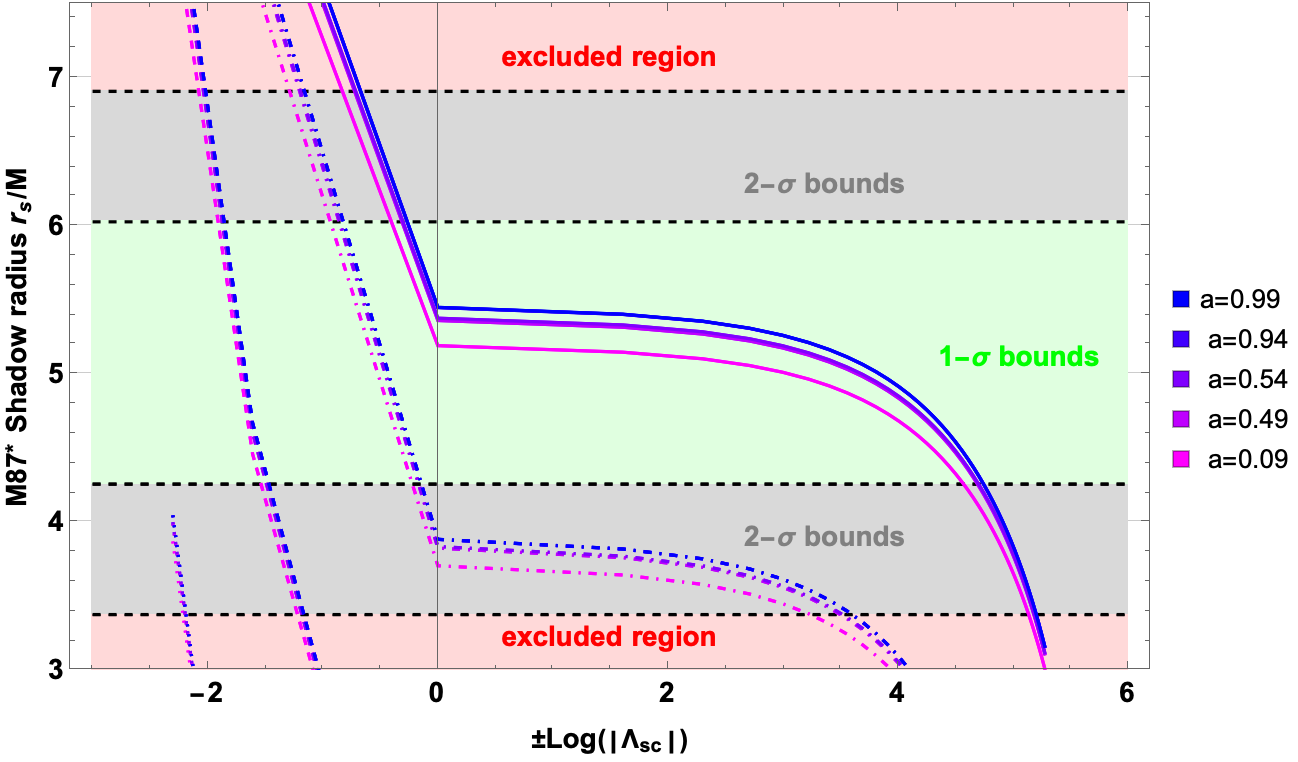} };
    \zoombox[color code=violet,dash dot,magnification=5]{0.256,0.62}
    \zoombox[color code=magenta,dotted,magnification=4]{0.14,0.32}
    \zoombox[color code=cyan,dashed,magnification=4]{0.2,0.48}
    \zoombox[magnification=3]{0.5,0.55}
    \node[black,rotate=-35] at (7.,.4)     {\tiny$\mathcal{A}= 0$};
      \node[black,rotate=-35] at (6.,.3)     {\tiny$\mathcal{A}= 0.007$};
      \node[black,rotate=-35] at (2.7,.3)     {\tiny$\mathcal{A}= 0.014$};
       \node[black,rotate=-35] at (1.5,.29)     {\tiny$\mathcal{A}= 0.02$};
       \draw[violet,dashed,very thick] (0.3,5.2) rectangle (.645,.4);
\end{tikzpicture}
\begin{tikzpicture}
 \node [] at (0,0    ) { \includegraphics[width=.5
\textwidth]{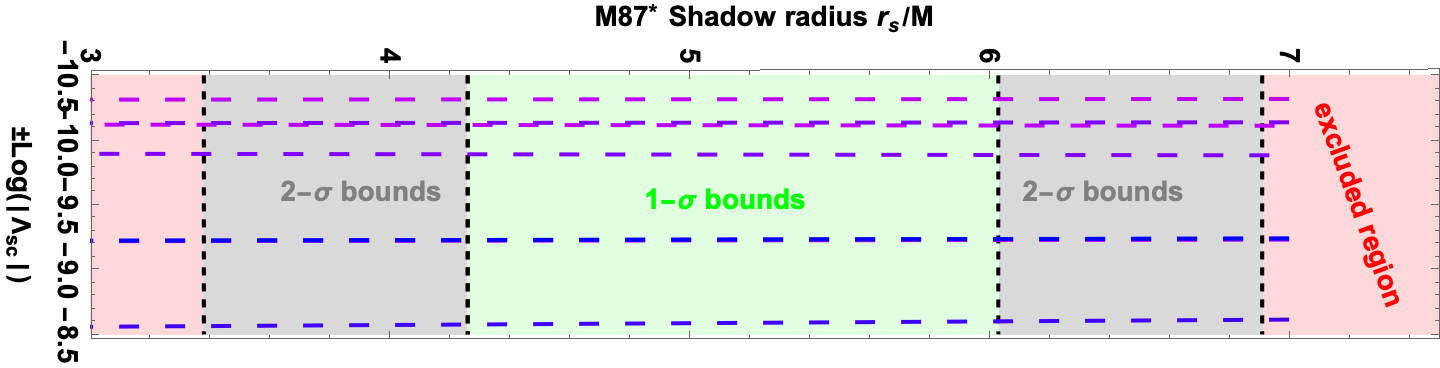} };
\draw[violet,dashed,very thick] (-4.6,1.3) rectangle (4.6,-1.3);
\node[black,rotate=-90] at (4.9,0)     {$\mathcal{A}\simeq 0.5$};
\end{tikzpicture}
\caption{ \it \footnotesize {\bf Left:} The shadow radii $r_\text{s}/M$ viewed on the equatorial plane (deviated from $\pi/2$) of the considered accelerating black hole  with cosmological constant as a function of 
the cosmological constant by varying the rotation and acceleration parameters. The olive/gray/red shaded regions refer to the areas that are (1-$\sigma$)\big/(2-$\sigma$) consistent$\big/$ inconsistent with the $\textrm{M87}^\star$ observations and highlight that the latter set constraints on the black hole parameters. {\bf Right:}  Deep zoom in different regions of the left panel.
{\bf Bottom}: A zoom in the negative cosmological constant domain where the quantity $r_\text{s}/M$ allow us to reach the acceleration value $\mathcal{A}\simeq0.5$.}
\label{rsm87}
\end{figure*}
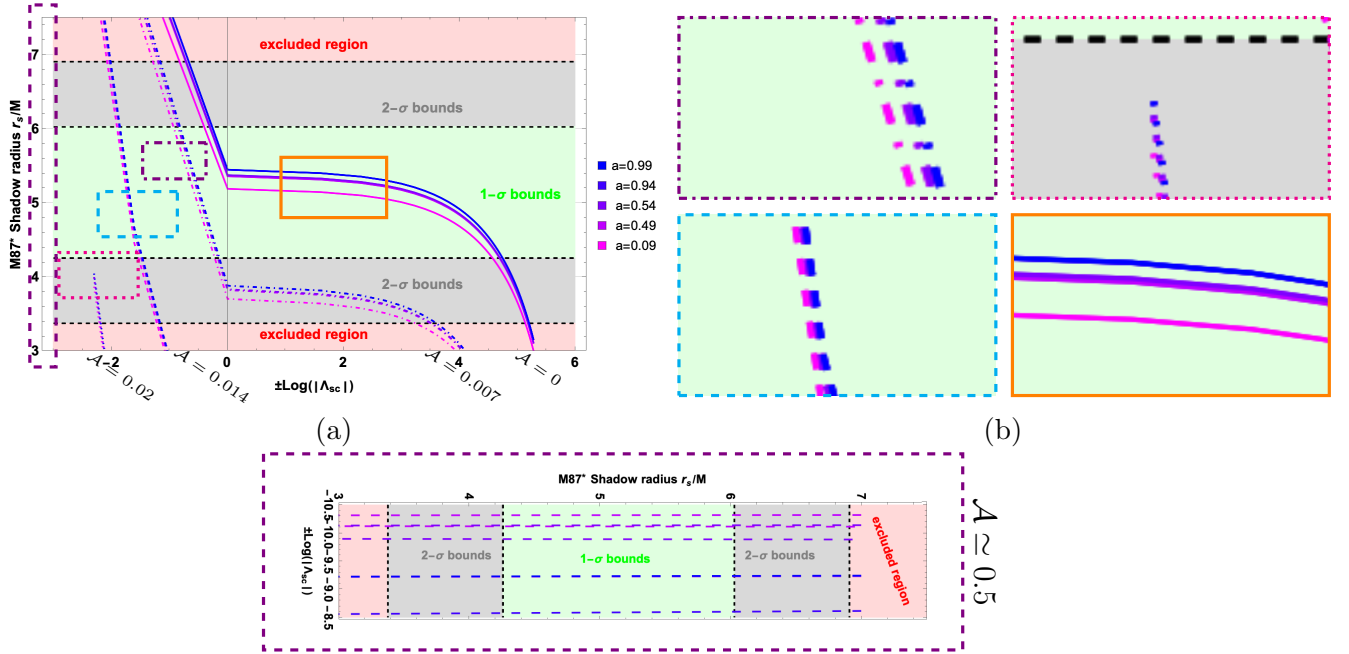

The next relevant step in our exploitation of the experimental constraints of Tab.\ref{m87_bounds} is considering the second and the third columns, namely, we graphically illustrate the constraints on the shadow radii $\frac{r_\text{s}}{M}$ viewed on the equatorial plane  in terms of 
 the cosmological constant in Fig.\ref{rsm87} and by varying the accelerating and the spin parameters. 
In our previous analysis, we have observed that the latitude of the circular orbit is influenced by the acceleration parameter, leading to variations in the observer's inclination angles and the maximum values of the shadow radius and distortion for the accelerating black hole (see Fig.\ref{fig3}). Therefore, we focus on the equatorial plane located at the maximum deviation from $\pi/2$.

Based on Fig.\ref{rsm87}, we can conclude that the considered parameters in describing the shadow cast fall within the upper and lower bounds at the 1-$\sigma$ and 2-$\sigma$ levels. Where this happens, the range in cosmological constant is more extensive in the positive interval (dS) than in the negative domain (AdS) for small acceleration parameter $\mathcal{A}$. while the AdS black hole configuration becomes predominant in the high acceleration parameter scheme $\mathcal{A}\gtrsim 0.01$. For more detailed constraints, some numerical intervals associated with a small range of acceleration are listed in the following Tab.\ref{dddeltam87}
\begin{table*}
\centering
\resizebox{.9\textwidth}{!}{%
\begin{tabular}{cl|llll|}
\cline{2-6}
\multicolumn{1}{l|}{}                                                                              & \multicolumn{1}{c|}{\multirow{2}{*}{$\mathcal{A}$}} & \multicolumn{4}{c|}{$\bm \Lambda$}                                                                                       \\ \cline{3-6} 
\multicolumn{1}{l|}{}                                                                              & \multicolumn{1}{c|}{}                               & \multicolumn{1}{c|}{lower bound limit} & \multicolumn{1}{c||}{upper bound limit} & \multicolumn{1}{l|}{lower bound limit} & upper bound limit \\ \hline\hline
\multicolumn{1}{|c|}{\multirow{4}{*}{\begin{tabular}[c]{@{}c@{}}1-$\sigma$\\ bounds\end{tabular}}} & 0                                                   & \multicolumn{1}{c|}{$9.57116\times10^{-5}$}            & \multicolumn{1}{c||}{$-0.000151048$}            & \multicolumn{1}{c|}{$0,000113619$}            &        \multicolumn{1}{c|}{$-0.000131152$}    \\ \cline{2-6} 
\multicolumn{1}{|c|}{}                                                                             & 0.007                                               & \multicolumn{1}{c|}{$-0.000127242$}            & \multicolumn{1}{c||}{$-0.000255243$}            & \multicolumn{1}{c|}{$-0.000116197$}            &     \multicolumn{1}{c|}{$-0.000228432$ }       \\ \cline{2-6} 
\multicolumn{1}{|c|}{}                                                                             & 0.014                                               & \multicolumn{1}{l|}{$-0.000472307$}            & \multicolumn{1}{c||}{$-0.000679142$}            & \multicolumn{1}{c|}{$-0.000440104$}            &   \multicolumn{1}{c|}{ $-0,000645729$ }        \\ \cline{2-6} 
\multicolumn{1}{|c|}{}                                                                             & 0.0196                                              & \multicolumn{1}{c|}{-\ -}            & \multicolumn{1}{c||}{-\ -}            & \multicolumn{1}{c|}{-\ -}            &      \multicolumn{1}{c|}{-\ -}       \\ \hline\hline
\multicolumn{1}{|c|}{\multirow{4}{*}{\begin{tabular}[c]{@{}c@{}}2-$\sigma$\\ bounds\end{tabular}}} & 0                                                   & \multicolumn{1}{c|}{$0.00016711$}            & \multicolumn{1}{c||}{$-0.000230748$}            & \multicolumn{1}{c|}{$0.000178224$}            &      \multicolumn{1}{c|}{$-0.000196352$}       \\ \cline{2-6} 
\multicolumn{1}{|c|}{}                                                                             & 0.007                                               & \multicolumn{1}{c|}{$2,50164\times10^{-5}$}            & \multicolumn{1}{c||}{$-0.000367018$}            & \multicolumn{1}{c|}{$3.63364\times10^{-5}$}            &       \multicolumn{1}{c|}{$-0.000318674$}      \\ \cline{2-6} 
\multicolumn{1}{|c|}{}                                                                             & 0.014                                               & \multicolumn{1}{c|}{$-0.000341993$}            & \multicolumn{1}{c||}{$-0.000806207$}            & \multicolumn{1}{c|}{$-0.000321906$}            &       \multicolumn{1}{c|}{$-0.000758847$}       \\ \cline{2-6} 
\multicolumn{1}{|c|}{}                                                                             & 0.0196                                              & \multicolumn{1}{c|}{$-0.000928509$}            & \multicolumn{1}{c||}{-\ -}            & \multicolumn{1}{l|}{$-0.00090996$}            &      \multicolumn{1}{c|}{-\ -}       \\ \hline\hline

\multicolumn{1}{l}{}                                                                               &                                                     & \multicolumn{2}{c||}{{\bf a=0.09}}                                         & \multicolumn{2}{c|}{{\bf a=0.99}}                    \\ \cline{3-6} 
\end{tabular}%
}
\caption{\it \footnotesize Some slow acceleration constraints on the black hole parameters using the shadow radius $r_s/M$ observational data listed in Tab.\ref{m87_bounds}.}
\label{dddeltam87}
\end{table*}
 from which, it is evident that the upper limit of the cosmological constant undergoes a shift to $(\Lambda_{\text{1-}\sigma} = 1.1361\times10^{-4}, \Lambda_{ \text{2-}\sigma} = 1.7822\times10^{-4})$. 
  Further, the bottom panel of Fig.\ref{rsm87} provides more information about the high acceleration ranges. Namely, for the acceleration parameter $\mathcal{A}\in[0.4,0.4875]$ the cosmological constant exhibits a range of variation from $\Lambda=-11.5844$ to $\Lambda=-0.5271$.

Once we have compared the obtained black hole shadow with the observational data of $\textrm{M87}^\star$ and identified the parameter regions that are consistent with the EHT measurements , our next step is to extend this analysis to include the measurements of $\textrm{Sgr~A}^\star$ by examining its shadow.  We aim to determine the allowed parameter regions that align with the $\textrm{Sgr~A}^\star$ observed results.

\subsection{Contact with $\textrm{Sgr~A}^\star$ data}

With the recent release of the black hole image of $\textrm{Sgr~A}^\star$ in addition to the image of $\textrm{M87}^\star$, we now have the opportunity to gain a better grasp of gravitational physics at the horizon scale. 
 Indeed, the observed emission ring of $\textrm{Sgr~A}^\star$ has an angular diameter of $51.8\pm 2.3 \mu as$, according to the Event Horizon Telescope collaboration \cite{EventHorizonTelescope:2022wkp}, whereas the angular diameter of its shadow is estimated to $48.7\pm 7 \mu as$. Measurements from various collaborations provide estimates for the mass and distance of $\textrm{Sgr~A}^\star$. Effectively,  Keck observatory team measurements reveal that  $\textrm{Sgr~A}^\star$ has a distance of $d=(7959\pm59\pm32)\:\rm pc$ and mass $m=(3.975\pm 0.058\pm 0.026)\times 10^{6} M_{\odot}$, which is derived by leaving the red-shift parameter-free, while distance $d=(7935\pm50)\;\rm pc$ and mass $m=(3.951\pm 0.047)\times 10^{6} M_{\odot}$, assuming the red-shift parameter sets to unity \cite{Do:2019txf}. 
Likewise, the Very Large Telescope together with the interferometer GRAVITY (VLTI) collaboration suggests that the $\textrm{Sgr~A}^\star$ possesses $m=(4.261\pm0.012)\times 10^{6} M_{\odot}$ and distance $d=(8246.7\pm9.3)\;\rm pc$ \cite{GRAVITY:2021xju,GRAVITY:2020gka}.   Otherwise, it further the black hole mass 
$m=(4.297\pm0.012\pm0.040)\times 10^{6} M_{\odot}$ and the distance $d=8277\pm9\pm33\; \rm pc$ by accounting for optical aberrations.  Based on comparisons between the observed image of $\textrm{Sgr~A}^\star$ and models derived from numerical simulations, it is inferred that the inclination angle $i$ of the source is greater than $50^\circ$. To calculate the theoretical angular diameter of $\textrm{Sgr~A}^\star$, a specific value of the inclination angle is chosen. In this work, we adopt $i\simeq 134^\circ$ (or equivalently $46^\circ$) \cite{refId0}. Additionally, the fractional deviation $\bm{\delta}$ and the dimensionless shadow radius $r_\text{s}/M$, obtained using the \textit{eht-img} algorithm \cite{EventHorizonTelescope:2022xqj}, are summarized in Tab.\ref{tab:bounds_sagA}.

\begin{table}
    \centering
    \resizebox{.7\textwidth}{!}{
    \begin{tabular}{ l l | c c c }
    
        & \multicolumn{4}{c}{$\textrm{Sgr~A}^\star$ estimates}\\[1mm]\hline\hline\\[-3.5mm]
        
        & &\hspace{\tabley} Deviation $\bm\delta$ &\hspace{\tablex} 1-$\sigma$ bounds &\hspace{\tablex} 2-$\sigma$ bounds\\[1mm]\hline
        
        \parbox[ht]{3mm}{\multirow{3}{*}{\rotatebox[origin=c]{90}{\textit{\scriptsize eht-img}\hspace{2mm}}}} & {VLTI}\hspace{\tabley} &\hspace{\tabley} $-0.08^{+0.09}_{-0.09}$ &\hspace{\tablex} {$4.31\le \frac{r_{\text{s}}}{M}\le 5.25$} &\hspace{\tablex} {$3.85\le \frac{r_{\text{s}}}{M}\le 5.72$}\\[3mm]
        
        & Keck\hspace{\tabley} &\hspace{\tabley} $-0.04^{+0.09}_{-0.10}$ &\hspace{\tablex} {$4.47\le \frac{r_{\text{s}}}{M}\le 5.46$} &\hspace{\tablex} {$3.95\le \frac{r_{\text{s}}}{M}\le 5.92$}\\[1mm]

        \hline
    \end{tabular}
    }
\caption{\it \footnotesize Sagittarius $A^\star$ bounds on the fractional deviation parameter $\bm{\delta}$ and the dimensionless shadow radius $r_\text{sh}/M$ estimated through the \textit{eht-img} algorithm.}
\label{tab:bounds_sagA}
\end{table}

\begin{figure*}\centering
\begin{tikzpicture}
\node[align=left] {
\includegraphics[width=0.5\textwidth]{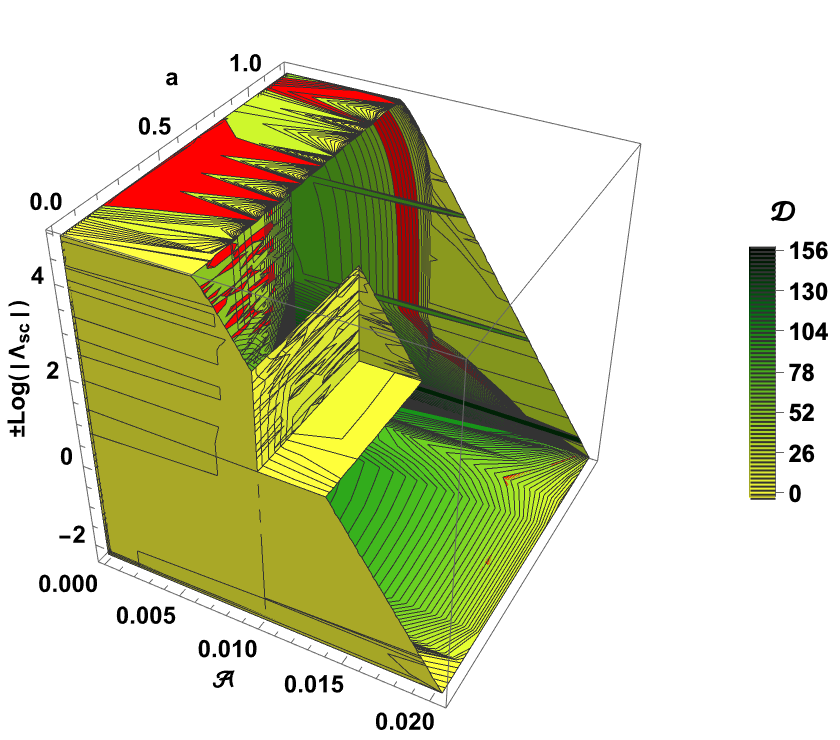}
\includegraphics[width=0.5\textwidth]{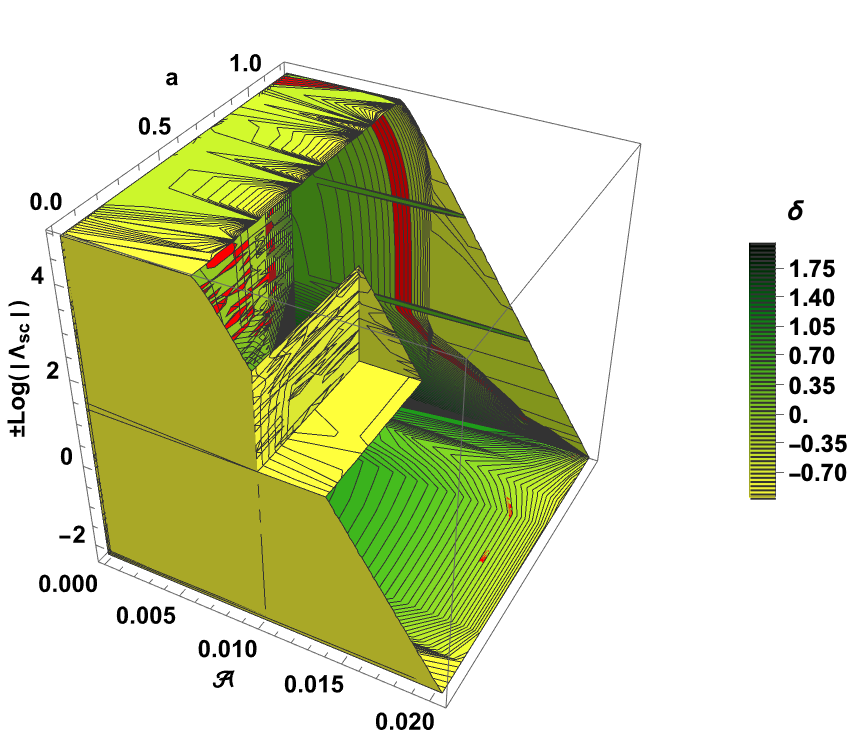}\\
\includegraphics[width=0.33\textwidth]{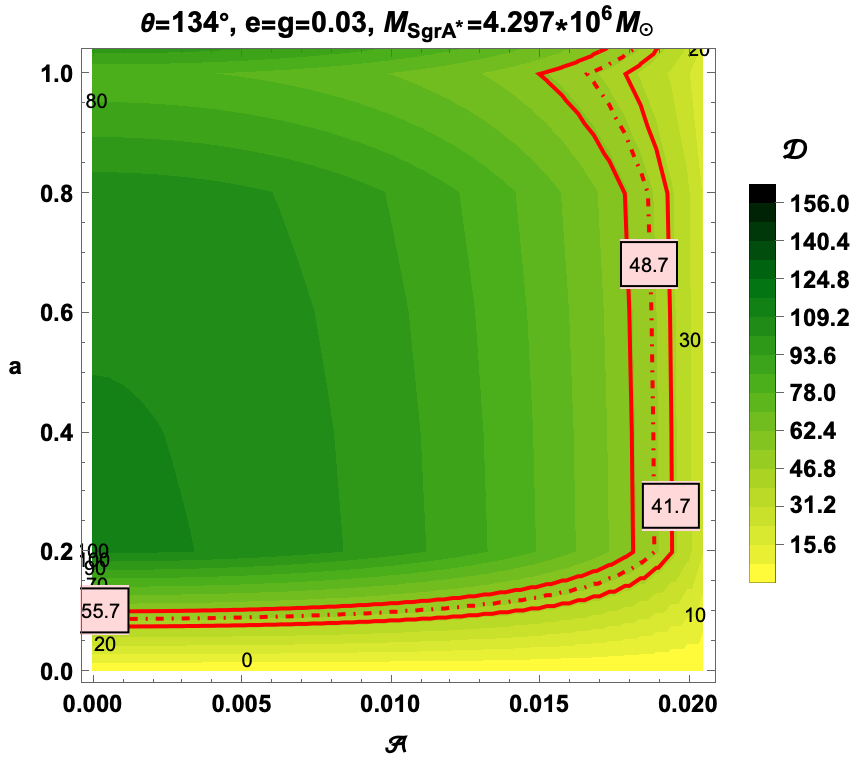}
\includegraphics[width=0.33\textwidth]{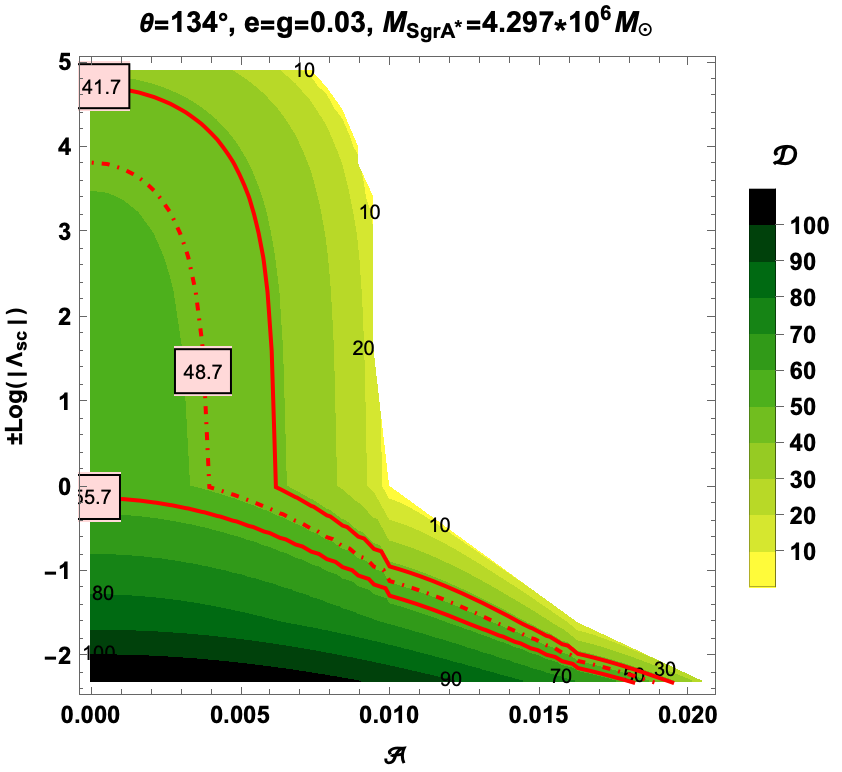}
\includegraphics[width=0.33\textwidth]{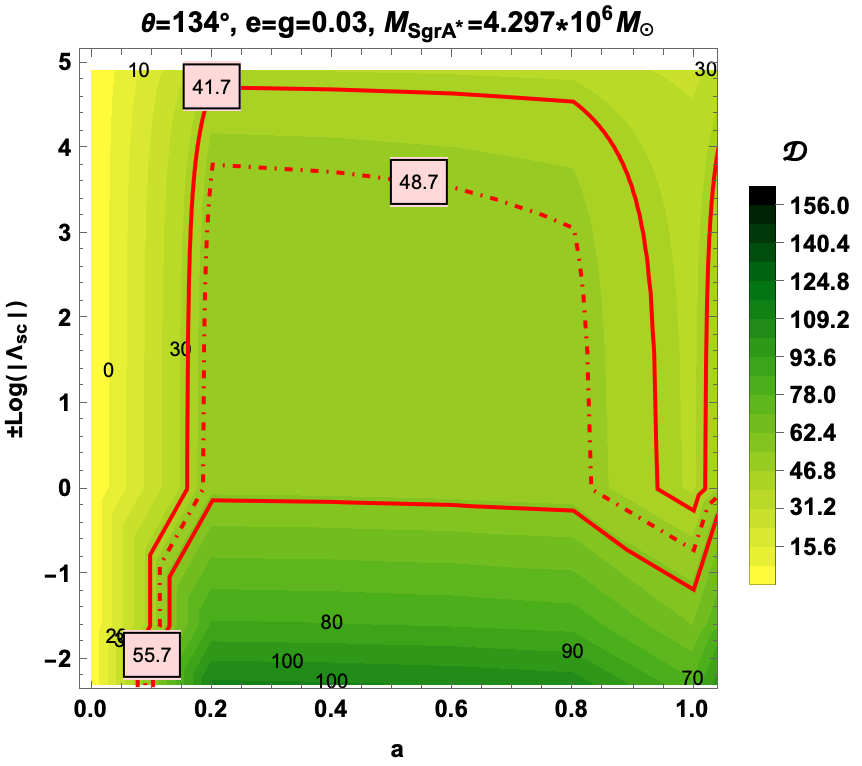}\\
\vspace{.2cm}
\includegraphics[width=0.33\textwidth]{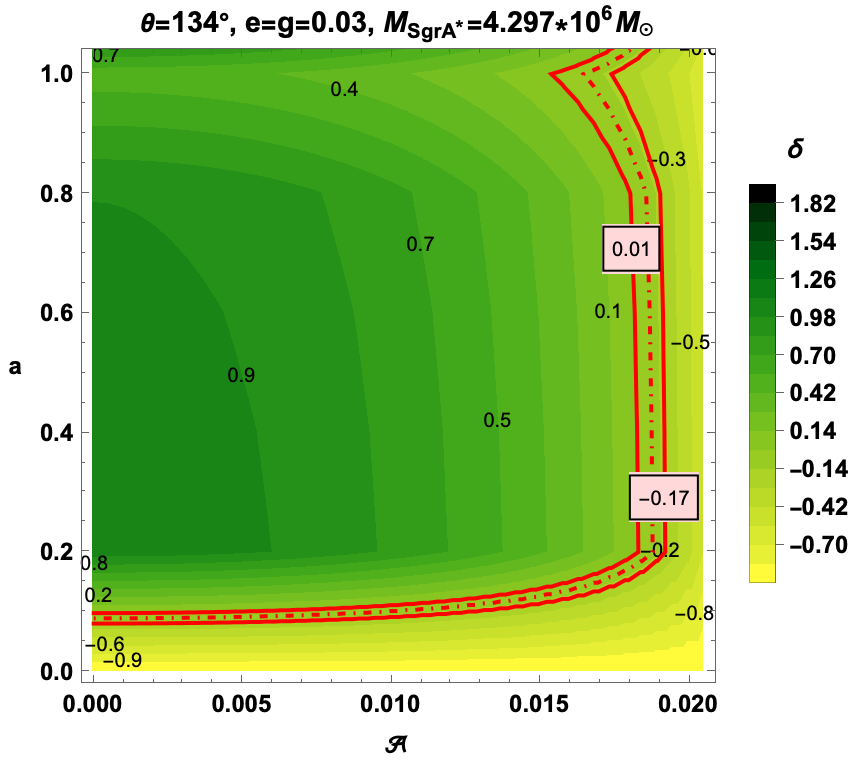}
\includegraphics[width=0.33\textwidth]{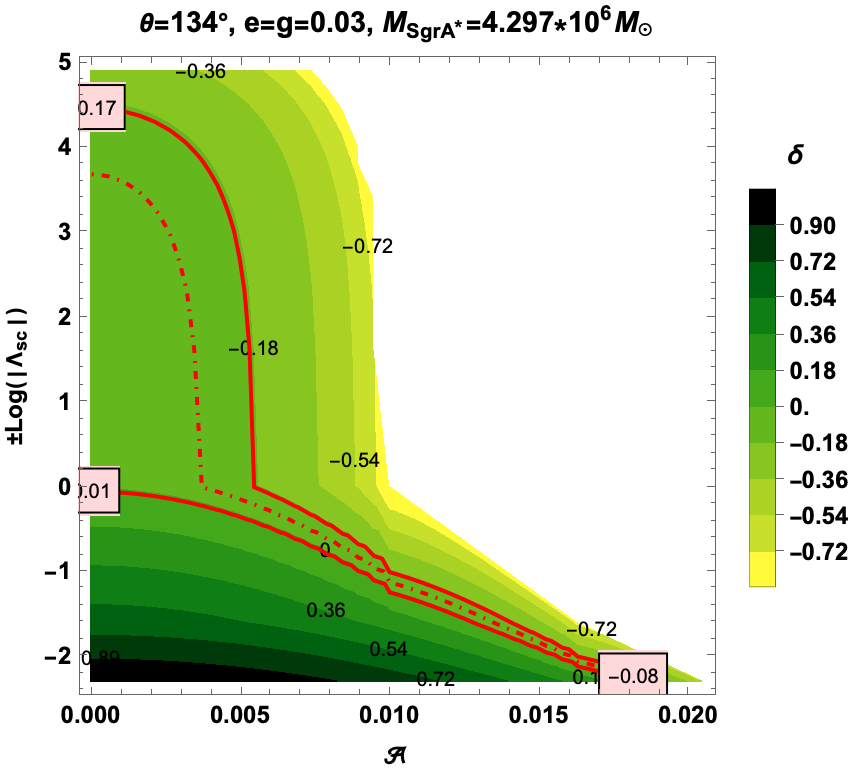}
\includegraphics[width=0.33\textwidth]{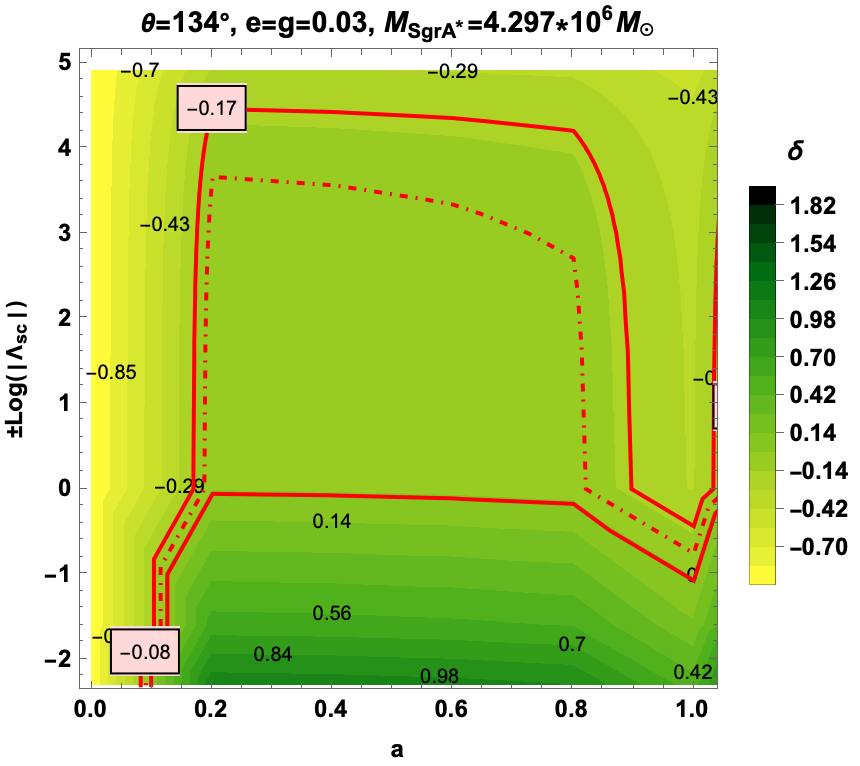}\\

\includegraphics[width=0.245\textwidth]{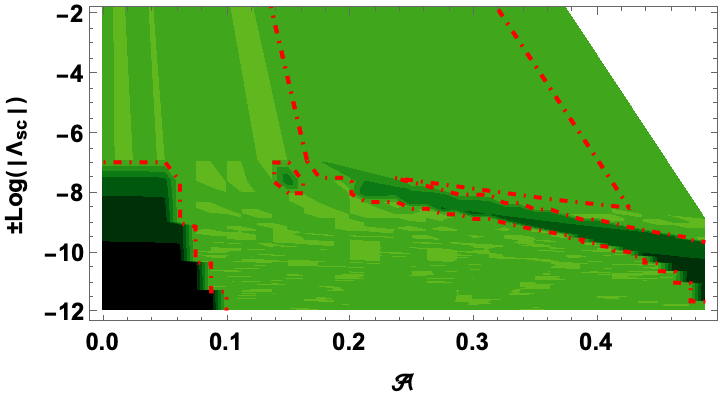}
\includegraphics[width=0.245\textwidth]{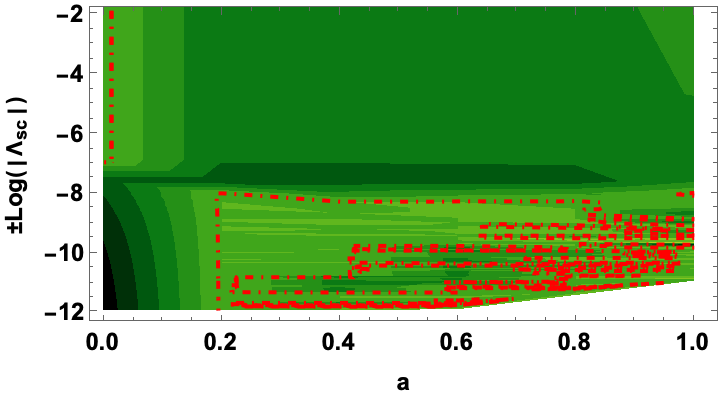}
\includegraphics[width=0.245\textwidth]{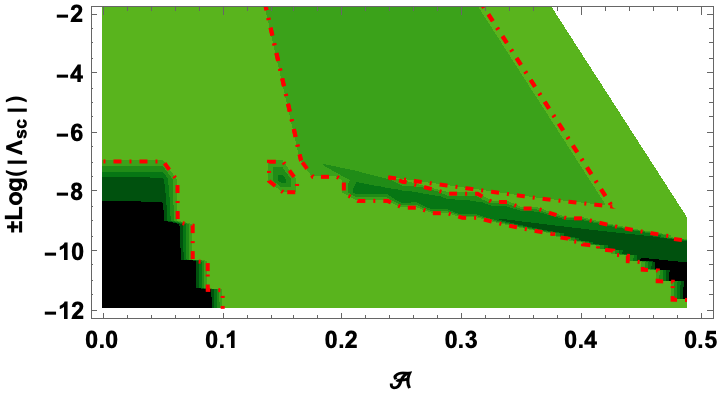}
\includegraphics[width=0.245\textwidth]{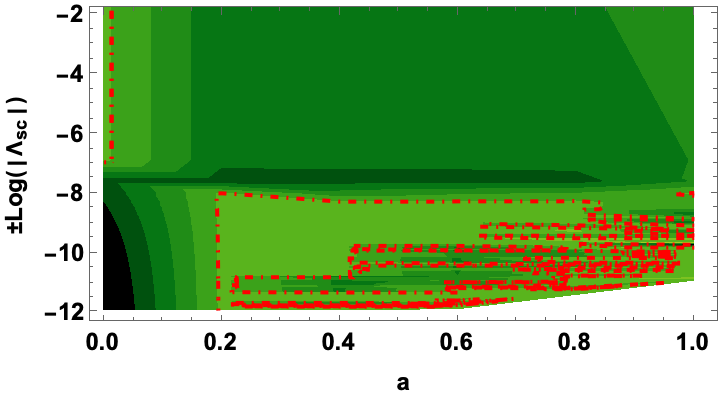}\\
};

\draw[violet, dashed,very thick] (-2.5,-1.4) rectangle (2.1,-1.9);
\draw[cyan, dashed,very thick] (-2.5,-6.3) rectangle (2.1,-6.85);
\draw[orange, dashed,very thick] (3.,-1.4) rectangle (7.6,-1.9);
\draw[magenta, dashed,very thick] (3.,-6.3) rectangle (7.6,-6.85);

\draw[violet, thick] (-8.4,-7.3) rectangle (-4.3,-9.6);
\draw[cyan, thick] (-4.25,-7.3) rectangle (-0.16,-9.6);
\draw[orange, thick] (-0.05,-7.3) rectangle (4.05,-9.6);
\draw[magenta, thick] (4.14,-7.3) rectangle (8.2,-9.6);
\end{tikzpicture}

\caption{\it \footnotesize Contours illustrating the dependence of the angular diameter $\mathcal{D}$,  and the fractional deviation $\bm \delta$ of the  $\textrm{Sgr~A}^\star$  on different planes of the accelerating black hole parameter. The solid violet line is associated with the VLTI measurements and the red dashed lines correspond estimated bounds.}
\label{figsgrAVLTI}
\end{figure*}

When mass and distance reported by the VLTI team are assumed, we depict in the Fig.\ref{figsgrAVLTI}, the angular diameter $\mathcal{D}$ and the fractional deviation $\bm{\delta}$ associated with the $\textrm{Sgr~A}^\star$ black hole in the $(\log(\Lambda_{sc}),\mathcal{A},a)$ diagram and the projection of such observables on each plane.

\begin{itemize}

\item In the $(a,\mathcal{A})$ plane, distinctive acceleration schemes appear among black holes (BHs) with low rotation rates, specifically for $a\lesssim 0.2$. BHs with slow acceleration ($\mathcal{A}\lesssim 0.01$) tend to exhibit an increase in both the angular diameter $\mathcal{D}$ and the fractional deviation $\bm{\delta}$ as the spin parameter $a$ increases. Whereas, when the acceleration becomes significant and reaches ($\mathcal{A}> 0.01$)  an increasing behavior appears in these observables due to a decrease in both $\mathcal{A}$ and $a$. In the case of highly rotating BHs with $a> 0.2$, the increase in observables is primarily driven by a decrease in both $\mathcal{A}$ and $a$.

Cross-referencing with experimental data in the context of angular diameter reveals that for slowly rotating black holes with $a\in[0.09^{+0.011}_{-0.012},0.2]$, the acceptable range of acceleration within the experimental band is $\mathcal{A} \in[0,0.0188^{+0.00061}_{-0.00071}]$. Meanwhile, for the range $a\in[0.2,1]$, the permissible band for the acceleration parameter is $\mathcal{A} \in[0.0188^{+0.00061}_{-0.00071},0.0166^{+0.00127}_{-0.00161}]$.

Regarding the fractional deviation, the acceptable acceleration ranges are estimated to be $\mathcal{A} \in[0,0.01875^{+0.00044}_{-0.00046}]$ and $\mathcal{A} \in[0.01875^{+0.00044}_{-0.00046},0.0165^{+0.00090}_{-0.00110}]$, corresponding to $a\in[0.09^{+0.010}_{-0.008},0.2]$ and $a\in[0.2,1]$, respectively.

\item The previously increasing scheme in the $(\Lambda, \mathcal{A})$ plane persists  as $\Lambda$ and $\mathcal{A}$  decrease. Otherwise, $\mathcal{D}$ allows the entire range of acceleration values $\mathcal{A}\in[0, 0.0187
^{+0.00062}_{-0.00067}]$. While, for the cosmological constant, it centered around $\Lambda = 4.62\times 10^{-5}$ (corresponding to the value $\mathcal{D} = 48.7\mu as$) with an absolute uncertainty margin between $1.14\times 10^{-4}$ and $-1.12\times 10^{-4}$ at the beginning of the acceleration interval and it ends at $\Lambda = -9.85\times 10^{-4}$, where an absolute margin of uncertainty between $-9.18\times 10^{-4}$ and $-9.85\times 10^{-4}$ have been taken.

Similarly, the fractional deviation $\bm{\delta}$ gives acceleration domain $\mathcal{A}\in[0, 0.0187
^{+0.00043}_{-0.00039}]$ and  the corresponding cosmological constant interval   begins from $\Lambda = 4.03\times 10^{-5}$ with an absolute uncertainty ranging from $8.75\times 10^{-5}$ to $-1.04\times 10^{-4}$, and ends at $\Lambda = -9.85\times 10^{-4}$ with uncertainties between $-9.34\times 10^{-4}$ and $-9.85\times 10^{-4}$. 
 Further investigations in the negative range of the cosmological constant for the observable $\mathcal{D}$ exhibit more intervals $\Lambda\in[-14.2914
, -0.3893
]$ and $\Lambda\in[-0.6863
, -0.0006
]$ within special acceleration values, evaluated to  $\mathcal{A}\in[0,0.1120]$ and $\mathcal{A}\in[0.3635, 0.4870]$ respectively.  Whereas the fraction deviation
 ${\bm \delta}$ gives  $\Lambda\in[-14.7266
, -0.3904
]$ and $\Lambda\in[-0.6870
, -0.1211
]$ associated with the acceleration ranges  $\mathcal{A}\in[0,0.1120]$ and $\mathcal{A}\in[0.3713, 0.4870]$ respectively.

\item The third line illustrating the $(\Lambda,a)$ plane discloses also three distinct behaviors. For BHs with low rotations ($a\lesssim 0.2$), both the angular diameter $\mathcal{D}$ and the fractional deviation $\bm{\delta}$ display an increasing trend with the rising spin parameter $a$ and cosmological constant $\Lambda$. Conversely, for BHs with high rotations ($0.2<a\lesssim 0.8$), the trend in the two observables primarily stems from the decrease in the cosmological constant $\Lambda$. Finally, for extremely rotating BHs ($a>0.8$), the observables $\mathcal{D}$ and $\bm{\delta}$ exhibit a decreasing trend with the increasing spin parameter $a$ and the increasing cosmological constant $\Lambda$.

Upon conducting a comparative analysis with the available experimental data on angular diameter, it has been observed that for slowly rotating black holes $a\in[0.088^{+0.012}_{-0.012},\\ 0.174^{+0.026}_{-0.028}]$, the cosmological constant is estimated to fall within a certain range $\Lambda\in[-9.83\times10^{-4},-1.13\times10^{-4}]$. However, for $a\in[0.2,1]$, the corresponding range of the cosmological constant is $\Lambda\in[(4.51\times10^{-5})^{+6.67\times10^{-5}}_{-1.58\times10^{-4}},(-2.02\times10^{-4})^{+7.43\times10^{-5}}_{-1.19\times10^{-4}}]$.

Such comparison in the case of the fractional deviation $\bm\delta$ indicates that for black holes with low rotations $a\in[0.088^{+0.009}_{-0.009},0.18^{+0.02}_{-0.02}]$, the experimental allowed cosmological constant is in $\Lambda\in[-9.88\times10^{-4},-1.05\times10^{-4}]$. Elsewhere, for all other physical values of the rotating parameter $a\in[0.2,1]$, the corresponding range of the cosmological constant is evaluated to be in $\Lambda\in[(3.92\times10^{-5})^{+4.75\times10^{-5}}_{-1.44\times10^{-4}},(-2.09\times10^{-4})^{+5.68\times10^{-5}}_{-8.00\times10^{-5}}]$.

Additional probes in negative cosmological domains uncover  within $\mathcal{D}$ and ${\bm \delta}$, $\Lambda\in[-0.1003 , -0.0006]$ and $\Lambda\in[-14.4350, -0.2869]$ for $a\in[0,0.012]$ and $a\in[0.19, 1]$ respectively.

\end{itemize}

\begin{figure*}\centering
\begin{tikzpicture}
\node[align=left] {
\includegraphics[width=0.5\textwidth]{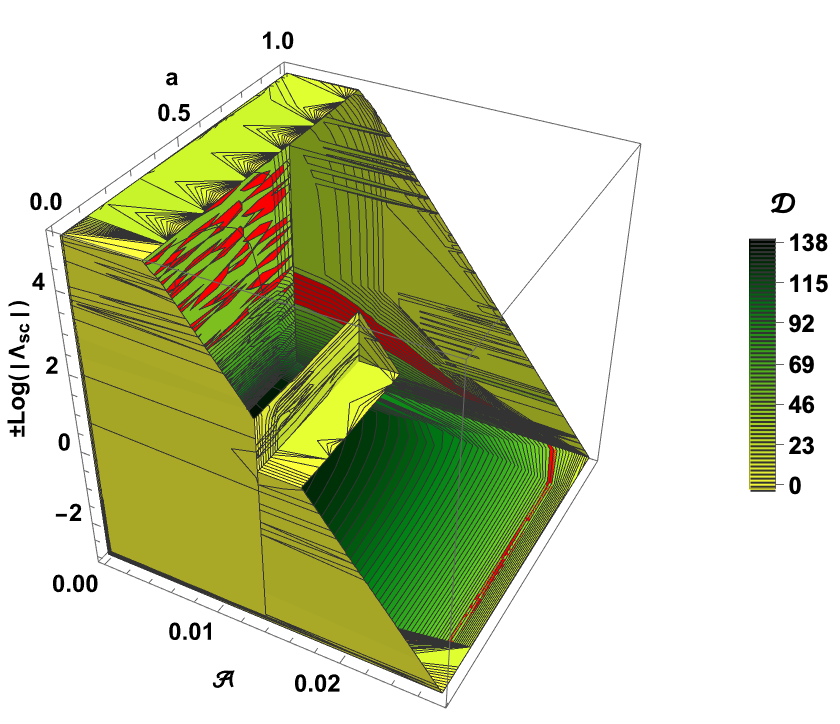}
\includegraphics[width=0.5\textwidth]{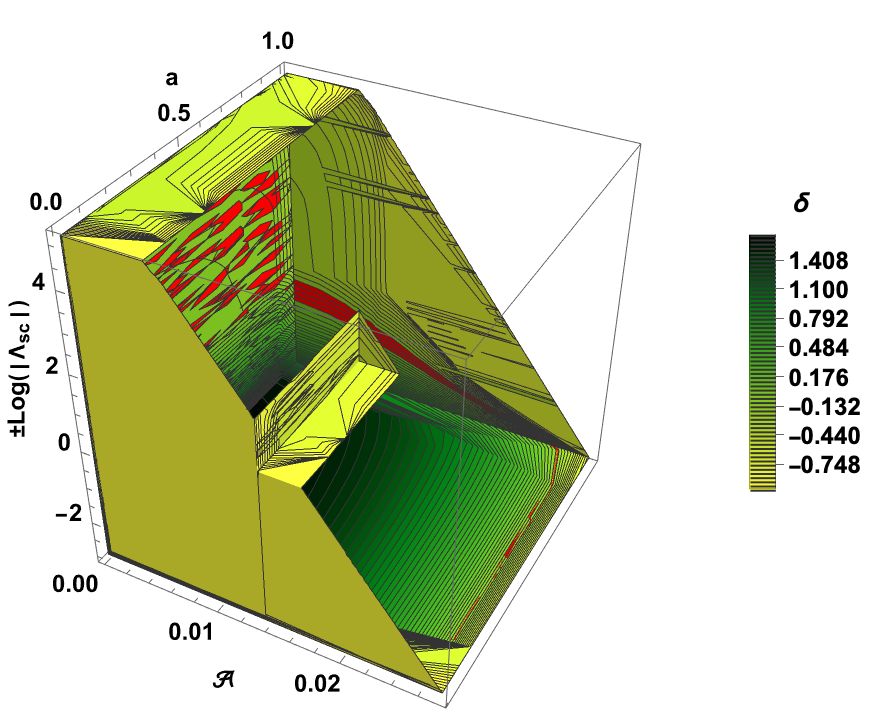}\\
\includegraphics[width=0.33\textwidth]{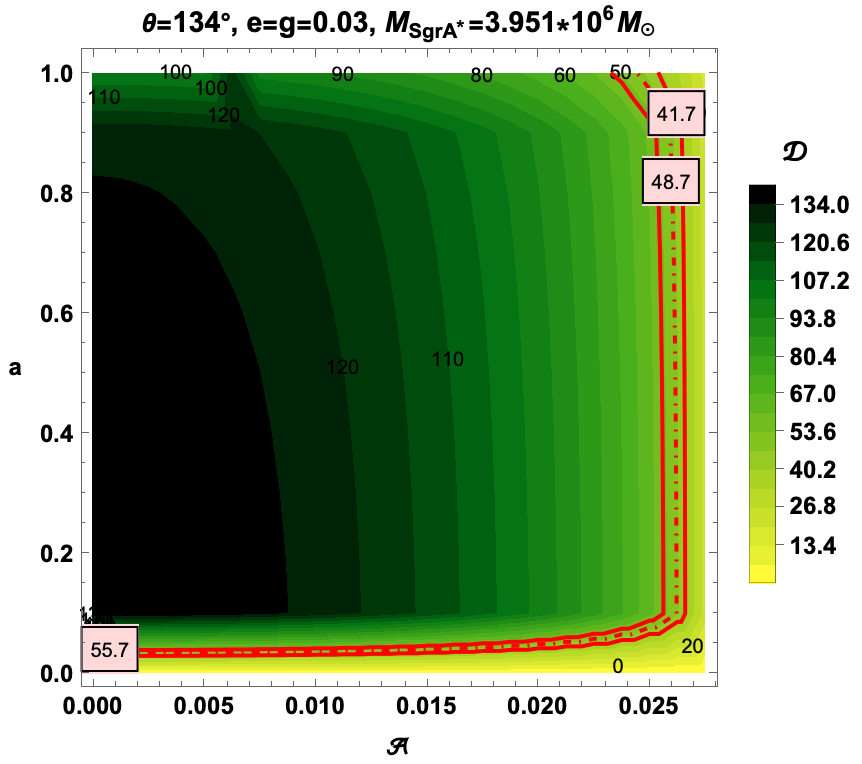}
\includegraphics[width=0.33\textwidth]{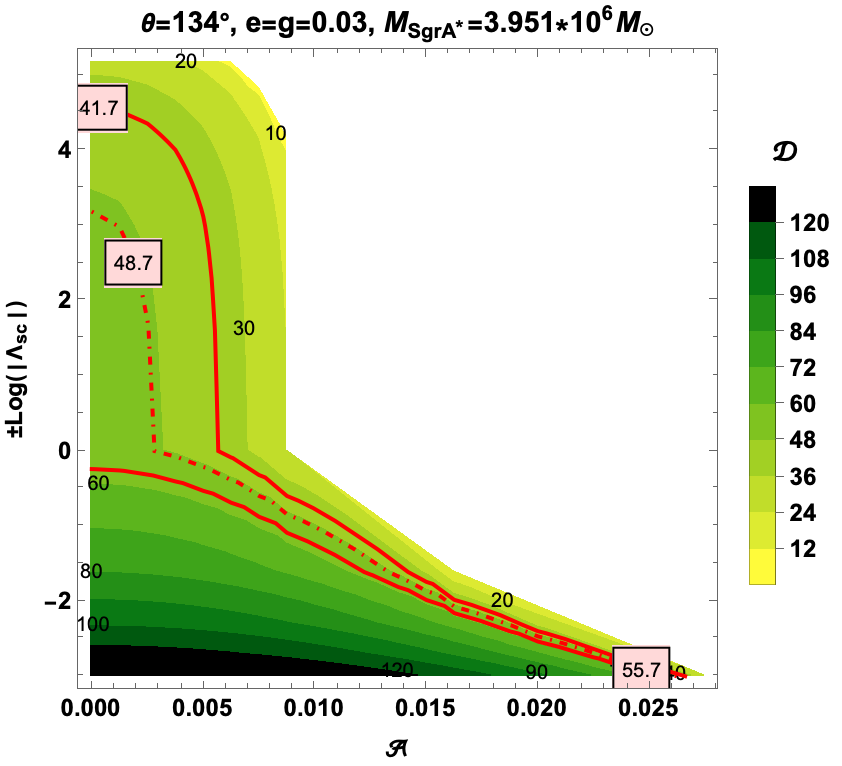}
\includegraphics[width=0.33\textwidth]{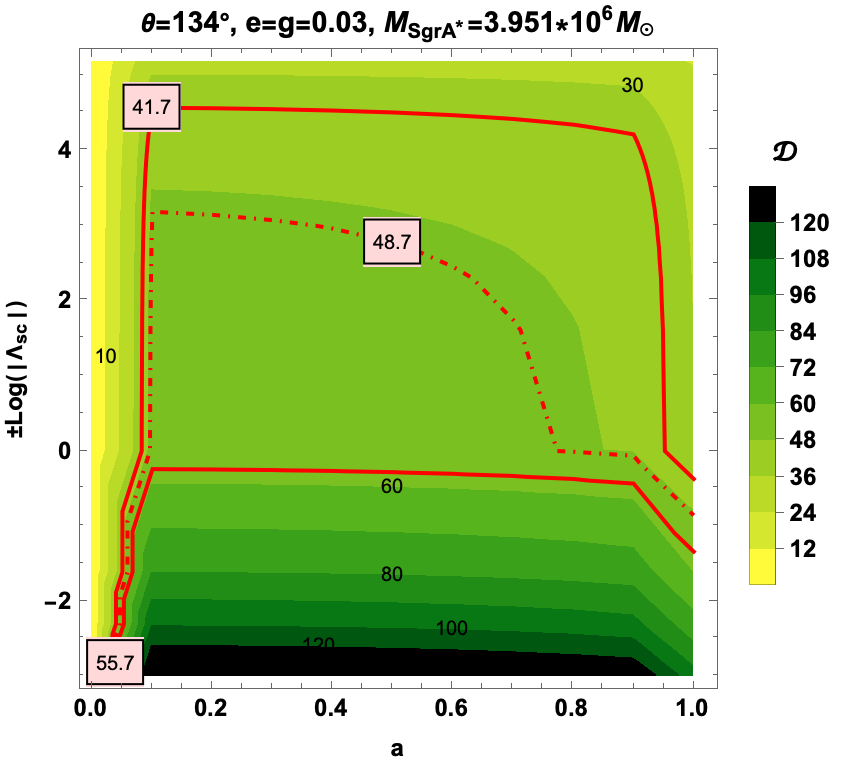}\\
\vspace{.2cm}
\includegraphics[width=0.33\textwidth]{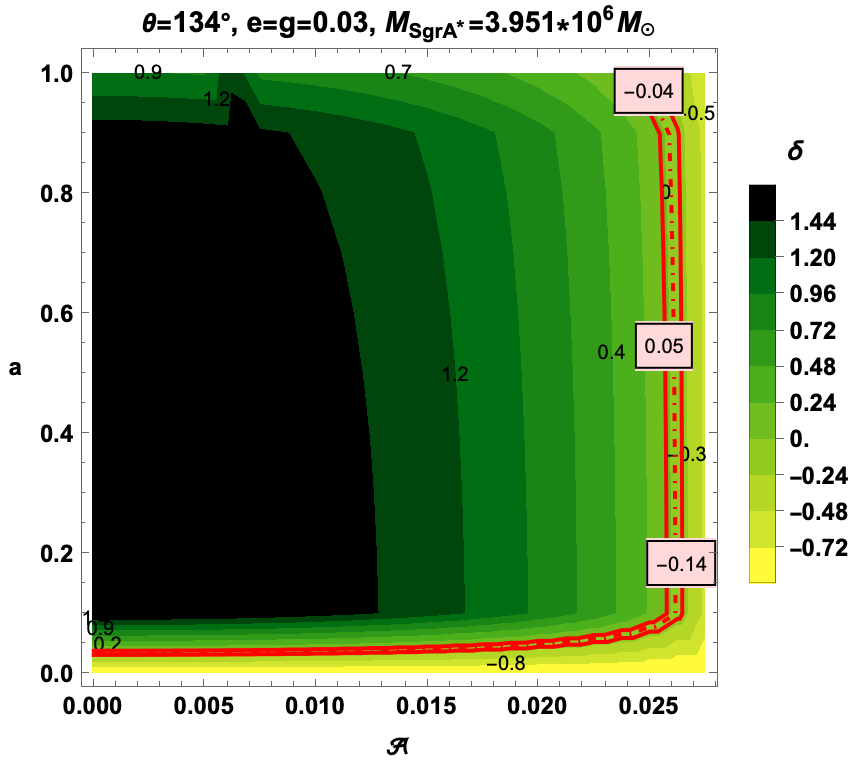}
\includegraphics[width=0.33\textwidth]{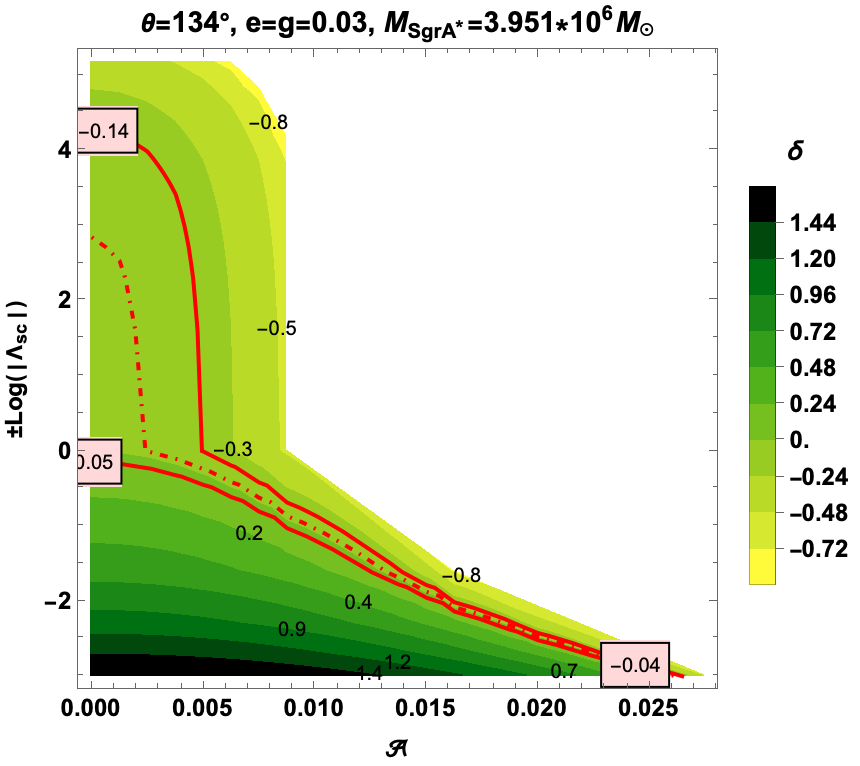}
\includegraphics[width=0.33\textwidth]{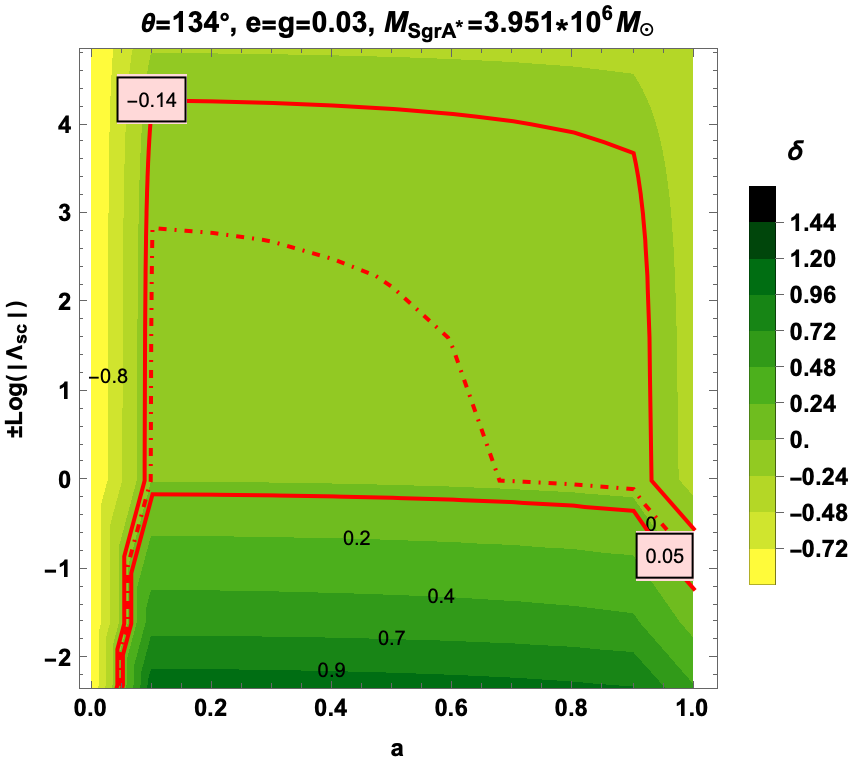}\\

\includegraphics[width=0.245\textwidth]{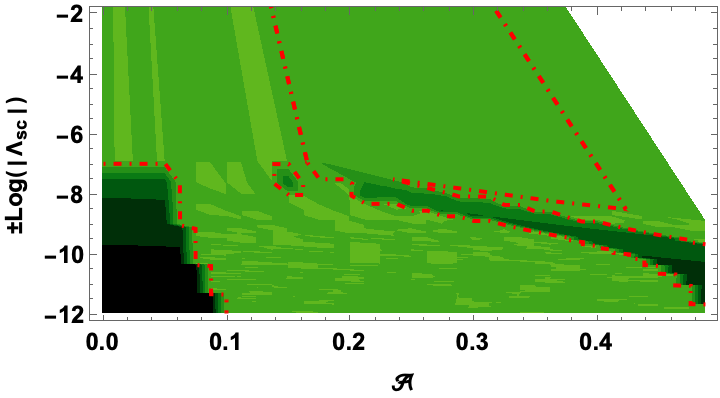}
\includegraphics[width=0.245\textwidth]{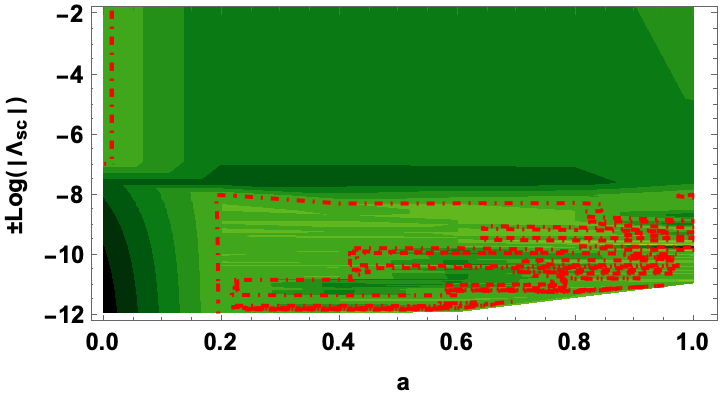}
\includegraphics[width=0.245\textwidth]{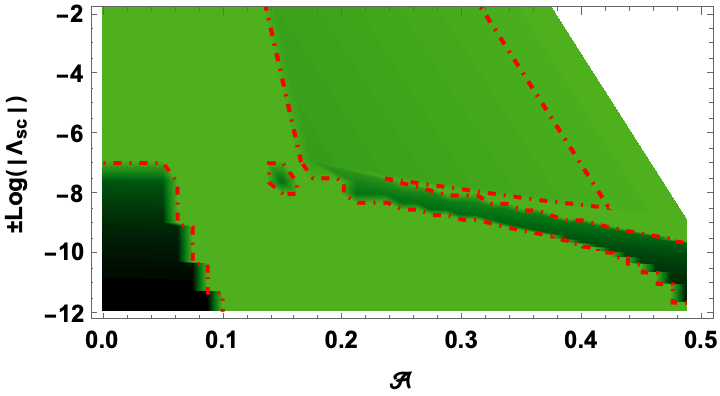}
\includegraphics[width=0.245\textwidth]{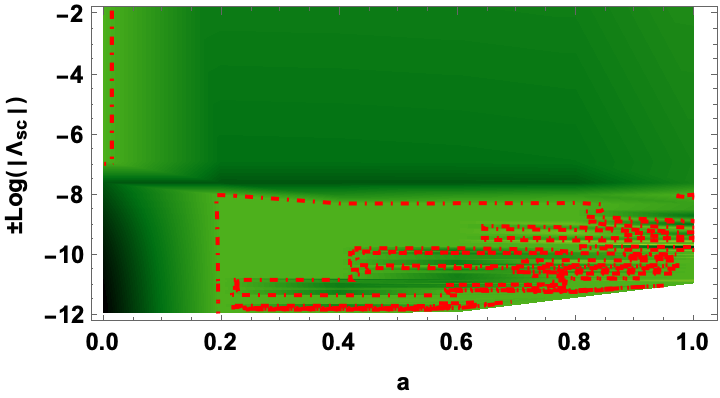}\\
};

\draw[violet, dashed,very thick] (-2.5,-1.4) rectangle (2.1,-1.9);
\draw[cyan, dashed,very thick] (-2.5,-6.3) rectangle (2.1,-6.85);
\draw[orange, dashed,very thick] (3.,-1.4) rectangle (7.6,-1.9);
\draw[magenta, dashed,very thick] (3.,-6.3) rectangle (7.6,-6.85);

\draw[violet, thick] (-8.4,-7.2) rectangle (-4.3,-9.5);
\draw[cyan, thick] (-4.25,-7.2) rectangle (-0.16,-9.5);
\draw[orange, thick] (-0.05,-7.2) rectangle (4.05,-9.5);
\draw[magenta, thick] (4.14,-7.2) rectangle (8.2,-9.5);
\end{tikzpicture}

\caption{\it \footnotesize Contours illustrating the dependence of the angular diameter $\mathcal{D}$,  and the fractional deviation $\bm \delta$ of the  $\textrm{Sgr~A}^\star$  on different planes of the accelerating black hole parameter. The solid violet line is associated with the Keck measurements and the red dashed lines correspond to estimated bounds.}
\label{figsgrAkeck}
\end{figure*}

The same graphical as Fig.\ref{figsgrAVLTI} illustrations are made in Fig.\ref{figsgrAkeck} when the Keck measurements are now considered.
 The behaviors of observables persist, and only the range parameter shifts slightly:
\begin{itemize}
\item In the $(a,\mathcal{A})$ plane, considering the low rotation values ($a\lesssim 0.1$), serval reduced regions compared to VLTI data within the acceleration parameter appear. Indeed, BHs with slow acceleration ($\mathcal{A}\lesssim 0.01$) show an increase in both the angular diameter $\mathcal{D}$ and the fractional deviation $\bm{\delta}$ as the spin parameter $a$ increases. Contrary, BHs with fast acceleration ($\mathcal{A}> 0.01$) exhibit a rise in the variation of these two quantities when $\mathcal{A}$ and $a$ increase.
 For $a> 0.1$, both observables increase when $\mathcal{A}$ and $a$ decrease. Concerning the angular diameter, it can be concluded that for slowly rotating black holes with $a\in[0.036^{+0.005}_{-0.005},0.1]$, the experimental band is permissible for the range of acceleration values $\mathcal{A} \in[0,0.02619^{+0.00039}_{-0.00060}]$. Meanwhile, for $a\in[0.1,1]$, the experimental band is allowed for $\mathcal{A} \in[0.02619^{+0.00039}{-0.00060},0.02447^{+0.00094}{-0.00108}]$ domain. 
 The specific ranges  $a\in[0.036^{+0.003}_{-0.003},0.1]$ and $a\in[0.1,1]$, corresponding to $\mathcal{A} \in[0,0.02614^{+0.00029}_{-0.00040}]$ and $\mathcal{A} \in[0.02614^{+0.00029}_{-0.00040},0.02438^{+0.00075}_{-0.00073}]$, respectively are associated with  the fractional deviation fitting.

\item Through $(\Lambda, \mathcal{A})$ projection, 
 the apparent diameter of the black hole shows an acceleration domain of $\mathcal{A}\in[0, 0.0185
 ^{+0.00069}_{-0.00078}]$. 
Whereas, the initial centered cosmological constant is
$\Lambda = 2.42\times 10^{-5}$ with absolute margins  $9.69\times 10^{-5}$ and $-1.25\times 10^{-4}$, while the ending value is estimated to be $\Lambda = -9.78\times 10^{-4}$ with absolute margins  $-9.05\times 10^{-4}$ and $-9.78\times 10^{-4}$.

The observable $\bm{\delta}$ presents an identical behavior to $\mathcal{D}$ by showing a range of acceleration parameter $\mathcal{A}\in[0, 0.0184
^{+0.00062}_{-0.00041}]$, the corresponding band of the cosmological constant is 
$[(1.72\times 10^{-5})^{+5.52\times 10^{-5}}_{-1.32\times 10^{-4}}, (-9.74\times 10^{-4})^{+6.04\times 10^{-5}}_{-0.00}]$.
 Additional probes in negative cosmological domains uncover  within $\mathcal{D}$, $\Lambda\in[-14.2914
, -0.3873
]$ and $\Lambda\in[-0.6904
, -0.0006
]$ for $\mathcal{A}\in[0,0.1120]$ and $\mathcal{A}\in[0.3613, 0.4870]$ respectively.

While the ${\bm \delta}$ discloses  $\Lambda\in[-14.7266
, -0.3908
]$ and $\Lambda\in[-0.6946
, -0.0362
]$ associated with $\mathcal{A}\in[0,0.1120]$ and $\mathcal{A}\in[0.3634, 0.4870]$ respectively.

\item  In the $(\Lambda,a)$ plane, once again, three distinct behaviors emerge. For BHs with low rotations $a\lesssim 0.1$, there is an increasing trend in the angular diameter $\mathcal{D}$ and the fractional deviation $\bm{\delta}$ as the spin parameter $a$ and cosmological constant $\Lambda$ increase. In contrast, BHs with high rotations $0.1<a\lesssim 0.9$ exhibit a decrease in the cosmological constant $\Lambda$.  Lastly, extremely rotating BHs ($a>0.9$) display a decreasing trend in the observables $\mathcal{D}$ and $\bm{\delta}$ with the increase in spin parameter ($a$) and cosmological constant ($\Lambda$).

The cross-comparison with the angular diameter observational data unveils that  for slow rotation parameter
 $a\in[0.034^{+0.005}_{-0.005},0.086^{+0.014}_{-0.014}]$, the cosmological constant is estimated to be in the interval $\Lambda\in[-1.98\times10^{-3},-1.26\times10^{-4}]$. While, for $a\in[0.1,1]$, the corresponding range is obtained to be $\Lambda\in[(2.41\times10^{-5})^{+7.28\times10^{-5}}_{-1.50\times10^{-4}},(-2.32\times10^{-4})^{+8.84\times10^{-5}}_{-1.45\times10^{-4}}]$.
While fractional deviation $\bm\delta$ suggests  $a\in[0.045^{+0.004}_{-0.004},0.18^{+0.01}_{-0.01}]$, the range of the cosmological constant is obtained to be $\Lambda\in[-9.31\times10^{-4},-1.15\times10^{-4}]$. Elsewhere, for all other possible values of the rotating parameter $a\in[0.1,1]$, the corresponding range is $\Lambda\in[(1.71\times10^{-5})^{+5.44\times10^{-5}}_{-1.33\times10^{-4}},(-2.38\times10^{-4})^{+6.75\times10^{-5}}_{-9.65\times10^{-5}}]$.

More zoom  in negative cosmological domains adds  through $\mathcal{D}$ and ${\bm \delta}$ the following intervals $\Lambda\in[-0.1015 , -0.0006]$ and $\Lambda\in[-14.4350, -0.2861]$ for $a\in[0,0.013]$ and $a\in[0.19, 1]$ respectively.

\end{itemize}

\begin{figure*}\centering
\begin{tikzpicture}
 \node [] at (-5,0    ) { \includegraphics[width=.48
\textwidth]{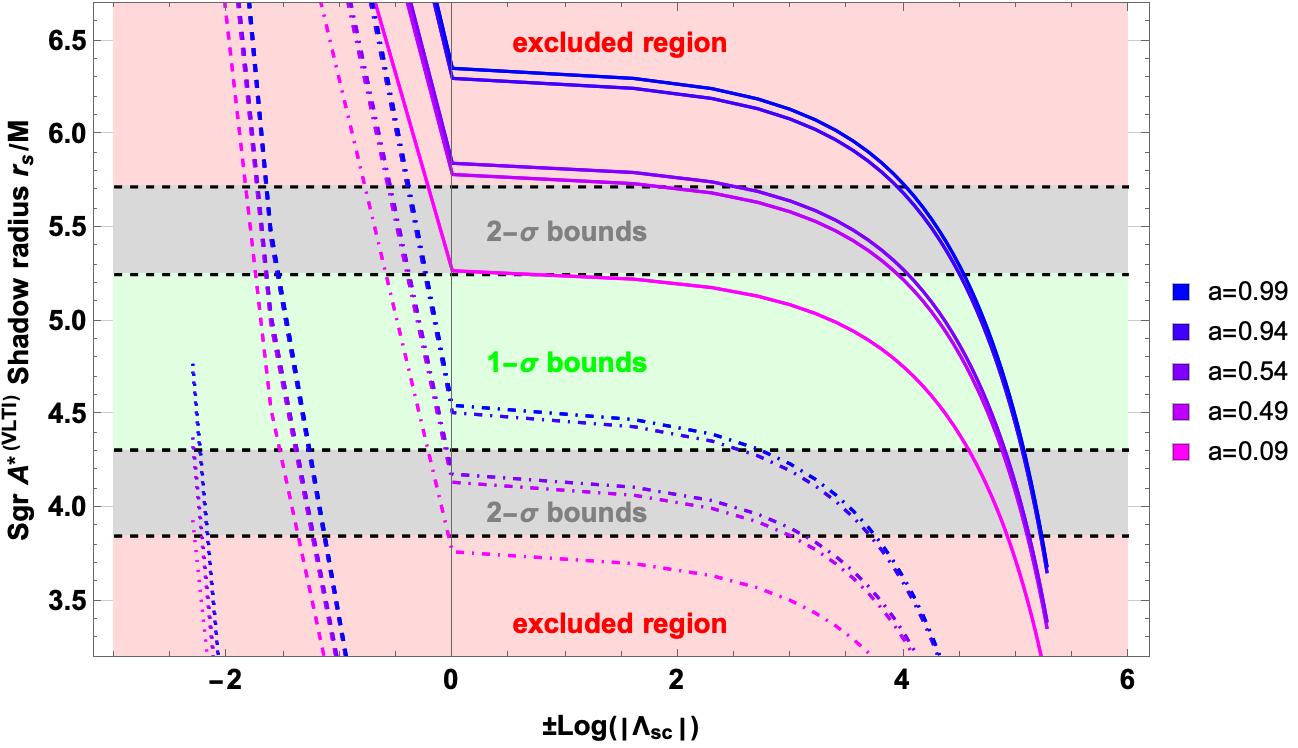} };
   \node[black,rotate=-35] at (-2.2,-2.1)     {\tiny$\mathcal{A}= 0$};
      \node[black,rotate=-35] at (-2.7,-2.2)     {\tiny$\mathcal{A}= 0.007$};
      \node[black,rotate=-35] at (-6.4,-2.2)     {\tiny$\mathcal{A}= 0.014$};
       \node[black,rotate=-35] at (-7.2,-2.29)     {\tiny$\mathcal{A}= 0.02$};
 \node [] at (4.2,0   ) { \includegraphics[width=.48
\textwidth]{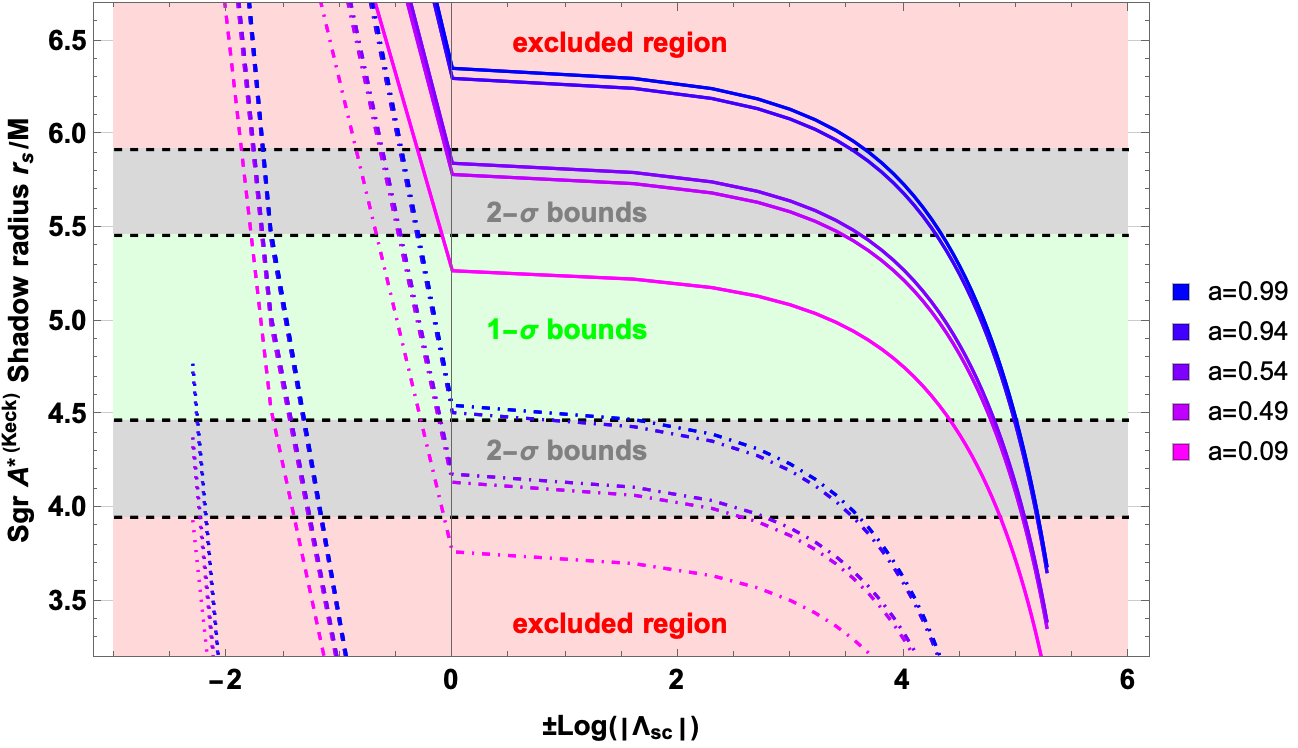} };
 \node[black,rotate=-35] at (7.1,-2.1)     {\tiny$\mathcal{A}= 0$};
      \node[black,rotate=-35] at (6.4,-2.2)     {\tiny$\mathcal{A}= 0.007$};
      \node[black,rotate=-35] at (2.75,-2.2)     {\tiny$\mathcal{A}= 0.014$};
       \node[black,rotate=-35] at (2.1,-2.29)     {\tiny$\mathcal{A}= 0.02$};
\node [] at (-5,-4   ) { \includegraphics[width=.5
\textwidth]{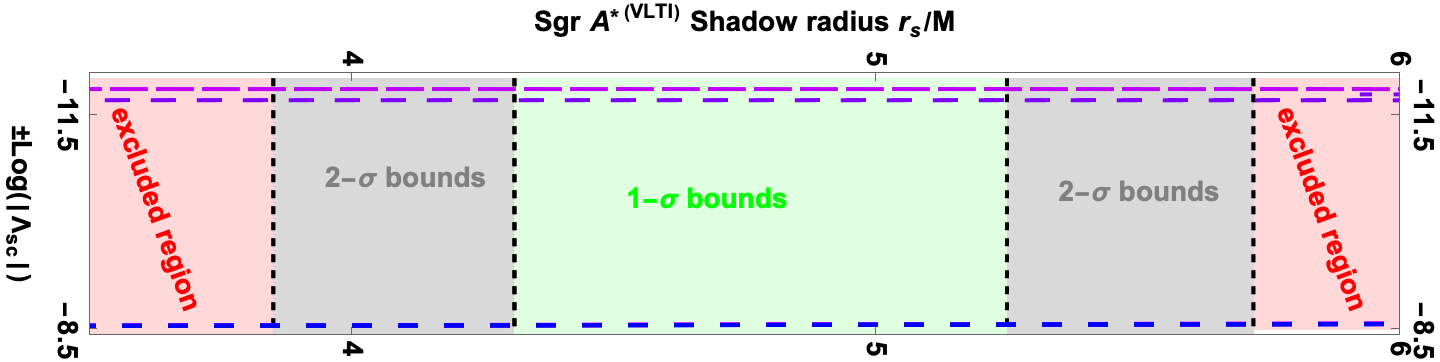} };
  \node[black,rotate=-0] at (0.8,-5.4)     {$\mathcal{A}\simeq 0.5$};
\node [] at (4.2,-4   ) { \includegraphics[width=.5
\textwidth]{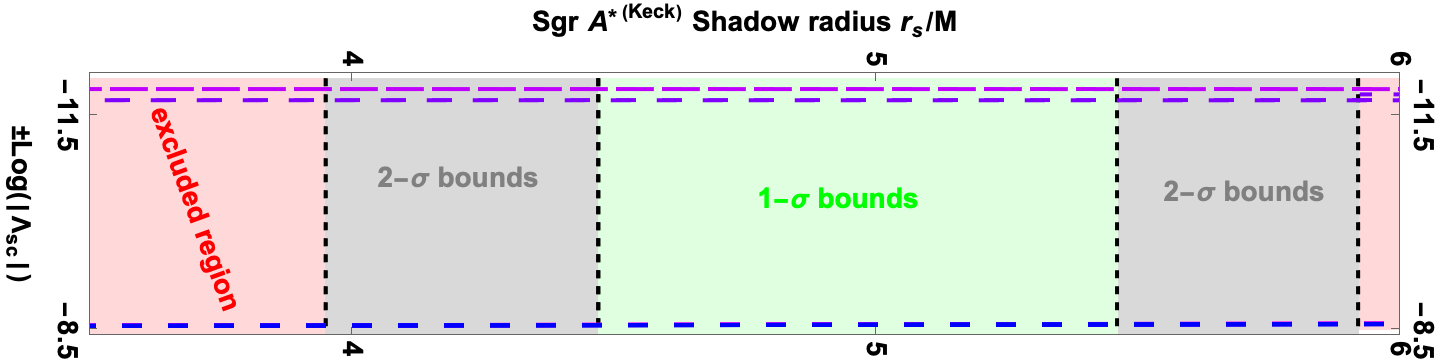} };
  \node[black,rotate=-0] at (-8.3,-5.4)     {$\mathcal{A}\simeq 0.5$};
\draw[violet,dashed,very thick] (-8.7,2.4) rectangle (-8.2,-2.1);
\draw[violet,dashed,very thick] (-9.2,-2.8) rectangle (-.8,-5.2);
\draw[olive,dashed,very thick] (0.5,2.4) rectangle (1.,-2.1);
\draw[olive,dashed,very thick] (.0,-2.8) rectangle (8.4,-5.2);
\end{tikzpicture}

\caption{ \it\footnotesize   The shadow radii $r_\text{s}/M$ viewed on the equatorial plane (deviated from $\pi/2$) of the considered accelerating black hole with cosmological constant as a function of the cosmological constant by varying the rotation and acceleration parameters. The olive/gray/red shaded regions refer to the areas that are (1-$\sigma$)\big/(2-$\sigma$) consistent$\big/$ inconsistent with the $\textrm{Sgr~A}^\star$ observations and highlight that the latter set constraints on the black hole parameters. {\bf Left:} The VLTI measurements are considered.
{\bf Right:} Through the Keck data.
{\bf Bottom}: Deep zoom in the negative cosmological constant domain where the quantity $r_\text{s}/M$ allows us to reach the acceleration value $\mathcal{A}\simeq0.5$.
}
\label{rssgrVLTI}
\end{figure*}
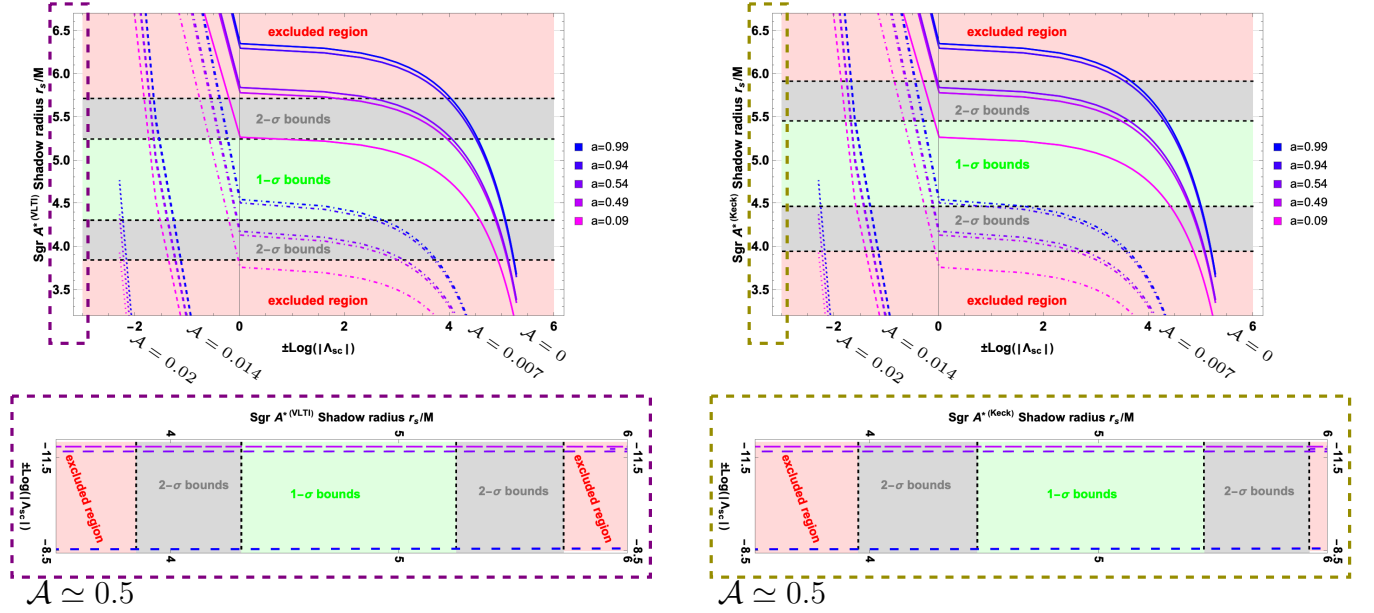

Now,  and like the $\textrm{M87}^\star$ scenario, the ensuing pivotal phase of our observational investigations involves representing the limitations on the shadow radii $\frac{r_\text{s}}{M}$ as observed from the equatorial plane based on data from the VLTI and Keck observatories. These constraints are depicted in Fig.\ref{rssgrVLTI}.

The top-left panel of Fig.\ref{rssgrVLTI} associated with the VLTI reveals that, under low accelerations, the upper constraints on the cosmological constant concerning shadow radii are $(\Lambda_{\text{1-}\sigma} = 1.5457\times10^{-4}, \Lambda_{ \text{2-}\sigma} = 1.8208\times10^{-4})$. In contrast, the bottom panel demonstrates that within the acceleration parameter range $\mathcal{A}\in[0.4,0.4875]$, the cosmological constant exhibits a range of variation from $\Lambda=-14.0084$ to $\Lambda=-0.5218$. 
 The right panel, depicting Keck observatory estimations, indicates similar constraints to those elucidated by the VLTI measurements for high accelerations. However, for low accelerations, the results reveal ($\Lambda_{\text{1-}\sigma} = 8.7774\times10^{-5}$, $\Lambda_{\text{2-}\sigma} = 1.8015\times10^{-4}$).

At present, we do not know any constraint on axis ratio $\Delta A$ and the deviation from circularity $\Delta \mathcal{C}$ quantities from the shadow of $\textrm{Sgr~A}^\star$, and thus no more constraints on the black hole parameters can be established regarding these observables.

\section{Conclusions}
\label{sec:Conclusions}

The observations of the supermassive black holes $\textrm{M87}^\star$ and $\textrm{Sgr~A}^\star$ by the Event Horizon Telescope present an unparalleled opportunity to scrutinize various theories of gravity while also unveiling crucial insights into the nature of black holes. In essence, the data collected from these observations enables rigorous examinations of gravitational theories at the horizon scale, where the strong gravity regime governs, providing a means to extract invaluable information about the fundamental attributes of different classes of black holes. Thus, investigating the signature of black holes in astrophysical observations is highly worthwhile.

In this study, we have delved into the shadow characteristics of $AdS_4\big/dS_4$ black hole solutions within the framework of Einstein-Maxwell theory. These black holes exhibit acceleration, rotation, and carry both electric and magnetic charges. To be more precise, we have meticulously analyzed the contour and dimensions of the shadow cast by such a black hole, while also investigating the impact of each parameter on these inherent properties.
Certainly, due to the black hole's acceleration, the circular orbits of photons with fixed radial and latitude coordinates deviate from the equatorial plane, leading to noteworthy alterations in the black hole shadow's characteristics. Moreover, the sign of the cosmological constant plays a pivotal role in modulating these shadow properties, in particular, its size
 diminishes as the acceleration parameter increases, with the sign of the cosmological constant exerting a substantial influence. Meanwhile, the magnitude of rotation predominantly influences the shape of the shadow. In contrast, the charges exhibit a modest impact on the shadow size, leading to a slight reduction as their values increase. Furthermore, we have also explored the implications of the observer's distance on the shadow configuration.

Subsequently, we have computed the essentials of the geometrical observables, namely the angular diameter $\mathcal{D}$, the fractional deviation $\bm{\delta}$, the deviation from circularity $\Delta\mathcal{C}$, the axis ratio $\Delta A$ and the shadow radius $r_\text{s}/M$. Then, we have drawn a comparison between our findings and the measurements obtained by the Event Horizon Telescope to construct some constraints on the black hole parameters.

After confronting the resulting shadow of the black hole with observational data of the $\textrm{M87}^\star$ to find the allowed regions of the black hole model parameters for which the obtained shadow is consistent with such data. We have obtained relevant constraints on the acceleration parameter and the cosmological constant. Specifically, the acceleration parameter $\mathcal{A}$ falls within the interval $[0,0.475^{+0.0125}_{-0.0125}]$, and the cosmological constant spans the range $\Lambda\in[-14.5801^{+1.1051\times10^{-6}}_{-1.1051\times10^{-6}},(1.0086\times10^{-6})^{+9.2121\times10^{-5}}_{-1.5396\times10^{-4}}]$.
 Next, realizing the same examinations but with the $\textrm{Sgr~A}^\star$ measurements summarizes the following constraints: the acceleration scheme maintains the same interval range as the $\textrm{M87}^\star$. However, the cosmological constant undergoes a slight shift to the domain $[-14.7266^{+1.1051\times10^{-6}}_{-1.1051\times10^{-6}},(4.6200\times10^{-5})^{+6.7890\times10^{-5}}_{-1.5888\times10^{-4}}]$ for the VLTI observatory. Additionally, within the Keck measurements, the cosmological constant is found in the range $\Lambda\in[-14.7266^{+1.1051\times10^{-6}}_{-1.1051\times10^{-6}},(2.4149\times10^{-5})^{+7.2836\times10^{-5}}_{-1.5094\times10^{-4}}]$. Regarding other parameters, both black holes exhibit a spin parameter $a$ within the range of $[0,1]$, and the electrical/magnetic charges are constrained to $\{e,g\}\in[0,0.09[$.
 Moreover, the sensitivity to $\frac{r_{\text{s}}}{M}$ measurements primarily resides within the cosmological constant intervals, particularly in the case of $\textrm{M87}^\star$ data. In this context, the upper bound undergoes a shift to $(\Lambda_{\text{1-}\sigma} = 1.1361\times10^{-4}, \Lambda_{ \text{2-}\sigma} = 1.7822\times10^{-4})$. The VLTI observations yield a range of $(\Lambda_{\text{1-}\sigma} = 1.5457\times10^{-4}, \Lambda_{ \text{2-}\sigma} = 1.8208\times10^{-4})$, and the Keck results reveal only $\Lambda_{\text{2-}\sigma} = 1.8015\times10^{-4}$.

In summary, this study highlights the dependence of the black hole shadow's geometrical properties on several parameters: the acceleration and rotation parameters. The sign and magnitude of the cosmological constant also influence the size of the shadow.  
Moreover, the Event Horizon Telescope images serve as a powerful tool for imposing constraints on black hole parameters and gravity theories. 
 Finally, it remains plausible that negative values of the cosmological constant could give rise to black hole solutions within the framework of minimal gauged supergravity, potentially serving as viable candidates for astrophysical black holes.

Going ahead, the Next Generation EHT (ngEHT) project will give us much sharper pictures of $\textrm{M87}^\star$ and $\textrm{Sgr~A}^\star$ black holes, as well as possible real-time footage of the evolution of the accretion disk surrounding the black hole horizon. This will usher certainly in a new era in strong gravity fundamental physics.
 Otherwise, this study raises other several intriguing questions, one of which revolves around the possibility of fingerprinting supersymmetric black hole solutions using data from the Event Horizon Telescope. Addressing this specific aspect is a subject we plan to delve into in future research endeavors.

\section*{Data availability}
Since this work is theoretical, there is no data used to support the findings of this study.

\bibliographystyle{
unsrt}
\bibliography{shad_acc.bib}

\begin{thebibliography}{100}

\bibitem{EventHorizonTelescope:2019dse}
Kazunori Akiyama et~al.
\newblock {First M87 Event Horizon Telescope Results. I. The Shadow of the
  Supermassive Black Hole}.
\newblock {\em Astrophys. J. Lett.}, 875:L1, 2019.

\bibitem{EventHorizonTelescope:2019ths}
Kazunori Akiyama et~al.
\newblock {First M87 Event Horizon Telescope Results. IV. Imaging the Central
  Supermassive Black Hole}.
\newblock {\em Astrophys. J. Lett.}, 875(1):L4, 2019.

\bibitem{EventHorizonTelescope:2022wkp}
Kazunori Akiyama et~al.
\newblock {First Sagittarius A* Event Horizon Telescope Results. I. The Shadow
  of the Supermassive Black Hole in the Center of the Milky Way}.
\newblock {\em Astrophys. J. Lett.}, 930(2):L12, 2022.

\bibitem{EventHorizonTelescope:2022apq}
Kazunori Akiyama et~al.
\newblock {First Sagittarius A* Event Horizon Telescope Results. II. EHT and
  Multiwavelength Observations, Data Processing, and Calibration}.
\newblock {\em Astrophys. J. Lett.}, 930(2):L13, 2022.

\bibitem{LIGOScientific:2016emj}
B.~P. Abbott et~al.
\newblock {GW150914: The Advanced LIGO Detectors in the Era of First
  Discoveries}.
\newblock {\em Phys. Rev. Lett.}, 116(13):131103, 2016.

\bibitem{Bardeen:1973tla}
J.~M. Bardeen.
\newblock {Timelike and null geodesics in the Kerr metric}.
\newblock In {\em {Les Houches Summer School of Theoretical Physics}: {Black
  Holes}}, pages 215--240, 1973.

\bibitem{Falcke:1999pj}
Heino Falcke, Fulvio Melia, and Eric Agol.
\newblock {Viewing the shadow of the black hole at the galactic center}.
\newblock {\em Astrophys. J. Lett.}, 528:L13, 2000.

\bibitem{Fathi:2023ccx}
Mohsen Fathi and Norman Cruz.
\newblock {Observational signatures of a static f(R) black hole with thin
  accretion disk}.
\newblock {\em Eur. Phys. J. C}, 83(12):1160, 2023.

\bibitem{Saadati:2023jym}
Reza Saadati and Fatimah Shojai.
\newblock {Geodetic precession and shadow of quantum extended black holes}.
\newblock {\em Class. Quant. Grav.}, 41(1):015032, 2024.

\bibitem{Hoshimov:2023tlz}
Husanboy Hoshimov, Odil Yunusov, Farruh Atamurotov, Mubasher Jamil, and
  Ahmadjon Abdujabbarov.
\newblock {Weak gravitational lensing and shadow of a GUP-modified
  Schwarzschild black hole in the presence of plasma}.
\newblock {\em Phys. Dark Univ.}, 43:101392, 2024.

\bibitem{Gao:2023mjb}
Xiao-Jun Gao, Tao-Tao Sui, Xiao-Xiong Zeng, Yu-Sen An, and Ya-Peng Hu.
\newblock {Investigating shadow images and rings of the charged Horndeski black
  hole illuminated by various thin accretions}.
\newblock {\em Eur. Phys. J. C}, 83:1052, 2023.

\bibitem{Gregoris:2023wrg}
Daniele Gregoris.
\newblock {On some new black hole, wormhole and naked singularity solutions in
  the free Dirac\textendash{}Born\textendash{}Infeld theory}.
\newblock {\em Eur. Phys. J. C}, 83(11):1056, 2023.

\bibitem{Zahid:2023csk}
Muhammad Zahid, Javlon Rayimbaev, Furkat Sarikulov, Saeed~Ullah Khan, and
  Jingli Ren.
\newblock {Shadow of rotating and twisting charged black holes with cloud of
  strings and quintessence}.
\newblock {\em Eur. Phys. J. C}, 83(9):855, 2023.

\bibitem{Ovgun:2023wmc}
Ali \"Ovg\"un, Lemuel John~F. Sese, and Reggie~C. Pantig.
\newblock {Constraints via the Event Horizon Telescope for Black Hole Solutions
  with Dark Matter under the Generalized Uncertainty Principle Minimal Length
  Scale Effect}.
\newblock {\em Annalen Phys.}, 2023:2300390, 9 2023.

\bibitem{Ovgun:2023ego}
Ali \"Ovg\"un, Reggie~C. Pantig, and \'Angel Rinc\'on.
\newblock {4D scale-dependent Schwarzschild-AdS/dS black holes: study of shadow
  and weak deflection angle and greybody bounding}.
\newblock {\em Eur. Phys. J. Plus}, 138(3):192, 2023.

\bibitem{Zubair:2023cor}
M.~Zubair, Muhammad~Ali Raza, and Eiman Maqsood.
\newblock {Rotating black hole in Kalb\textendash{}Ramond gravity: Constraining
  parameters by comparison with EHT observations of Sgr A* and M87*}.
\newblock {\em Phys. Dark Univ.}, 42:101334, 2023.

\bibitem{Magos:2023nnb}
Daniela Magos, Nora Bret\'on, and Alfredo Mac\'\i{}as.
\newblock {Orbits in static magnetically and dyonically charged
  Einstein-Euler-Heisenberg black hole spacetimes}.
\newblock {\em Phys. Rev. D}, 108(6):064014, 2023.

\bibitem{Lambiase:2023zeo}
Gaetano Lambiase, Leonardo Mastrototaro, Reggie~C. Pantig, and Ali Ovgun.
\newblock {Probing Schwarzschild-like black holes in metric-affine bumblebee
  gravity with accretion disk, deflection angle, greybody bounds, and neutrino
  propagation}.
\newblock {\em JCAP}, 12:026, 2023.

\bibitem{Yang:2023agi}
Yi~Yang, Dong Liu, Ali \"Ovg\"un, Gaetano Lambiase, and Zheng-Wen Long.
\newblock {Rotating black hole mimicker surrounded by the string cloud}.
\newblock {\em Phys. Rev. D}, 109(2):024002, 2024.

\bibitem{Weyl:1917gp}
H.~Weyl.
\newblock {The theory of gravitation}.
\newblock {\em Annalen Phys.}, 54:117--145, 1917.

\bibitem{newman1961new}
ET~Newman and LA~Tamburino.
\newblock New approach to einstein's empty space field equations.
\newblock {\em Journal of Mathematical Physics}, 2(5):667--674, 1961.

\bibitem{Robinson:1962zz}
I.~Robinson and A.~Trautman.
\newblock {Some spherical gravitational waves in general relativity}.
\newblock {\em Proc. Roy. Soc. Lond. A}, 265:463--473, 1962.

\bibitem{PhysRevD.2.1359}
William Kinnersley and Martin Walker.
\newblock Uniformly accelerating charged mass in general relativity.
\newblock {\em Phys. Rev. D}, 2:1359--1370, Oct 1970.

\bibitem{Griffiths:2006tk}
J.~B. Griffiths, P.~Krtous, and J.~Podolsky.
\newblock {Interpreting the C-metric}.
\newblock {\em Class. Quant. Grav.}, 23:6745--6766, 2006.

\bibitem{Plebanski:1976gy}
J.~F. Plebanski and M.~Demianski.
\newblock {Rotating, charged, and uniformly accelerating mass in general
  relativity}.
\newblock {\em Annals Phys.}, 98:98--127, 1976.

\bibitem{Podolsky:2003gm}
Jiri Podolsky, Marcello Ortaggio, and Pavel Krtous.
\newblock {Radiation from accelerated black holes in an anti-de Sitter
  universe}.
\newblock {\em Phys. Rev. D}, 68:124004, 2003.

\bibitem{Krtous:2003tc}
Pavel Krtous and Jiri Podolsky.
\newblock {Radiation from accelerated black holes in de Sitter universe}.
\newblock {\em Phys. Rev. D}, 68:024005, 2003.

\bibitem{Kinnersley:1970zw}
W.~Kinnersley and M.~Walker.
\newblock {Uniformly accelerating charged mass in general relativity}.
\newblock {\em Phys. Rev. D}, 2:1359--1370, 1970.

\bibitem{Dias:2002mi}
Oscar J.~C. Dias and Jose P.~S. Lemos.
\newblock {Pair of accelerated black holes in anti-de Sitter background: AdS C
  metric}.
\newblock {\em Phys. Rev. D}, 67:064001, 2003.

\bibitem{Dias:2003xp}
Oscar J.~C. Dias and Jose P.~S. Lemos.
\newblock {Pair of accelerated black holes in a de Sitter background: The dS C
  metric}.
\newblock {\em Phys. Rev. D}, 67:084018, 2003.

\bibitem{Griffiths:2005qp}
J.~B. Griffiths and J.~Podolsky.
\newblock {A New look at the Plebanski-Demianski family of solutions}.
\newblock {\em Int. J. Mod. Phys. D}, 15:335--370, 2006.

\bibitem{EslamPanah:2022ihg}
B.~Eslam~Panah.
\newblock {Charged Accelerating BTZ Black Holes}.
\newblock {\em Fortsch. Phys.}, 71(8-9):2300012, 2023.

\bibitem{EslamPanah:2023ypz}
B.~Eslam~Panah.
\newblock {Three-dimensional energy-dependent C-metric: Black hole solutions}.
\newblock {\em Phys. Lett. B}, 844:138111, 2023.

\bibitem{EslamPanah:2024dfq}
B.~Eslam~Panah, S.~Zare, and H.~Hassanabadi.
\newblock {Accelerating AdS black holes in gravity\textquoteright{}s rainbow}.
\newblock {\em Eur. Phys. J. C}, 84(3):259, 2024.

\bibitem{Dowker:1993bt}
Fay Dowker, Jerome~P. Gauntlett, David~A. Kastor, and Jennie~H. Traschen.
\newblock {Pair creation of dilaton black holes}.
\newblock {\em Phys. Rev. D}, 49:2909--2917, 1994.

\bibitem{Gregory:1995hd}
Ruth Gregory and Mark Hindmarsh.
\newblock {Smooth metrics for snapping strings}.
\newblock {\em Phys. Rev. D}, 52:5598--5605, 1995.

\bibitem{Eardley:1995au}
Douglas~M. Eardley, Gary~T. Horowitz, David~A. Kastor, and Jennie~H. Traschen.
\newblock {Breaking cosmic strings without monopoles}.
\newblock {\em Phys. Rev. Lett.}, 75:3390--3393, 1995.

\bibitem{Emparan:2001wn}
Roberto Emparan and Harvey~S. Reall.
\newblock {A Rotating black ring solution in five-dimensions}.
\newblock {\em Phys. Rev. Lett.}, 88:101101, 2002.

\bibitem{Appels:2016uha}
Michael Appels, Ruth Gregory, and David Kubiznak.
\newblock {Thermodynamics of Accelerating Black Holes}.
\newblock {\em Phys. Rev. Lett.}, 117(13):131303, 2016.

\bibitem{Astorino:2016ybm}
Marco Astorino.
\newblock {Thermodynamics of Regular Accelerating Black Holes}.
\newblock {\em Phys. Rev. D}, 95(6):064007, 2017.

\bibitem{Cassani:2021dwa}
Davide Cassani, Jerome~P. Gauntlett, Dario Martelli, and James Sparks.
\newblock {Thermodynamics of accelerating and supersymmetric AdS4 black holes}.
\newblock {\em Phys. Rev. D}, 104(8):086005, 2021.

\bibitem{Anabalon:2018ydc}
Andr\'es Anabal\'on, Michael Appels, Ruth Gregory, David Kubiz\v{n}\'ak,
  Robert~B. Mann, and Al\"\i{} Ovg\"un.
\newblock {Holographic Thermodynamics of Accelerating Black Holes}.
\newblock {\em Phys. Rev. D}, 98(10):104038, 2018.

\bibitem{Anabalon:2018qfv}
Andr\'es Anabal\'on, Finnian Gray, Ruth Gregory, David Kubiz\v{n}\'ak, and
  Robert~B. Mann.
\newblock {Thermodynamics of Charged, Rotating, and Accelerating Black Holes}.
\newblock {\em JHEP}, 04:096, 2019.

\bibitem{Zhang:2018vqs}
Jialin Zhang, Yanjun Li, and Hongwei Yu.
\newblock {Accelerating AdS black holes as the holographic heat engines in a
  benchmarking scheme}.
\newblock {\em Eur. Phys. J. C}, 78(8):645, 2018.

\bibitem{Zhang:2018hms}
Jialin Zhang, Yanjun Li, and Hongwei Yu.
\newblock {Thermodynamics of charged accelerating AdS black holes and
  holographic heat engines}.
\newblock {\em JHEP}, 02:144, 2019.

\bibitem{Rostami:2019ivr}
M.~Rostami, J.~Sadeghi, S.~Miraboutalebi, A.~A. Masoudi, and B.~Pourhassan.
\newblock {Charged accelerating AdS black hole of $f(R)$ gravity and the
  Joule\textendash{}Thomson expansion}.
\newblock {\em Int. J. Geom. Meth. Mod. Phys.}, 17(09):2050136, 2020.

\bibitem{Belhaj:2020lai}
Adil Belhaj, Hasan El~Moumni, and Karima Masmar.
\newblock {$f(R)$ Gravity Effects on Charged Accelerating AdS Black Holes using
  Holographic Tools}.
\newblock {\em Adv. High Energy Phys.}, 2020:4092730, 2020.

\bibitem{Barrientos:2023tqb}
Jose Barrientos and Adolfo Cisterna.
\newblock {Ehlers transformations as a tool for constructing accelerating NUT
  black holes}.
\newblock {\em Phys. Rev. D}, 108(2):024059, 2023.

\bibitem{Barrientos:2023dlf}
Jos\'e Barrientos, Adolfo Cisterna, and Konstantinos Pallikaris.
\newblock {Pleban\'ski-Demia\'nski \`a la Ehlers-Harrison: Exact Rotating and
  Accelerating Type I Black Holes}.
\newblock 9 2023.

\bibitem{Barrientos:2022bzm}
Jose Barrientos, Adolfo Cisterna, David Kubiznak, and Julio Oliva.
\newblock {Accelerated black holes beyond Maxwell's electrodynamics}.
\newblock {\em Phys. Lett. B}, 834:137447, 2022.

\bibitem{Vilenkin:2018zol}
Alexander Vilenkin, Yuri Levin, and Andrei Gruzinov.
\newblock {Cosmic strings and primordial black holes}.
\newblock {\em JCAP}, 11:008, 2018.

\bibitem{Gussmann:2021mjj}
Alexander Gu\ss{}mann.
\newblock {Polarimetric signatures of the photon ring of a black hole that is
  pierced by a cosmic axion string}.
\newblock {\em JHEP}, 08:160, 2021.

\bibitem{Maldacena:1997re}
Juan~Martin Maldacena.
\newblock {The Large N limit of superconformal field theories and
  supergravity}.
\newblock {\em Adv. Theor. Math. Phys.}, 2:231--252, 1998.

\bibitem{Benini:2015eyy}
Francesco Benini, Kiril Hristov, and Alberto Zaffaroni.
\newblock {Black hole microstates in AdS$_{4}$ from supersymmetric
  localization}.
\newblock {\em JHEP}, 05:054, 2016.

\bibitem{Benini:2016rke}
Francesco Benini, Kiril Hristov, and Alberto Zaffaroni.
\newblock {Exact microstate counting for dyonic black holes in AdS4}.
\newblock {\em Phys. Lett. B}, 771:462--466, 2017.

\bibitem{Benini:2018ywd}
Francesco Benini and Paolo Milan.
\newblock {Black Holes in 4D $\mathcal{N}$=4 Super-Yang-Mills Field Theory}.
\newblock {\em Phys. Rev. X}, 10(2):021037, 2020.

\bibitem{Nian:2019pxj}
Jun Nian and Leopoldo~A. Pando~Zayas.
\newblock {Microscopic entropy of rotating electrically charged AdS$_{4}$ black
  holes from field theory localization}.
\newblock {\em JHEP}, 03:081, 2020.

\bibitem{Hosseini:2017fjo}
Seyed~Morteza Hosseini, Kiril Hristov, and Achilleas Passias.
\newblock {Holographic microstate counting for AdS$_{4}$ black holes in massive
  IIA supergravity}.
\newblock {\em JHEP}, 10:190, 2017.

\bibitem{Benini:2017oxt}
Francesco Benini, Hrachya Khachatryan, and Paolo Milan.
\newblock {Black hole entropy in massive Type IIA}.
\newblock {\em Class. Quant. Grav.}, 35(3):035004, 2018.

\bibitem{Bobev:2019zmz}
Nikolay Bobev and P.~Marcos Crichigno.
\newblock {Universal spinning black holes and theories of class $ \mathcal{R}
  $}.
\newblock {\em JHEP}, 12:054, 2019.

\bibitem{Peebles:2002gy}
P.~J.~E. Peebles and Bharat Ratra.
\newblock {The Cosmological Constant and Dark Energy}.
\newblock {\em Rev. Mod. Phys.}, 75:559--606, 2003.

\bibitem{Kubiznak:2012wp}
David Kubiznak and Robert~B. Mann.
\newblock {P-V criticality of charged AdS black holes}.
\newblock {\em JHEP}, 07:033, 2012.

\bibitem{Cai:2013qga}
Rong-Gen Cai, Li-Ming Cao, Li~Li, and Run-Qiu Yang.
\newblock {P-V criticality in the extended phase space of Gauss-Bonnet black
  holes in AdS space}.
\newblock {\em JHEP}, 09:005, 2013.

\bibitem{Belhaj:2015hha}
A.~Belhaj, M.~Chabab, H.~El~Moumni, K.~Masmar, M.~B. Sedra, and A.~Segui.
\newblock {On Heat Properties of AdS Black Holes in Higher Dimensions}.
\newblock {\em JHEP}, 05:149, 2015.

\bibitem{Belhaj:2015uwa}
A.~Belhaj, M.~Chabab, H.~El~Moumni, K.~Masmar, and M.~B. Sedra.
\newblock {On Thermodynamics of AdS Black Holes in M-Theory}.
\newblock {\em Eur. Phys. J. C}, 76(2):73, 2016.

\bibitem{Belhaj:2020nqy}
A.~Belhaj, L.~Chakhchi, H.~El~Moumni, J.~Khalloufi, and K.~Masmar.
\newblock {Thermal Image and Phase Transitions of Charged AdS Black Holes using
  Shadow Analysis}.
\newblock {\em Int. J. Mod. Phys. A}, 35(27):2050170, 2020.

\bibitem{ElMoumni:2018fml}
H.~El~Moumni.
\newblock {Revisiting the phase transition of
  AdS-Maxwell\textendash{}power-Yang\textendash{}Mills black holes via AdS/CFT
  tools}.
\newblock {\em Phys. Lett. B}, 776:124--132, 2018.

\bibitem{EventHorizonTelescope:2019uob}
Kazunori Akiyama et~al.
\newblock {First M87 Event Horizon Telescope Results. II. Array and
  Instrumentation}.
\newblock {\em Astrophys. J. Lett.}, 875(1):L2, 2019.

\bibitem{EventHorizonTelescope:2019jan}
Kazunori Akiyama et~al.
\newblock {First M87 Event Horizon Telescope Results. III. Data Processing and
  Calibration}.
\newblock {\em Astrophys. J. Lett.}, 875(1):L3, 2019.

\bibitem{EventHorizonTelescope:2019pgp}
Kazunori Akiyama et~al.
\newblock {First M87 Event Horizon Telescope Results. V. Physical Origin of the
  Asymmetric Ring}.
\newblock {\em Astrophys. J. Lett.}, 875(1):L5, 2019.

\bibitem{EventHorizonTelescope:2019ggy}
Kazunori Akiyama et~al.
\newblock {First M87 Event Horizon Telescope Results. VI. The Shadow and Mass
  of the Central Black Hole}.
\newblock {\em Astrophys. J. Lett.}, 875(1):L6, 2019.

\bibitem{EventHorizonTelescope:2021bee}
Kazunori Akiyama et~al.
\newblock {First M87 Event Horizon Telescope Results. VII. Polarization of the
  Ring}.
\newblock {\em Astrophys. J. Lett.}, 910(1):L12, 2021.

\bibitem{EventHorizonTelescope:2021srq}
Kazunori Akiyama et~al.
\newblock {First M87 Event Horizon Telescope Results. VIII. Magnetic Field
  Structure near The Event Horizon}.
\newblock {\em Astrophys. J. Lett.}, 910(1):L13, 2021.

\bibitem{Konoplya:2019xmn}
R.~A. Konoplya.
\newblock {Quantum corrected black holes: quasinormal modes, scattering,
  shadows}.
\newblock {\em Phys. Lett. B}, 804:135363, 2020.

\bibitem{Chael:2021rjo}
Andrew Chael, Michael~D. Johnson, and Alexandru Lupsasca.
\newblock {Observing the Inner Shadow of a Black Hole: A Direct View of the
  Event Horizon}.
\newblock {\em Astrophys. J.}, 918(1):6, 2021.

\bibitem{Belhaj:2020okh}
A.~Belhaj, H.~Belmahi, M.~Benali, W.~El~Hadri, H.~El~Moumni, and
  E.~Torrente-Lujan.
\newblock {Shadows of 5D black holes from string theory}.
\newblock {\em Phys. Lett. B}, 812:136025, 2021.

\bibitem{Belhaj:2020rdb}
A.~Belhaj, M.~Benali, A.~El~Balali, H.~El~Moumni, and S.~E. Ennadifi.
\newblock {Deflection angle and shadow behaviors of quintessential black holes
  in arbitrary dimensions}.
\newblock {\em Class. Quant. Grav.}, 37(21):215004, 2020.

\bibitem{Ghasemi-Nodehi:2021ipd}
M.~Ghasemi-Nodehi, Chandrachur Chakraborty, Qingjuan Yu, and Youjun Lu.
\newblock {Investigating the existence of gravitomagnetic monopole in M87*}.
\newblock {\em Eur. Phys. J. C}, 81(10):939, 2021.

\bibitem{Bacchini:2021fig}
Fabio Bacchini, Daniel~R. Mayerson, Bart Ripperda, Jordy Davelaar, H\'ector
  Olivares, Thomas Hertog, and Bert Vercnocke.
\newblock {Fuzzball Shadows: Emergent Horizons from Microstructure}.
\newblock {\em Phys. Rev. Lett.}, 127(17):171601, 2021.

\bibitem{Atamurotov:2021hoq}
Farruh Atamurotov, Ahmadjon Abdujabbarov, and Wen-Biao Han.
\newblock {Effect of plasma on gravitational lensing by a Schwarzschild black
  hole immersed in perfect fluid dark matter}.
\newblock {\em Phys. Rev. D}, 104(8):084015, 2021.

\bibitem{Belhaj:2021rae}
A.~Belhaj, H.~Belmahi, and M.~Benali.
\newblock {Superentropic AdS black hole shadows}.
\newblock {\em Phys. Lett. B}, 821:136619, 2021.

\bibitem{Afrin:2021imp}
Misba Afrin, Rahul Kumar, and Sushant~G. Ghosh.
\newblock {Parameter estimation of hairy Kerr black holes from its shadow and
  constraints from M87*}.
\newblock {\em Mon. Not. Roy. Astron. Soc.}, 504:5927--5940, 2021.

\bibitem{Konoplya:2021slg}
R.~A. Konoplya and A.~Zhidenko.
\newblock {Shadows of parametrized axially symmetric black holes allowing for
  separation of variables}.
\newblock {\em Phys. Rev. D}, 103(10):104033, 2021.

\bibitem{Rayimbaev:2022mrk}
Javlon Rayimbaev, Bushra Majeed, Mubasher Jamil, Kimet Jusufi, and Anzhong
  Wang.
\newblock {Quasiperiodic oscillations, quasinormal modes and shadows of
  Bardeen\textendash{}Kiselev Black Holes}.
\newblock {\em Phys. Dark Univ.}, 35:100930, 2022.

\bibitem{Guo:2021bhr}
Sen Guo, Guan-Ru Li, and En-Wei Liang.
\newblock {Influence of accretion flow and magnetic charge on the observed
  shadows and rings of the Hayward black hole}.
\newblock {\em Phys. Rev. D}, 105(2):023024, 2022.

\bibitem{Cimdiker:2021cpz}
\.Irfan \c{C}imdiker, Durmu\c{s} Demir, and Ali \"Ovg\"un.
\newblock {Black hole shadow in symmergent gravity}.
\newblock {\em Phys. Dark Univ.}, 34:100900, 2021.

\bibitem{Okyay:2021nnh}
Mert Okyay and Ali \"Ovg\"un.
\newblock {Nonlinear electrodynamics effects on the black hole shadow,
  deflection angle, quasinormal modes and greybody factors}.
\newblock {\em JCAP}, 01(01):009, 2022.

\bibitem{Bronzwaer:2021lzo}
Thomas Bronzwaer and Heino Falcke.
\newblock {The Nature of Black Hole Shadows}.
\newblock {\em Astrophys. J.}, 920(2):155, 2021.

\bibitem{Liu:2021yev}
Cheng Liu, Sen Yang, Qiang Wu, and Tao Zhu.
\newblock {Thin accretion disk onto slowly rotating black holes in
  Einstein-\AE{}ther theory}.
\newblock {\em JCAP}, 02(02):034, 2022.

\bibitem{Shaikh:2021yux}
Rajibul Shaikh, Kunal Pal, Kuntal Pal, and Tapobrata Sarkar.
\newblock {Constraining alternatives to the Kerr black hole}.
\newblock {\em Mon. Not. Roy. Astron. Soc.}, 506(1):1229--1236, 2021.

\bibitem{Zhao:2015fya}
Fan Zhao and Jianfeng Tang.
\newblock {Gravitational lensing effects of Schwarzschild\textendash{}de Sitter
  black hole}.
\newblock {\em Phys. Rev. D}, 92(8):083011, 2015.

\bibitem{Ishihara:2016sfv}
Asahi Ishihara, Yusuke Suzuki, Toshiaki Ono, and Hideki Asada.
\newblock {Finite-distance corrections to the gravitational bending angle of
  light in the strong deflection limit}.
\newblock {\em Phys. Rev. D}, 95(4):044017, 2017.

\bibitem{Faraoni:2016wae}
Valerio Faraoni and Marianne Lapierre-Leonard.
\newblock {Beyond lensing by the cosmological constant}.
\newblock {\em Phys. Rev. D}, 95(2):023509, 2017.

\bibitem{He:2017alg}
Hong-Jian He and Zhen Zhang.
\newblock {Direct Probe of Dark Energy through Gravitational Lensing Effect}.
\newblock {\em JCAP}, 08:036, 2017.

\bibitem{Belhaj:2020kwv}
A.~Belhaj, M.~Benali, A.~El Balali, W.~El Hadri, and H.~El~Moumni.
\newblock {Cosmological constant effect on charged and rotating black hole
  shadows}.
\newblock {\em Int. J. Geom. Meth. Mod. Phys.}, 18(12):2150188, 2021.

\bibitem{Firouzjaee_2019}
Javad~T. Firouzjaee and Alireza Allahyari.
\newblock Black hole shadow with a cosmological constant for cosmological
  observers.
\newblock {\em The European Physical Journal C}, 79(11), nov 2019.

\bibitem{Afrin:2021ggx}
Misba Afrin and Sushant~G. Ghosh.
\newblock {Estimating the Cosmological Constant from Shadows of
  Kerr\textendash{}de Sitter Black Holes}.
\newblock {\em Universe}, 8(1):52, 2022.

\bibitem{Hou:2021okc}
Yehui Hou, Minyong Guo, and Bin Chen.
\newblock {Revisiting the shadow of braneworld black holes}.
\newblock {\em Phys. Rev. D}, 104(2):024001, 2021.

\bibitem{Maluf:2020kgf}
R.~V. Maluf and Juliano C.~S. Neves.
\newblock {Black holes with a cosmological constant in bumblebee gravity}.
\newblock {\em Phys. Rev. D}, 103(4):044002, 2021.

\bibitem{Vagnozzi:2019apd}
Sunny Vagnozzi and Luca Visinelli.
\newblock {Hunting for extra dimensions in the shadow of M87*}.
\newblock {\em Phys. Rev. D}, 100(2):024020, 2019.

\bibitem{Allahyari:2019jqz}
Alireza Allahyari, Mohsen Khodadi, Sunny Vagnozzi, and David~F. Mota.
\newblock {Magnetically charged black holes from non-linear electrodynamics and
  the Event Horizon Telescope}.
\newblock {\em JCAP}, 02:003, 2020.

\bibitem{Chabab:2019kfs}
M.~Chabab, H.~El~Moumni, S.~Iraoui, and K.~Masmar.
\newblock {Probing correlation between photon orbits and phase structure of
  charged AdS black hole in massive gravity background}.
\newblock {\em Int. J. Mod. Phys. A}, 34(35):1950231, 2020.

\bibitem{Khodadi:2020jij}
Mohsen Khodadi, Alireza Allahyari, Sunny Vagnozzi, and David~F. Mota.
\newblock {Black holes with scalar hair in light of the Event Horizon
  Telescope}.
\newblock {\em JCAP}, 09:026, 2020.

\bibitem{Chowdhuri:2020ipb}
Abhishek Chowdhuri and Arpan Bhattacharyya.
\newblock {Shadow analysis for rotating black holes in the presence of plasma
  for an expanding universe}.
\newblock {\em Phys. Rev. D}, 104(6):064039, 2021.

\bibitem{Vagnozzi:2022moj}
Sunny Vagnozzi et~al.
\newblock {Horizon-scale tests of gravity theories and fundamental physics from
  the Event Horizon Telescope image of Sagittarius A}.
\newblock {\em Class. Quant. Grav.}, 40(16):165007, 2023.

\bibitem{Ghosh:2022kit}
Sushant~G. Ghosh and Misba Afrin.
\newblock {An Upper Limit on the Charge of the Black Hole Sgr A* from EHT
  Observations}.
\newblock {\em Astrophys. J.}, 944(2):174, 2023.

\bibitem{KumarWalia:2022aop}
Rahul Kumar~Walia, Sushant~G. Ghosh, and Sunil~D. Maharaj.
\newblock {Testing Rotating Regular Metrics with EHT Results of Sgr A*}.
\newblock {\em Astrophys. J.}, 939(2):77, 2022.

\bibitem{Banerjee:2022iok}
Indrani Banerjee, Subhadip Sau, and Soumitra SenGupta.
\newblock {Signatures of regular black holes from the shadow of Sgr A* and
  M87*}.
\newblock {\em JCAP}, 09:066, 2022.

\bibitem{Khodadi:2022pqh}
Mohsen Khodadi and Gaetano Lambiase.
\newblock {Probing Lorentz symmetry violation using the first image of
  Sagittarius A*: Constraints on standard-model extension coefficients}.
\newblock {\em Phys. Rev. D}, 106(10):104050, 2022.

\bibitem{Uniyal:2022vdu}
Akhil Uniyal, Reggie~C. Pantig, and Ali \"Ovg\"un.
\newblock {Probing a non-linear electrodynamics black hole with thin accretion
  disk, shadow, and deflection angle with M87* and Sgr A* from EHT}.
\newblock {\em Phys. Dark Univ.}, 40:101178, 2023.

\bibitem{Afrin:2022ztr}
Misba Afrin, Sunny Vagnozzi, and Sushant~G. Ghosh.
\newblock {Tests of Loop Quantum Gravity from the Event Horizon Telescope
  Results of Sgr A*}.
\newblock {\em Astrophys. J.}, 944(2):149, 2023.

\bibitem{Shaikh:2022ivr}
Rajibul Shaikh.
\newblock {Testing black hole mimickers with the Event Horizon Telescope image
  of Sagittarius A*}.
\newblock {\em Mon. Not. Roy. Astron. Soc.}, 523(1):375--384, 2023.

\bibitem{Banerjee:2022jog}
Indrani Banerjee, Sumanta Chakraborty, and Soumitra SenGupta.
\newblock {Hunting extra dimensions in the shadow of Sgr A*}.
\newblock {\em Phys. Rev. D}, 106(8):084051, 2022.

\bibitem{Ghosh:2022mka}
Saptaswa Ghosh and Arpan Bhattacharyya.
\newblock {Analytical study of gravitational lensing in Kerr-Newman
  black-bounce spacetime}.
\newblock {\em JCAP}, 11:006, 2022.

\bibitem{Islam:2020xmy}
Shafqat~Ul Islam, Rahul Kumar, and Sushant~G. Ghosh.
\newblock {Gravitational lensing by black holes in the $4D$
  Einstein-Gauss-Bonnet gravity}.
\newblock {\em JCAP}, 09:030, 2020.

\bibitem{Kumar:2020owy}
Rahul Kumar and Sushant~G. Ghosh.
\newblock {Rotating black holes in $4D$ Einstein-Gauss-Bonnet gravity and its
  shadow}.
\newblock {\em JCAP}, 07:053, 2020.

\bibitem{Chandrasekhar:1985kt}
Subrahmanyan Chandrasekhar.
\newblock {\em {The mathematical theory of black holes}}.
\newblock Oxford Classic Texts in the Physical Sciences. Clarendon Press,
  November 5, 1998.

\bibitem{Kumar:2020yem}
Rahul Kumar, Amit Kumar, and Sushant~G. Ghosh.
\newblock {Testing Rotating Regular Metrics as Candidates for Astrophysical
  Black Holes}.
\newblock {\em Astrophys. J.}, 896(1):89, 2020.

\bibitem{Grenzebach:2015oea}
Arne Grenzebach, Volker Perlick, and Claus L\"ammerzahl.
\newblock {Photon Regions and Shadows of Accelerated Black Holes}.
\newblock {\em Int. J. Mod. Phys. D}, 24(09):1542024, 2015.

\bibitem{Zhang:2020xub}
Ming Zhang and Jie Jiang.
\newblock {Shadows of accelerating black holes}.
\newblock {\em Phys. Rev. D}, 103(2):025005, 2021.

\bibitem{Carter:1968rr}
Brandon Carter.
\newblock {Global structure of the Kerr family of gravitational fields}.
\newblock {\em Phys. Rev.}, 174:1559--1571, 1968.

\bibitem{Griffiths:2009dfa}
Jerry~B. Griffiths and Jiri Podolsky.
\newblock {\em {Exact Space-Times in Einstein's General Relativity}}.
\newblock Cambridge Monographs on Mathematical Physics. Cambridge University
  Press, Cambridge, 2009.

\bibitem{Cunha:2016bpi}
Pedro V.~P. Cunha, Carlos A.~R. Herdeiro, Eugen Radu, and Helgi~F. Runarsson.
\newblock {Shadows of Kerr black holes with and without scalar hair}.
\newblock {\em Int. J. Mod. Phys. D}, 25(09):1641021, 2016.

\bibitem{Hioki:2009na}
Kenta Hioki and Kei-ichi Maeda.
\newblock {Measurement of the Kerr Spin Parameter by Observation of a Compact
  Object's Shadow}.
\newblock {\em Phys. Rev. D}, 80:024042, 2009.

\bibitem{Banerjee:2019nnj}
Indrani Banerjee, Sumanta Chakraborty, and Soumitra SenGupta.
\newblock {Silhouette of M87*: A New Window to Peek into the World of Hidden
  Dimensions}.
\newblock {\em Phys. Rev. D}, 101(4):041301, 2020.

\bibitem{Johannsen:2010ru}
Tim Johannsen and Dimitrios Psaltis.
\newblock {Testing the No-Hair Theorem with Observations in the Electromagnetic
  Spectrum: II. Black-Hole Images}.
\newblock {\em Astrophys. J.}, 718:446--454, 2010.

\bibitem{Bambi:2019tjh}
Cosimo Bambi, Katherine Freese, Sunny Vagnozzi, and Luca Visinelli.
\newblock {Testing the rotational nature of the supermassive object M87* from
  the circularity and size of its first image}.
\newblock {\em Phys. Rev. D}, 100(4):044057, 2019.

\bibitem{EventHorizonTelescope:2022xqj}
Kazunori Akiyama et~al.
\newblock {First Sagittarius A* Event Horizon Telescope Results. VI. Testing
  the Black Hole Metric}.
\newblock {\em Astrophys. J. Lett.}, 930(2):L17, 2022.

\bibitem{EventHorizonTelescope:2021dqv}
Prashant Kocherlakota et~al.
\newblock {Constraints on black-hole charges with the 2017 EHT observations of
  M87*}.
\newblock {\em Phys. Rev. D}, 103(10):104047, 2021.

\bibitem{Do:2019txf}
Tuan Do et~al.
\newblock {Relativistic redshift of the star S0-2 orbiting the Galactic center
  supermassive black hole}.
\newblock {\em Science}, 365(6454):664--668, 2019.

\bibitem{GRAVITY:2021xju}
R.~Abuter et~al.
\newblock {Mass distribution in the Galactic Center based on interferometric
  astrometry of multiple stellar orbits}.
\newblock {\em Astron. Astrophys.}, 657:L12, 2022.

\bibitem{GRAVITY:2020gka}
R.~Abuter et~al.
\newblock {Detection of the Schwarzschild precession in the orbit of the star
  S2 near the Galactic centre massive black hole}.
\newblock {\em Astron. Astrophys.}, 636:L5, 2020.

\bibitem{refId0}
{The GRAVITY Collaboration}, {Abuter, R.}, {Amorim, A.}, {Baub\"ock, M.},
  {Berger, J. P.}, {Bonnet, H.}, {Brandner, W.}, {Cl\'enet, Y.}, {Coud\'e du
  Foresto, V.}, {de Zeeuw, P. T.}, {Dexter, J.}, {Duvert, G.}, {Eckart, A.},
  {Eisenhauer, F.}, {F\"orster Schreiber, N. M.}, {Garcia, P.}, {Gao, F.},
  {Gendron, E.}, {Genzel, R.}, {Gerhard, O.}, {Gillessen, S.}, {Habibi, M.},
  {Haubois, X.}, {Henning, T.}, {Hippler, S.}, {Horrobin, M.},
  {Jim\'enez-Rosales, A.}, {Jocou, L.}, {Kervella, P.}, {Lacour, S.},
  {Lapeyr\`ere, V.}, {Le Bouquin, J.-B.}, {L\'ena, P.}, {Ott, T.}, {Paumard,
  T.}, {Perraut, K.}, {Perrin, G.}, {Pfuhl, O.}, {Rabien, S.}, {Rodriguez
  Coira, G.}, {Rousset, G.}, {Scheithauer, S.}, {Sternberg, A.}, {Straub, O.},
  {Straubmeier, C.}, {Sturm, E.}, {Tacconi, L. J.}, {Vincent, F.}, {von
  Fellenberg, S.}, {Waisberg, I.}, {Widmann, F.}, {Wieprecht, E.}, {Wiezorrek,
  E.}, {Woillez, J.}, and {Yazici, S.}
\newblock A geometric distance measurement to the galactic center black hole
  with 0.3\% uncertainty.
\newblock {\em A\&A}, 625:L10, 2019.

\end{thebibliography}

\end{document}